\documentclass[10pt,twocolumn,letterpaper]{article}

\usepackage[pagenumbers]{cvpr} 

\usepackage{graphicx}
\usepackage{float}
\usepackage{amsmath}
\usepackage{amssymb}
\usepackage{booktabs}
\usepackage{longtable}

\usepackage[pagebackref,breaklinks,colorlinks]{hyperref}

\usepackage[capitalize]{cleveref}
\crefname{section}{Sec.}{Secs.}
\Crefname{section}{Section}{Sections}
\Crefname{table}{Table}{Tables}
\crefname{table}{Tab.}{Tabs.}

\begin{document}

\onecolumn
\title{\textbf{\textit{Space AI}} \\ Leveraging Artificial Intelligence for Space to Improve Life on Earth 
}
\author{Ziyang Wang\\
{\tt\small ziyangwang@ieee.org}
}
\maketitle

\begin{figure*}[htbp]
  \centering
  \includegraphics[width=\linewidth]{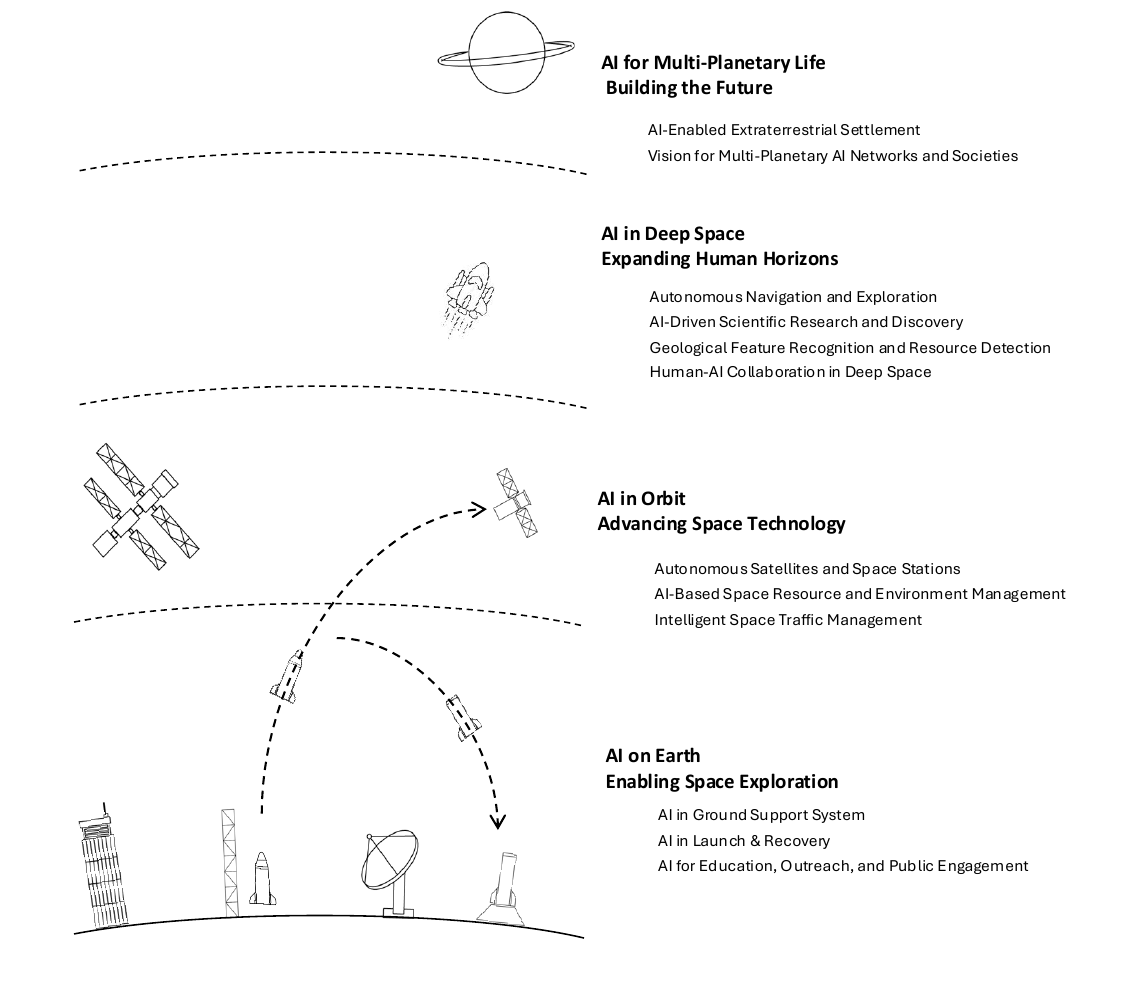}
  \caption{Space AI: Leveraging Artificial Intelligence for Space to Improve Life on Earth.}
  \label{fig:brief}
\end{figure*}
\clearpage
\tableofcontents

\clearpage
\twocolumn
\begin{abstract}

Artificial Intelligence (AI) is rapidly transforming a broad range of human activities, from healthcare and agriculture to finance, biotechnology, and industry. As Earth’s technological frontiers begin to plateau, the next inevitable leap for humanity lies in the exploration and utilization of outer space—a domain offering virtually limitless resources, territory, and opportunities. In this grand transition, AI emerges not merely as a tool but as a central enabler, capable of addressing the extreme complexity, uncertainty, and autonomy required for sustainable space operations.

This paper introduces \textbf{\textit{Space AI}} as a unified interdisciplinary field at the intersection of artificial intelligence and space science and technology. We consolidate historical developments and contemporary progress, and propose a systematic framework that organises Space AI into four mission contexts: (1) \textbf{AI on Earth}, covering intelligent mission planning, spacecraft design optimisation, simulation, and ground-based data analytics; (2) \textbf{AI in Orbit}, focusing on satellite and station autonomy, space robotics, on-board/near-real-time data processing, communication optimisation, and orbital safety; (3) \textbf{AI in Deep Space}, enabling autonomous navigation, adaptive scientific discovery, resource mapping, and long-duration human--AI collaboration under communication constraints; and (4) \textbf{AI for Multi-Planetary Life}, supporting in-situ resource utilisation, habitat and infrastructure construction, life-support and ecological management, and resilient interplanetary networks.

By providing a clear scope, taxonomy, and mission-aligned viewpoint, this work positions Space AI as both a practical engineering agenda and a coherent research field. We also outline key open challenges spanning robust autonomy, verification and validation, safety and cybersecurity, and ethical and regulatory governance. Ultimately, Space AI can accelerate humanity's capability to explore and operate in space, while translating advances in sensing, robotics, optimisation, and trustworthy AI into broad societal impact on Earth.

\end{abstract}

\begin{figure*}[htbp]
  \centering
  \includegraphics[width=\linewidth]{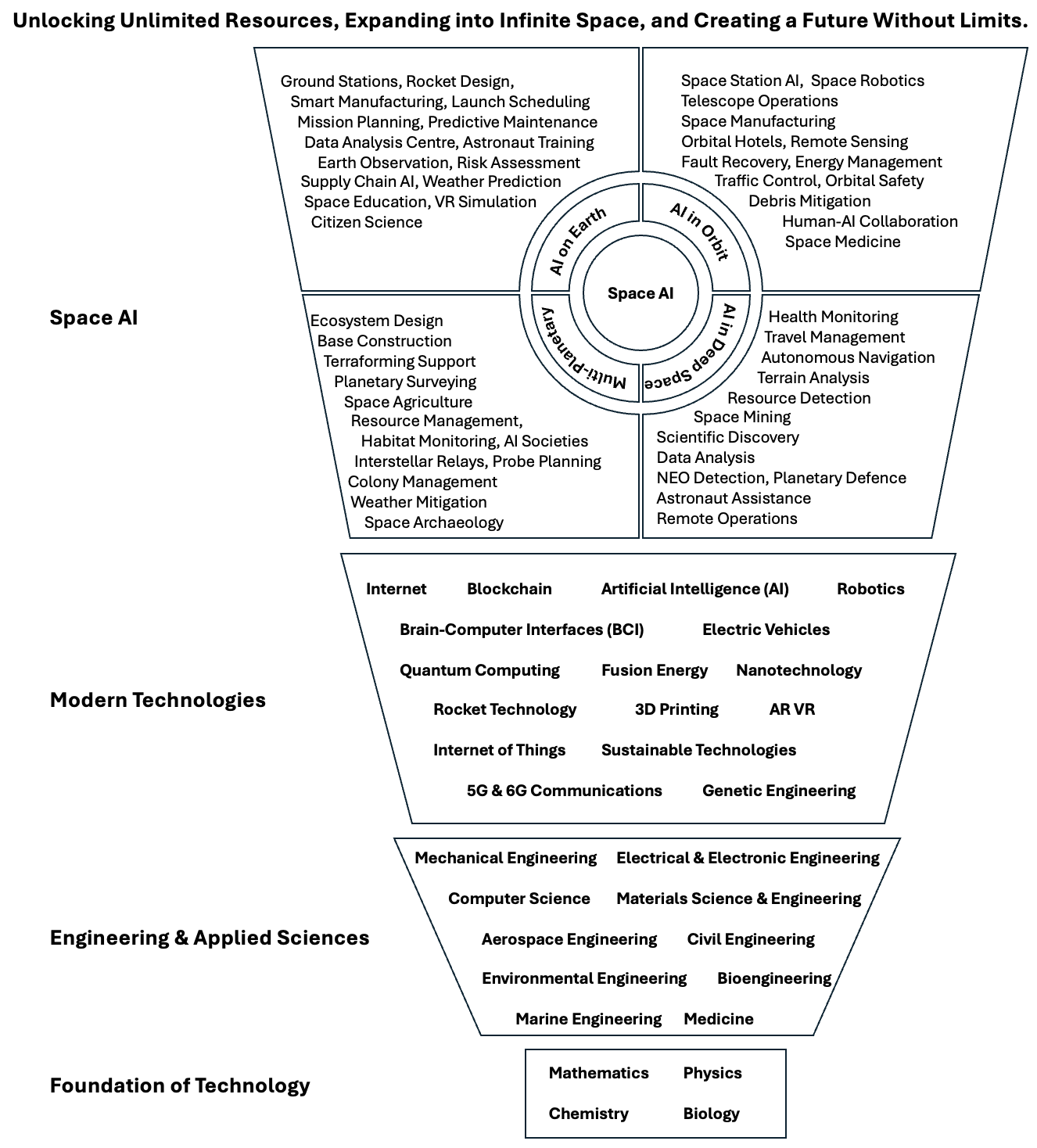}
  \caption{Space AI: A Hierarchical Framework of Knowledge, Technology, and Exploration.}
  \label{fig:risingfigure}
\end{figure*}

\section{Introduction}

Artificial Intelligence (AI) has become a foundational technology for intelligent perception, decision-making, and control in complex environments. Broadly, AI refers to systems and algorithms capable of performing tasks typically associated with human intelligence, including reasoning, learning from experience, problem solving, and adaptive interaction with their environment \cite{turing2009computing,russell2016artificial,goertzel2014artificial,lake2017building}. Over the past decade, advances in machine learning---particularly deep learning---have enabled major progress across a range of domains. In healthcare, AI supports diagnostics, medical imaging, and personalised treatment \cite{bajwa2021artificial,alowais2023revolutionizing,shen2017deep}; in industry, it drives autonomous manufacturing, quality control, and predictive maintenance \cite{wang2019intelligence,arinez2020artificial}; in agriculture, it enables precision farming, yield forecasting, and resource optimisation \cite{talaviya2020implementation,sood2022artificial,pandey2024towards}; and in creative practice, generative models increasingly contribute to art, music, and design workflows \cite{natale2024lovelace,hernandez2022music}. These advances have also fostered specialised subfields---such as \textit{Medical AI}, \textit{Industrial AI}, \textit{Agricultural AI}, \textit{Embodied AI}, \textit{Creative/Artistic AI}, and \textit{AI for Science}---each developing distinct research communities, benchmarks, and toolchains.

Despite this rapid progress, many established AI subfields have been primarily shaped by terrestrial assumptions: abundant opportunities for data collection, accessible maintenance and upgrades, high-bandwidth communication, and frequent human-in-the-loop oversight. Space missions, by contrast, operate under fundamentally different constraints, including extreme environments (radiation, vacuum, thermal cycling), limited energy budgets, restricted communication bandwidth with long latency, and prolonged requirements for autonomous operation. Space science and technology---encompassing areas such as astrophysics, planetary science, astronautics, astrobiology, and satellite systems \cite{zombeck2006handbook,borner2005astronomy}---seeks to answer fundamental questions about the universe while advancing capabilities for sustained operation beyond Earth. Achieving these goals increasingly demands intelligent, adaptive, and verifiable autonomy at scale.

Although AI has been used in a variety of space-related systems, a unified and systematic field---\textbf{\textit{Space AI}}---has not yet been clearly defined or institutionalised. Existing efforts are often mission-specific and fragmented, developed independently across agencies and industrial organisations without a coherent disciplinary framework \cite{oche2021applications,russo2022using,gao2017review,richards2023safely,jasiobedzki1998computer,matthies2007computer,kodheli2020satellite,fourati2021artificial,cuna2025application,rezaei2025bridging,fotopoulou2024review}. As contemporary missions grow in complexity and ambition, there is a timely need to define Space AI as a distinct, integrated research and engineering domain.

This paper addresses the gap by formally introducing \textbf{\textit{Space AI}} as an interdisciplinary paradigm at the intersection of artificial intelligence and space science/technology. We clarify its motivation, define its scope and boundaries, synthesise key application areas, and articulate its strategic significance for future space capabilities and Earth benefits. We also propose a structured framework that organises Space AI into four mission contexts, as illustrated in Figure~\ref{fig:brief}.

\subsection{Motivation}

Space activities are entering an era characterised by rapidly increasing scale, autonomy demands, and operational complexity. A visual snapshot of the growing importance and impact of global space activity is summarised in Figure~\ref{fig:risingfigure}. Contemporary and planned missions---including autonomous planetary rovers, dense satellite constellations, on-orbit servicing, and prospective crewed deep-space exploration---often exceed what can be managed through continuous human oversight alone \cite{zilberstein2003decision,thangavel2024artificial}. Communication delays, uncertain environments, and time-critical decision-making make purely manual control or rigid pre-programmed behaviours inadequate for many next-generation scenarios. Under these conditions, AI becomes a practical enabler for robust autonomy, adaptive decision-making, dynamic resource allocation, and sustained operations.

AI-enabled autonomy is already evident in several influential mission capabilities. NASA’s Mars rovers (e.g., Curiosity and Perseverance) employ machine vision, onboard hazard assessment, and autonomous navigation to traverse unstructured terrain and pursue scientific targets under significant communication latency \cite{agrawal2021analyzing}. In Earth orbit, large constellations (e.g., Starlink) increasingly rely on automated conjunction assessment and collision-avoidance workflows to operate safely in congested orbital regimes \cite{mcdowell2020low}. Reusable launch systems also demonstrate the value of advanced guidance and control: for example, Falcon 9 booster recovery integrates real-time sensing, navigation, and closed-loop control for precision landing; recent studies further suggest that learning-based guidance can achieve competitive performance in simulated descent and landing scenarios \cite{sanchez2018real}. Collectively, these examples illustrate that intelligence is becoming a core requirement for safety, efficiency, and scalability across space operations.

Beyond engineering considerations, Space AI also has strategic and governance implications. As autonomous capabilities expand, questions of reliability assurance, cybersecurity, accountability, orbital sustainability, and the regulation of autonomous systems become increasingly urgent \cite{carlo2023importance,hoadley2018artificial,horowitz2022artificial}. A coherent field definition can therefore support not only technical progress but also shared standards, evaluation practices, and responsible deployment.

Accordingly, we argue that Space AI should be established as a formal interdisciplinary field rather than treated as a collection of isolated ``AI in space'' applications. A unified research agenda can accelerate progress in trustworthy autonomy, onboard intelligence, multi-agent coordination, verification and validation, and human--AI teaming across mission phases. This paper provides a systematic conceptual framework and an operational taxonomy (Figure~\ref{fig:brief}) intended to support coherent education, sustained funding, and cross-community collaboration.

\subsection{Objectives}

This paper defines \textit{Space AI} as a distinct domain that integrates AI methodologies, space-mission requirements, and governance considerations into a cohesive framework. Our objectives are fourfold.

\textbf{First}, we articulate a clear definition, scope, and boundary conditions for Space AI as a standalone discipline, distinguishing it from ad-hoc ``AI + aerospace'' workstreams.

\textbf{Second}, we survey and synthesise AI’s roles across the full spectrum of mission contexts, from ground systems and launch operations to orbital platforms, deep-space exploration, and long-duration human--AI collaboration. This end-to-end viewpoint emphasises transferability and synergy across contexts, rather than treating each application in isolation.

\textbf{Third}, we identify the major challenges and future opportunities that determine whether Space AI can become reliable and responsible in practice. Key issues include robustness under radiation and extreme environments, limited energy and compute, verification and validation of autonomous decision-making, cybersecurity, and the ethics of delegating high-stakes decisions to machines. We also discuss the need for mature standards and regulatory frameworks aligned with the realities of autonomous space systems.

\textbf{Fourth}, we outline strategies for cross-disciplinary collaboration and global governance to support sustainable growth of the field, spanning academia, industry, and governmental stakeholders. We argue that coordinated efforts are required to develop common benchmarks, reliability standards, safety cases, and governance principles for AI-enabled space operations.

\subsection{Definition and Scope of Space AI}

Space AI (also referred to as \textit{AI for Space}) concerns the systematic research, development, deployment, and governance of AI technologies tailored to the constraints and opportunities of space environments. It is not merely the application of AI to isolated aerospace tasks; rather, it aims to integrate machine intelligence with space systems engineering to enable autonomous, adaptive, and trustworthy operations across mission lifecycles.

We define Space AI as:

\begin{quote}
\emph{``Space AI is an interdisciplinary field that develops and assures AI-enabled autonomy for space systems, spanning mission design, onboard decision-making, and operations across Earth, orbit, deep space, and multi-planetary infrastructure, with an emphasis on safety, reliability, and sustainability.''}
\end{quote}

To provide a structured view, we organise Space AI into four mission contexts:

\begin{itemize}
    \item \textbf{AI on Earth}: terrestrial AI capabilities that support space activities, including simulation, mission planning and scheduling, launch and recovery operations, spacecraft design optimisation, large-scale space data analytics, and education/outreach platforms.
    \item \textbf{AI in Orbit}: onboard and near-real-time intelligence for satellites and orbital platforms, including autonomous operations, fault detection and recovery, communication and spectrum optimisation, remote sensing and Earth observation, debris monitoring and collision avoidance, and robotic assembly/maintenance.
    \item \textbf{AI in Deep Space}: autonomy for deep-space probes and planetary systems under communication latency, including navigation, terrain understanding, adaptive scientific discovery, resource mapping, planetary defence (e.g., NEO tracking), and decision support in uncertain environments.
    \item \textbf{AI for Multi-Planetary Life}: AI systems enabling sustained off-world habitation, including habitat construction, in-situ resource utilisation (ISRU), life-support and environmental control, ecological modelling, resilient interplanetary networks, and long-term societal and heritage considerations.
\end{itemize}

Technically, Space AI spans machine learning, deep learning, reinforcement learning, robotics and autonomous systems, computer vision, natural language processing, multi-agent systems, edge AI, and trustworthy/explainable AI. It also intersects with aerospace engineering, planetary science, human factors, systems engineering, international space law and policy, and ethics. By articulating Space AI as a coherent domain, this paper aims to support interdisciplinary collaboration and provide a strategic reference point for academia, industry, and government.

\subsection{Structure of the Paper}

This paper is organised into seven sections:
\begin{itemize}
    \item \textbf{Section 2} reviews key milestones in space exploration and AI, motivating their convergence.
    \item \textbf{Section 3} surveys \textbf{AI on Earth} for mission planning, launch/recovery, ground support, and outreach.
    \item \textbf{Section 4} surveys \textbf{AI in Orbit} for autonomous satellites/stations, communications, robotics, and orbital safety.
    \item \textbf{Section 5} surveys \textbf{AI in Deep Space} for autonomous navigation, scientific discovery, geological analysis, and human--AI collaboration.
    \item \textbf{Section 6} surveys \textbf{AI for Multi-Planetary Life} for settlement, ISRU, infrastructure, and interplanetary networks.
    \item \textbf{Section 7} concludes with trends, regulations and standards, open challenges, and future opportunities for Space AI.
\end{itemize}

Ultimately, this work aims to position Space AI as a practical engineering agenda and an emerging research field with clear scope, taxonomy, and governance considerations, accelerating space capabilities while translating advances into societal impact on Earth.

To keep up with the rapid development in this field, we will regularly update this review with the latest relevant papers and develop an open-source implementation at \url{https://github.com/ziyangwang007/AI4Space}.

\begin{figure*}[htbp]
    \centering
    \includegraphics[width=\linewidth]{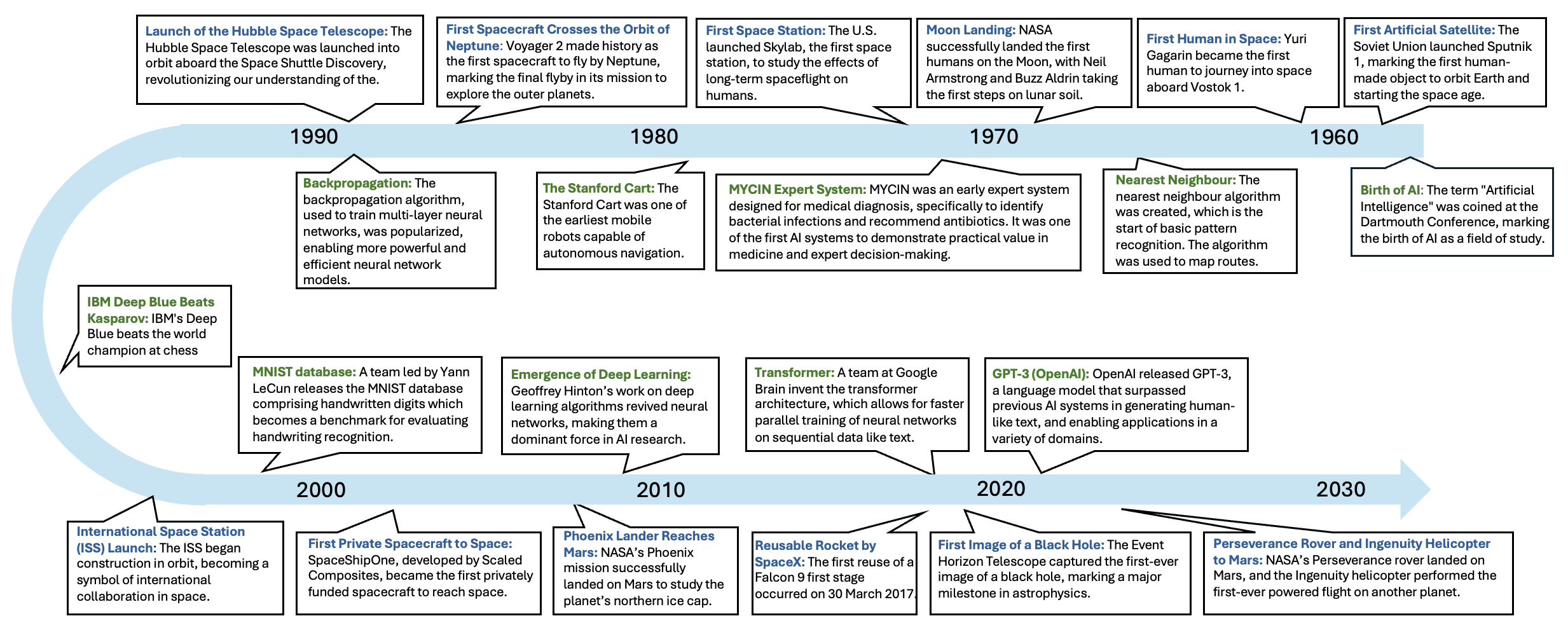}
    \caption{The Overview of Key Milestones in \textcolor{blue}{Space Exploration} \& \textcolor{green}{Artificial Intelligence}.}
    \label{fig:milestone}
\end{figure*}

\begin{table*}[!h]\centering
\caption{Selected key milestones in the history of space exploration and artificial intelligence, showing the parallel progress of these two domains.\label{tab:space-ai-milestones}}
\begin{tabular}{p{1cm}p{7.3cm}p{1cm}p{7.3cm}}
\hline
\textbf{Year} & \textbf{Space Exploration Milestone} & \textbf{Year} & \textbf{Artificial Intelligence Milestone} \\
\hline
1957 & Sputnik 1 launches, becoming the first artificial Earth satellite & 1950 & Alan Turing proposes a test for machine intelligence (the “Turing test”) \\
1961 & Yuri Gagarin orbits Earth, becoming the first human in space & 1952 & Arthur Samuel programs a computer to play checkers and learn to improve, an early breakthrough in machine learning \\
1963 & Valentina Tereshkova flies on Vostok 6 as the first woman in space & 1956 & The Dartmouth workshop coins the term “artificial intelligence,” inaugurating AI as a formal field of study \\
1965 & Aleksey Leonov performs the first spacewalk (extravehicular activity) & 1957 & Frank Rosenblatt introduces the perceptron, the first artificial neural network capable of learning \\
1969 & Apollo 11 lands astronauts on the Moon (first human steps on another celestial body) & 1966 & Joseph Weizenbaum’s ELIZA program debuts as the first chatbot simulating human-like conversation \\
1970 & Venera 7 becomes the first probe to soft-land on another planet (Venus), transmitting data from the surface & 1969 & SRI’s Shakey robot navigates and plans actions autonomously — the first mobile robot to reason about its own actions \\
1971 & Salyut 1, the world’s first space station, is launched into Earth orbit & 1972 & Development begins on MYCIN, an early expert system for medical diagnosis \\
1976 & Viking 1 lands on Mars and sends back the first photographs from its surface & 1974 & The Lighthill Report critiques AI research in the UK, triggering the first “AI winter” \\
1981 & Space Shuttle \textit{Columbia} completes its first flight, inaugurating reusable crewed spacecraft & 1979 & Stanford Cart navigates a cluttered room using computer vision — a self-driving precursor \\
1989 & Voyager 2 completes the “Grand Tour” of the outer planets with a flyby of Neptune & 1980 & XCON expert system is deployed to configure VAX computers, a landmark in commercial AI \\
1990 & Hubble Space Telescope is deployed, offering unprecedented astronomical imaging & 1986 & Backpropagation is popularized for training deep neural networks, reviving interest in connectionism \\
2000 & Expedition 1 boards the ISS, beginning continuous human habitation in space & 1997 & IBM’s Deep Blue defeats world chess champion Garry Kasparov \\
2004 & \textit{SpaceShipOne} conducts the first privately funded human spaceflight & 1998 & NASA’s Remote Agent becomes the first AI system to autonomously control a spacecraft (Deep Space 1) \\
2005 & ESA’s Huygens probe lands on Titan — the first landing on an outer solar system moon & 2005 & Stanford’s Stanley wins the DARPA Grand Challenge, driving 132 miles autonomously \\
2014 & Philae lander performs the first soft landing on a comet (67P/Churyumov–Gerasimenko) & 2011 & IBM’s Watson defeats human champions on \textit{Jeopardy!}, demonstrating deep NLP capabilities \\
2015 & New Horizons performs the first Pluto flyby, returning high-resolution images & 2012 & AlexNet wins ImageNet, sparking the deep learning revolution in computer vision \\
2015 & SpaceX lands a Falcon 9 first-stage booster after orbital launch, achieving reusability & 2015 & Tesla releases Autopilot, bringing AI-assisted driving to consumer vehicles \\
2019 & Chang’e 4 lands on the Moon’s far side — the first mission to explore this region & 2016 & DeepMind’s AlphaGo defeats Go world champion Lee Sedol, marking a major AI milestone \\
2020 & Crew Dragon \textit{Demo-2} launches astronauts to the ISS — first crewed flight by a private company & 2020 & AlphaFold 2 predicts protein structures with atomic accuracy, solving a grand challenge in biology \\
2021 & Ingenuity helicopter achieves the first powered flight on another planet (Mars) & 2020 & OpenAI’s GPT-3.5 model demonstrates advanced natural language generation with 175B parameters \\
2021 & SpaceX’s Inspiration4 conducts the first all-civilian orbital spaceflight & 2022 & ChatGPT (built on GPT-3) brings generative AI into mainstream use \\
2022 & Artemis I loops around the Moon and returns, marking the start of NASA’s new lunar program & 2023 & GPT-4 achieves human-level performance on professional exams, redefining the limits of general-purpose AI \\
\hline
\end{tabular}
\label{tab:milestone}
\end{table*}

\section{Key Milestones in Space Exploration \& Artificial Intelligence}

The evolution of space exploration and AI has proceeded in parallel for over seven decades, with breakthroughs in each field often shaping and inspiring progress in the other. This section highlights pivotal moments that defined the trajectory of space science and AI research, from early orbital missions and lunar landings to foundational algorithms and intelligent systems. Table~\ref{tab:space-ai-milestones} and Figure \ref{fig:milestone} briefly summarizes these milestones chronologically, illustrating the intertwined development of these two transformative domains. In the following subsections, we provide a structured overview of this progression: first, by reviewing major historical, technological, and strategic goals in space exploration; then by examining the evolution of AI; and finally by tracing the growing intersection between the two in past, current, and emerging space missions.

\subsection{Key Milestones in Space Exploration}

\subsubsection{Historical Milestones}

In October 1957, the Soviet Union launched \textit{Sputnik 1}, the first artificial satellite to orbit Earth, marking the dawn of the Space Age \cite{cospar1958sputnik,kuznetsov2015yakov}. This milestone catalyzed the creation of new institutions for space research. In the United States, President Eisenhower signed the legislation establishing NASA in 1958, with the agency officially beginning operations on October 1, 1958 \footnote{\url{https://www.nasa.gov/organizations/national-aeronautics-and-space-act POLICY}}. The early space race quickly progressed to human spaceflight: on April 12, 1961, cosmonaut Yuri Gagarin became the first person to orbit Earth aboard \textit{Vostok 1} \footnote{\url{https://www.nasa.gov/image-article/april-1961-first-human-entered-space/}}. In 1969, NASA’s Apollo 11 mission landed astronauts Neil Armstrong and Buzz Aldrin on the Moon, fulfilling humanity’s ancient aspiration to walk on another celestial body \footnote{\url{https://www.nasa.gov/mission/apollo-11/}}. 

During this transformative period, major space agencies were established to consolidate expertise: the European Space Agency (ESA) was founded in 1975 \footnote{\url{https://www.esa.int/Newsroom/About_the_European_Space_Agency}}, uniting European efforts in space, and the China National Space Administration (CNSA) was established in 1993 to guide China’s expanding space capabilities \footnote{\url{https://www.cnsa.gov.cn/english/}}. These early milestones established the foundational infrastructure, mission architectures, and technological capabilities that modern space programs build upon—laying the groundwork for a future in which artificial intelligence will play a critical role in supporting the complexity and autonomy of next-generation missions.

\subsubsection{Technological Breakthroughs}

The steady evolution of launch systems, spacecraft, and mission operations has enabled humanity to extend its reach deeper into space. Rocket technology progressed from early converted ICBMs to advanced launch vehicles like the Saturn V (which remains one of the most powerful rockets ever flown) \footnote{\url{https://science.nasa.gov/3d-resources/saturn-v-rocket/}}. The 1980s introduced the Space Shuttle (first launched in 1981) \footnote{\url{https://www.nasa.gov/space-shuttle/}}, the world’s first partially reusable crewed spacecraft, which flew 135 missions over 30 years and played a key role in satellite deployment and space station construction. Today, modern reusable rockets such as SpaceX’s Falcon 9 \footnote{\url{https://www.spacex.com/vehicles/falcon-9}} and the forthcoming Starship \footnote{\url{https://www.spacex.com/vehicles/starship}} are further reducing costs and increasing launch cadence. In parallel, mastery of orbital mechanics and interplanetary trajectories has allowed missions to reach distant planets via techniques like gravity assists and trans-lunar injections.

Advances in satellite technology were equally transformative. Early weather and communications satellites of the 1960s laid the foundation for today’s GPS \footnote{\url{https://www.nasa.gov/directorates/somd/space-communications-navigation-program/gps/}}, remote sensing, and broadband networks. These platforms increasingly rely on onboard autonomy for station-keeping, fault detection, and network coordination, paving the way for AI-enhanced constellations. Meanwhile, planetary exploration became more ambitious: the Soviet Luna 9 \footnote{\url{https://lunarexploration.esa.int/explore/missions/239?ia=334}} achieved the first soft landing on the Moon in 1966, and NASA’s Viking 1 \footnote{\url{https://science.nasa.gov/mission/viking-1/}} became the first successful Mars lander in 1976. The 1977 launches of Voyager 1 and Voyager 2 \footnote{\url{https://science.nasa.gov/mission/voyager/}} opened the era of deep-space probes, with onboard systems that could even be reprogrammed en route — an early form of adaptive spacecraft intelligence.

By the late 1990s, autonomous spacecraft control moved from concept to reality. In 1998, NASA’s Deep Space 1 \footnote{\url{https://science.nasa.gov/mission/deep-space-1/}} mission successfully tested the Remote Agent \cite{bernard1999spacecraft,muscettola1998remote}, the first artificial intelligence software to autonomously command a spacecraft. This breakthrough foreshadowed the sophisticated onboard decision-making now being developed for future probes. By the early 2000s, planetary rovers demonstrated advanced autonomy on the Martian surface. NASA’s twin Spirit and Opportunity rovers (landed 2004), and later the larger Curiosity rover (landed 2012), were equipped with systems to avoid hazards, plan routes, and even conduct limited science observations without real-time human input. These advances were made possible by the miniaturization of hardware, improved onboard computing, and more robust software architectures. Autonomy matured from simple reactive control loops to intelligent, goal-driven behavior — a critical precursor to deploying fully capable AI in space missions. The increasing need for autonomy in harsh, distant, or communication-delayed environments created a technical imperative for AI to move from an Earth-bound support role to an in situ decision-maker.

\subsubsection{Future Goals}

Today, national and international space agencies are embarking on a new era of ambitious, long-duration, and multi-planetary missions. NASA’s Artemis program \footnote{\url{https://www.nasa.gov/humans-in-space/artemis/}}, developed in collaboration with ESA, JAXA, and other partners, aims to return humans to the Moon and lay the groundwork for future crewed missions to Mars \cite{bains2022has}. The program includes establishing lunar infrastructure — such as surface habitats, rovers, and the Gateway lunar-orbit space station — and utilizing lunar resources to support long-term exploration. ESA is contributing key components (for example, the Orion spacecraft’s service module) and jointly planning robotic lunar missions. Meanwhile, China’s lunar program is progressing through the Chang’e series of robotic probes, with plans for a crewed Moon landing by 2030 and a Mars mission by 2033.

In Earth orbit, the long-term occupation of the ISS continues, while China’s new space station Tiangong \footnote{\url{https://www.planetary.org/space-missions/chinese-space-station}} serves as an independent laboratory for human spaceflight and microgravity science. Both platforms increasingly depend on intelligent systems to manage life support, navigation, and experiments. Proposed commercial stations and the Lunar Gateway will further increase operational complexity, necessitating higher degrees of autonomy and resilience — natural use cases for AI.

Private industry also plays a growing role. SpaceX’s fully reusable Starship vehicle is envisioned to carry people and cargo to Mars. Elon Musk has described a long-term goal of establishing a self-sustaining Martian colony, which would demand intelligent systems for everything from habitat maintenance to agriculture and communications. Even in near-term exploration, AI is crucial: autonomous guidance, hazard avoidance, and real-time decision support will be essential for Mars surface operations, where communication delays make Earth-based control impractical.

Resource extraction from asteroids and other bodies represents another frontier. Missions like NASA’s OSIRIS-REx, which returned samples from asteroid Bennu in 2023, have demonstrated the feasibility of robotic rendezvous and sample return. Commercial ventures — such as Planetary Resources and Deep Space Industries — have proposed asteroid mining operations to supply propellant and raw materials for in-space infrastructure. Such scenarios depend on highly autonomous systems capable of operating without continuous Earth supervision in hazardous, unstructured environments.

Across all these goals, a common theme emerges: the scale, complexity, and remoteness of future missions will necessitate a paradigm shift from human-in-the-loop operations to AI-augmented autonomy. Tomorrow’s missions cannot rely solely on pre-scripted instructions or delayed ground control; they require systems that perceive, learn, adapt, and make decisions in real time. In short, “Space AI” will be indispensable — not only as a technological upgrade, but as a structural enabler of the next phase of space exploration. The cumulative achievements of past decades have laid the groundwork; the future will be shaped by how successfully we embed intelligence into the core of space systems.

\subsection{Evolution of Artificial Intelligence} 

AI has progressed through distinct paradigms since its inception, each era characterized by different approaches and breakthroughs. Broadly, the evolution can be delineated into four phases: the \textit{Symbolic AI Era}, the \textit{Machine Learning Era}, the \textit{Modern AI Era}, and anticipated \textit{Future Directions}. In this section, we review these phases and their defining features, highlight key AI-driven breakthroughs in medicine, agriculture, and science, and discuss ongoing technical and philosophical challenges facing the field.

\subsubsection{Brief History and Evolution of AI}

\paragraph{Symbolic AI Era (1950s--1980s).}  
Early AI systems were built on symbolic logic and rule-based reasoning. Programs such as the \textit{Logic Theorist} and \textit{MYCIN} exemplified expert systems that codified human knowledge into if-then rules. These systems performed well in narrow domains but lacked adaptability, scalability, and robustness in uncertain or dynamic environments. Limitations in knowledge engineering and the brittleness of logic-based systems led to the first and second ``AI winters'' due to unmet expectations.

\paragraph{Machine Learning Era (1990s--2010s).}  
AI transitioned toward statistical learning and data-driven models. This era introduced techniques such as neural networks (via backpropagation), support vector machines, and ensemble methods like random forests. The rise of big data and increased computing power fueled predictive modeling. Notable achievements included IBM’s \textit{Deep Blue} defeating a world chess champion and the widespread adoption of AI in language, vision, and recommendation systems.

\paragraph{Modern AI Era (2010s--present).}  
The resurgence of neural networks, powered by deep learning, transformed AI capabilities. Convolutional neural networks (CNNs) revolutionized vision tasks, while recurrent networks and transformers redefined natural language processing. Reinforcement learning successes such as \textit{AlphaGo} demonstrated strategic AI. The emergence of foundation models like GPT-3 and BERT exemplify the power of pretraining and transfer learning across domains. Generative models (GANs, diffusion models) enable content synthesis, while transformers underpin state-of-the-art performance in multiple modalities.

\paragraph{Future Directions.}  
Research is expanding into neuromorphic computing, quantum machine learning, continual learning, and AGI. Neuromorphic chips aim to emulate biological efficiency and adaptability. Quantum algorithms may one day accelerate training and inference. Cognitive architectures and hybrid symbolic-neural systems seek to achieve broader, more generalizable intelligence with interpretability and adaptability.

\subsubsection{Key Advances of AI in Related Domains}

\paragraph{Medical AI.}  
AI now supports diagnostic imaging, electronic health records, and personalized medicine. Deep learning models identify anomalies in radiology and pathology with expert-level accuracy. Transfer learning enables effective model performance with limited medical data, particularly for rare conditions. AI also accelerates drug discovery, supports robotic-assisted surgery, and enables continuous health monitoring in resource-constrained or remote environments.

\paragraph{Agricultural AI.}  
Precision agriculture benefits from AI in crop monitoring, disease detection, soil analysis, and yield prediction. Autonomous farm equipment leverages computer vision and sensor fusion for navigation, planting, spraying, and harvesting. These advances support sustainable agriculture by optimizing resource usage—such as water, fertilizer, and energy—while enhancing resilience to climate variability and improving food security.

\paragraph{Industrial AI.}  
In manufacturing and industry, AI is central to predictive maintenance, quality control, production scheduling, and robotic automation. Computer vision systems detect product defects in real time, while reinforcement learning agents optimize factory throughput. Digital twins and AI-driven simulations improve design, monitoring, and lifecycle management of complex engineering systems. Industrial AI plays a key role in the development of Industry 4.0 and cyber-physical systems.

\paragraph{Artistic AI.}  
AI is increasingly used in creative fields such as music, visual art, literature, and design. Generative models—including GANs, diffusion models, and large language models—are capable of producing human-like artworks, music compositions, and poetry. These systems not only enable novel forms of artistic expression, but also support human creators through co-creative tools, style transfer, and real-time content generation.

\paragraph{Embodied AI.}  
Embodied AI refers to systems where intelligence is grounded in physical interaction with the world, such as robots and autonomous agents. Applications range from household assistants and warehouse logistics robots to quadrupeds and humanoid machines. Embodied AI integrates perception, action, and reasoning in dynamic environments, often leveraging reinforcement learning and sim-to-real transfer. This domain bridges AI with robotics and is foundational to applications in exploration, mobility, and human-robot collaboration.

\paragraph{AI for Science.}  
AI accelerates scientific discovery across disciplines. In molecular biology, systems like AlphaFold have solved protein folding with near-experimental accuracy. In physics and materials science, AI helps discover new compounds, optimize experiments, and simulate complex systems. Climate science benefits from transformer-based and graph-based models for weather forecasting and climate modeling. More broadly, AI assists in data analysis, anomaly detection, hypothesis generation, and experimental design across astronomy, chemistry, neuroscience, and economics.

\paragraph{AI for Security and Defense.}  
AI technologies are increasingly integrated into national security, cybersecurity, and defense operations. Applications include autonomous surveillance, threat detection, decision support systems, and cyberattack prevention. In military contexts, AI is used for situational awareness, battlefield robotics, and strategic planning—though such applications raise ethical and regulatory concerns that continue to be debated internationally.

\paragraph{AI for Education.}  
AI is transforming education through personalized learning platforms, intelligent tutoring systems, and automated assessment. Natural language processing enables real-time feedback, question answering, and adaptive content delivery tailored to each learner’s pace and performance. AI also supports teacher workloads through grading assistants, curriculum generation, and educational data analytics.

\paragraph{AI for Sustainability and Energy.}  
In environmental monitoring and energy systems, AI contributes to optimizing power grids, forecasting renewable energy output, and modeling ecosystems. AI is also used for smart building control, carbon accounting, water management, and circular economy systems, aligning with goals in climate resilience and sustainable development.

\subsubsection{Current Limitations and Challenges of AI Technologies}

\begin{itemize}
    \item \textbf{Computational and Energy Constraints:} Large-scale models demand significant computing resources and power, making deployment difficult in constrained environments. Model compression, edge AI, and neuromorphic hardware are emerging solutions.
    
    \item \textbf{Data Dependency and Generalization:} AI models often require large, labeled datasets and struggle with rare or out-of-distribution cases. Techniques such as few-shot learning, synthetic data, and domain adaptation address these gaps.

    \item \textbf{Explainability and Trust:} Most high-performance AI systems lack interpretability, complicating deployment in safety-critical domains. Explainable AI (XAI), human-in-the-loop systems, and hybrid models are proposed mitigations.

    \item \textbf{Robustness in Extreme Environments:} AI models degrade under radiation, thermal stress, or sensor noise. Solutions include radiation-hardened processors, self-healing algorithms, and robust training against environmental perturbations.

    \item \textbf{Latency and Autonomy:} Communication delays in remote operations require real-time AI inference and decision-making. Efficient, autonomous edge AI systems with embedded safety constraints are essential in such scenarios.

    \item \textbf{Ethical and Regulatory Issues:} Bias in datasets, lack of governance, and risks of misuse challenge safe AI deployment. International standards and ethical frameworks are critical for responsible innovation.
\end{itemize}

In summary, AI has evolved from symbolic reasoning to powerful learning systems capable of perception, prediction, and generation. While transformative, current AI technologies face challenges in efficiency, transparency, and reliability. Future directions in AI research seek to address these limitations and expand the boundaries of what intelligent systems can achieve.

\subsection{When AI Meet with Space} 

AI has increasingly become integral to space exploration, enabling spacecraft and astronauts to operate more autonomously and efficiently. From early robotic probes to upcoming interplanetary missions, AI techniques are being used to navigate alien terrains, manage complex spacecraft systems, analyze vast streams of data, and even assist human crews. This overview surveys representative examples of AI in space missions – first looking at historical applications by agencies like NASA, ESA, and CNSA – and then examining near-future missions where AI is poised to play a major role. The trend across these examples underlines the emergence of "Space AI" as a coherent field combining space engineering and advanced autonomy.

\subsubsection{Historical Applications of AI in Space Missions}

\begin{table*}[htbp]
\small
\setlength{\tabcolsep}{3pt}
\renewcommand{\arraystretch}{0.95}
\caption{Historical Applications of AI in Space Missions (Chronological), Part I.}
\label{tab:space_ai_missions}\label{tab:ai4space}
\centering
\begin{tabular}{p{3.3cm}p{0.9cm}p{1.9cm}p{1.5cm}p{5cm}p{2cm}p{1.2cm}}
\toprule
\textbf{Mission Name} & \textbf{Year} & \textbf{Org.} & \textbf{Type} & \textbf{Info} & \textbf{AI Domain} & \textbf{Ref.}\\
\midrule
Hubble Telescope (SPIKE) & 1990 & NASA/ESA & Government & AI-based scheduling / constraint optimisation for observation planning & Earth & \cite{johnston1990spike,johnston1992spike}\\

Space Shuttle Diagnostics & 1990s & NASA & Government & Expert-system-based fault diagnosis for launch/engine monitoring (recommend adding a primary ref) & Earth & \cite{duyar1992fault} \\

Sojourner (Mars Pathfinder) & 1997 & NASA/JPL & Government & Reactive autonomous obstacle avoidance for surface traversal & Deep Space & \cite{matijevic1998autonomous,sherwood2001autonomously} \\

Deep Space 1 & 1998 & NASA/ JPL & Government & Autonomous planning / execution / fault response (Remote Agent flight validation in 1999) & Deep Space & \cite{taylor2016deep,johnston2017ai} \\

Earth Observing-1 (EO-1) & 2003--2004 & NASA & Government & Onboard event detection and autonomous replanning/retargeting (ASE) & Orbit & \cite{ungar2003overview} \\

ISS Environmental Control & 2000s & NASA & Government & Automated monitoring and control for life-support/environmental systems (control/FDIR emphasis) & Orbit & \cite{gratius2024digital,jones2019controls} \\

PROBA-1 & 2001 & ESA & Government & Autonomous onboard operations / mission automation for a small satellite & Orbit & \cite{duca2008hyperspectral,bermyn2000proba} \\

Spirit \& Opportunity & 2004 & NASA/ JPL & Government & Autonomous navigation (AutoNav) for traversability analysis and hazard avoidance & Deep Space & \cite{lemmon2004atmospheric} \\

Mars Reconnaissance Orbiter & 2005/ 2006 & NASA/JPL & Government & Semi-autonomous fault protection and operations automation & Orbit & \cite{platt2024noise2noise,zurek2007overview} \\

Kepler Telescope & 2009 & NASA & Government & Large-scale exoplanet detection/classification from mission datasets (primarily ground-based ML/AI) & Earth & \cite{audenaert2023artificial,nasaArtificialIntelligence} \\

Curiosity Rover & 2012 & NASA/JPL & Government & Autonomous target selection (AEGIS/ChemCam) and onboard navigation support & Deep Space & \cite{astronomyDoesMars,Matthew2024,tashakori2019trajectory} \\

Falcon 9 Booster & 2015 & SpaceX & Commercial & Autonomous landing / guidance and control for first-stage recovery & Earth & \cite{blackmore2016autonomous} \\
\bottomrule
\end{tabular}
\end{table*}

\begin{table*}[htbp]
\small
\setlength{\tabcolsep}{3pt}
\renewcommand{\arraystretch}{0.95}
\caption{Historical Applications of AI in Space Missions (Chronological), Part II (continued).}
\centering
\begin{tabular}{p{3.3cm}p{0.9cm}p{1.9cm}p{1.5cm}p{5cm}p{2cm}p{1.2cm}}
\toprule
\textbf{Mission Name} & \textbf{Year} & \textbf{Org.} & \textbf{Type} & \textbf{Info} & \textbf{AI Domain} & \textbf{Ref.}\\
\midrule
CIMON Robot (ISS) & 2018 & ESA/DLR/IBM & Government & Crew-assistant robot for voice interaction and cognitive support demonstrations & Multi-Planetary & \cite{airbusHelloCIMON,schmitz2020towards} \\

Ejenta Health Monitoring (ISS) & 2018 & NASA/Ejenta & Government & AI-enabled health monitoring / decision support concepts for astronaut operations & Multi-Planetary & \cite{smith2023off} \\

TESS/Gaia Missions & 2013/ 2018 & NASA/ESA & Government & AI-based celestial classification / cataloguing from large-scale survey data & Earth & \cite{huang5261669machine,andrae2023gaia} \\

Yutu-2 (Chang'e-4) & 2019 & CNSA & Government & Onboard autonomy for lunar rover operations under remote command constraints & Deep Space & \cite{ma2020precise,wang2020computer,zhong2023deep} \\

Starlink Constellation & 2019 & SpaceX & Commercial & Automated/ autonomous conjunction risk mitigation and collision avoidance workflows & Orbit & \cite{spaceSpaceXStarlink,spacePerformingEvasive,hejduk2023conjunction} \\

Crew Dragon & 2020 & SpaceX & Commercial & Autonomous rendezvous and docking with ISS (proximity operations) & Orbit & \cite{collectspaceAstronautTurned} \\

PhiSat-1 ($\Phi$-Sat-1) & 2020 & ESA & Government & Onboard AI inference for cloud detection to reduce downlink volume & Orbit & \cite{giuffrida2021varphi,guerrisi2023artificial,murphy2021machine} \\

Perseverance Landing & 2021 & NASA/JPL & Government & Autonomous terrain-relative navigation during entry, descent, and landing & Deep Space & \cite{nasaMars2020,verma2023autonomous,paton2024inferred,TheNASAPerseveranceMarsLanding,johnson2017lander} \\

Perseverance Rover & 2021 & NASA/JPL & Government & Enhanced autonomous driving and adaptive sampling for in situ science & Deep Space & \cite{nasaPerseverancesSuperCam,nasaHeresChanging,maki2020mars} \\

Zhurong Rover (Tianwen-1) & 2021 & CNSA & Government & Autonomous navigation and hazard avoidance for Martian surface exploration & Deep Space & \cite{tian2021zhurong,wu2022landing,wan2025visual,jia2022high} \\

James Webb Telescope & 2021 & NASA/ESA & Government & AI-enabled ground analysis (e.g., galaxy morphology classification) and anomaly detection in pipelines & Earth & \cite{gardner2006james,robertson2023morpheus} \\

Chandrayaan-3 (Vikram/Pragyan) & 2023 & ISRO & Government & Vision-based hazard detection and avoidance for autonomous landing/rover ops (algorithmic autonomy) & Deep Space & \cite{sganalyticsDataAnalytics} \\
\bottomrule
\end{tabular}
\end{table*}

Over the past several decades, space agencies have progressively increased the level of autonomy used in mission operations. The trajectory is not a sudden leap, but a steady expansion from narrowly scoped onboard behaviours to mission-critical functions spanning mobility, fault management, science triage, and crew support. Historical examples across planetary rovers, spacecraft operations, observatories, and human spaceflight illustrate how autonomy has been adopted to compensate for communication delay, limited contact windows, and the practical limits of human attention.

Planetary rovers and landers provide some of the most visible demonstrations. Early surface missions established a baseline of reactive autonomy, where vehicles could detect obstacles and adjust motion locally under severe compute and sensing constraints \cite{maimone2007overview,matthies2007computer,goldberg2002stereo,maki2003mars}. The Mars Exploration Rover era extended these capabilities with vision-based navigation and hazard avoidance that allowed rovers to interpret stereo imagery, estimate traversability, and plan safer routes without waiting for step-by-step commands. This did not remove humans from the loop, but it meaningfully increased the amount of productive mobility per communication cycle and reduced the operational risk associated with uncertain terrain.

Subsequent rover generations increasingly coupled mobility with science productivity. Curiosity introduced onboard target selection concepts that allowed instruments to acquire additional measurements when the ground team was out of contact, using local imagery to select candidate targets according to uploaded priorities. Perseverance \footnote{\url{https://science.nasa.gov/mission/mars-2020-perseverance/}} further expanded the idea by integrating more capable autonomous driving with instrument-level decision logic. Adaptive measurement strategies, for example, allow an instrument to allocate effort toward the most informative regions of a target surface during a single activity block, rather than treating all points equally and relying on later human review to request follow-up. In parallel, vision-based landing enhancements have demonstrated how onboard perception can reduce landing hazard risk by comparing real-time imagery with preloaded maps and diverting away from unsafe terrain during descent \cite{wang2020vision,jie2007precise}. Beyond NASA, other agencies have also fielded rovers that operate under communication constraints and therefore incorporate local hazard assessment and autonomy to support remote exploration on the Moon and Mars.

Autonomy has also been demonstrated in spacecraft operations beyond surface mobility. A landmark milestone was the flight validation of goal-based onboard commanding in the late 1990s, where an onboard system could plan activities from high-level objectives, diagnose injected faults, and reconfigure spacecraft subsystems in closed loop without continuous ground control. This class of demonstration clarified what autonomy must do in deep space: detect problems, reason about plausible causes, and execute conservative recovery actions when immediate operator intervention is not possible. A second influential example is the integration of onboard science analysis with autonomous replanning in Earth-observing satellites, where acquired data could be evaluated in flight, events of interest detected, and observation plans adjusted to capture transient phenomena. These experiments established an enduring template for autonomous science: local triage and opportunistic follow-up constrained by safety and resource envelopes.

Routine mission automation is also evident in small satellite programmes that were designed from the start for a higher degree of onboard autonomy, including day-to-day operations management, payload scheduling, and navigation- and attitude-related functions. In parallel, decision-support and optimisation have long contributed to mission operations on the ground. Scheduling systems for space telescopes and deep-space communications must satisfy complex constraints on visibility, instrument configuration, and limited resources; algorithmic planning and constraint-solving approaches improved the feasibility and efficiency of long-term schedules even when the final execution remained human-supervised.

Scientific discovery pipelines provide a complementary perspective. Large survey missions in astronomy and Earth observation produce data volumes that exceed what can be inspected manually. Automated classification, triage, and anomaly detection therefore became essential for extracting scientific value. Exoplanet detection from space telescope light curves provides a clear case: learning-based classifiers enabled rapid screening of candidate transit signals at scale and helped surface weak signals that might be missed in purely manual workflows. While much of this analysis has historically occurred on the ground, a clear trend has emerged toward onboard or near-real-time inference for triage, particularly when downlink is limited or when responsiveness to transient events matters. Earth-observation demonstrations have shown that onboard screening can reduce downlink burden by prioritising only the most informative scenes, improving end-to-end efficiency without requiring transmission of all raw imagery.

Human spaceflight adds a final dimension: human–machine collaboration. On the International Space Station, assistant robots and interactive systems have been used to support procedure guidance, information retrieval, and demonstrations of collaborative workflows in microgravity (Examples seen in Figure \ref{fig:robothuman}). Health monitoring is another important direction, where physiological data streams from sensors can be aggregated and analysed to detect deviations from individual baselines and provide early warning \cite{abdelghafar2019intelligent}. The operational motivation is consistent with other domains: support decision-making locally when continuous expert supervision is not available, and reduce the workload associated with monitoring complex systems over long durations.

Overall, these examples show that autonomy in space has progressed from isolated demonstrations to an increasingly integrated set of capabilities. The common pattern is conservative, bounded autonomy that improves productivity and safety under communication and resource constraints, while preserving clear escalation paths and accountability. This historical progression sets the stage for the broader framing of Space AI as a field concerned with assured autonomy across mission lifecycles.

\subsubsection{AI in Near-Future and Planned Space Missions}

\begin{table*}[htbp]
\small
\setlength{\tabcolsep}{3pt}
\renewcommand{\arraystretch}{0.95}
\centering
\caption{Potential Applications of AI in Future Space Missions, Part I.}
\label{tab:future_ai_missions_p1}
\resizebox{\textwidth}{!}{%
\begin{tabular}{p{3.15cm}p{1.25cm}p{1.55cm}p{1.35cm}p{6.05cm}p{1.25cm}p{2.05cm}p{0.85cm}}
\toprule
\textbf{Mission / System} & \textbf{Target/NET} & \textbf{Lead} & \textbf{Type} &
\textbf{AI-relevant capability (examples)} & \textbf{AI Domain} & \textbf{Status} & \textbf{Ref.} \\
\midrule

\multicolumn{8}{l}{\textit{Cislunar infrastructure}}\\
Gateway (PPE+HALO initial elements) & 2027 (target) & NASA-led & Int'l / Gov &
Long-duration uncrewed ops; autonomy for system management/FDIR; robotics-enabled inspection/maintenance (see Canadarm3) &
Orbit & Planned & \footnote{\url{https://www.nasa.gov/missions/artemis/gateway/a-powerhouse-in-deep-space-gateways-power-and-propulsion-element/}} \\

Canadarm3 (on Gateway) & 2027+ & CSA & Gov &
Vision-enabled robotic inspection and autonomous/remote maintenance to support uncrewed Gateway operations &
Orbit & In development & \footnote{\url{https://www.asc-csa.gc.ca/eng/canadarm3/about.asp}} \\

\addlinespace[2pt]
\multicolumn{8}{l}{\textit{Planetary exploration and sample return}}\\
VIPER rover & TBD & NASA & Gov / PPP &
Autonomous navigation and hazard assessment for polar volatile prospecting &
Orbit/ Surface & Discontinued (2024); partnership options & \footnote{\url{https://science.nasa.gov/mission/viper/}} \\

ExoMars (Rosalind Franklin) & 2028 & ESA & Gov &
Autonomous traverse support; safe driving; science ops sequencing for drilling campaign &
Deep Space & Planned & \footnote{\url{https://www.esa.int/Science_Exploration/Human_and_Robotic_Exploration/Exploration/ExoMars}} \\

Dragonfly (Titan) & Jul 2028 (window) & NASA & Gov &
Rotorcraft autonomy: navigation, hazard avoidance, safe landing selection under high latency &
Deep Space & Planned & \footnote{\url{https://science.nasa.gov/mission/dragonfly/}}
 \\

Mars Sample Return (campaign) & TBD (2030s) & NASA/ESA & Gov &
Multi-vehicle autonomy: surface retrieval, precision rendezvous/capture, mission-level fault handling (architecture under review) &
Deep Space & Architecture under review & \footnote{\url{https://science.nasa.gov/mission/mars-sample-return/}} \footnote{\url{https://www.esa.int/Science_Exploration/Human_and_Robotic_Exploration/Exploration/Mars_sample_return}}
 \\

Tianwen-3 (Mars sample return) & $\sim$2028 & CNSA & Gov &
Autonomy for EDL/ascent and Mars-orbit rendezvous/docking; onboard decision support under latency &
Deep Space & Planned & \footnote{\url{https://english.www.gov.cn/news/202507/23/content_WS68803af3c6d0868f4e8f45d3.html}} \\

\multicolumn{8}{l}{\textit{Space telescopes and science pipelines}}\\
Nancy Grace Roman Telescope & NET 2027 & NASA & Gov &
ML-assisted transient/event triage in large survey pipelines; automated artefact rejection &
Deep Space & Planned & \footnote{\url{https://roman.gsfc.nasa.gov/}} \\

PLATO & Jan 2027 & ESA & Gov &
ML-based transit/variability classification at survey scale (ground pipeline emphasis) &
Deep Space & Planned & \footnote{\url{https://www.esa.int/Science_Exploration/Space_Science/Plato}} \\

ARIEL & 2029 & ESA & Gov &
AI/ML atmospheric retrieval and spectral interpretation at scale (ground pipeline emphasis) &
Deep Space & Planned & \footnote{\url{https://science.nasa.gov/uranus/moons/ariel/}} \\

\addlinespace[2pt]
\multicolumn{8}{l}{\textit{In-orbit servicing and debris removal}}\\
ClearSpace-1 & 2H 2026 (NET) & ESA/ ClearSpace & PPP &
Autonomous RPO + vision-based target tracking and capture for debris removal demonstration &
Orbit & Planned & \footnote{\url{https://www.esa.int/Space_Safety/ClearSpace-1}} \\

OSAM-1 (Restore-L) & -- & NASA & Gov &
Robotic servicing RPO/manipulation demo (refuel concept) &
Orbit & Cancelled (2024) & \footnote{\url{https://www.nasa.gov/mission/on-orbit-servicing-assembly-and-manufacturing-1/}} \\

\bottomrule
\end{tabular}%
}
\end{table*}

In the coming years, autonomy is set to become a defining element of ambitious space missions. Programmes spanning lunar return and sustained operations, planetary exploration, next-generation observatories, and in-orbit servicing are increasingly designed around the assumption that continuous human oversight is not always available. The practical drivers are well known: communication delays and intermittent contact, growing mission complexity, tighter power and thermal margins, and the need to manage fleets of assets at scale. In response, mission concepts are progressively embedding onboard decision support, resilient fault handling, and more capable local execution so that vehicles and habitats can remain productive and safe between ground contacts. The following examples illustrate how these ideas are shaping near-future architectures.

Lunar exploration and base operations provide a clear motivation because the Moon combines operational proximity with harsh environmental constraints and periods of limited visibility and power \cite{BENAROYA201714}. Artemis-era concepts emphasise long-lived infrastructure that can continue operating during uncrewed intervals. A cis-lunar outpost such as Gateway is intended to function as more than a rendezvous point: it is expected to support logistics, communications, and systems management over extended periods when crews are not present. This shifts emphasis toward autonomous vehicle and subsystem management, onboard fault detection and recovery, and goal-based operations supervised periodically from Earth rather than executed through continuous teleoperation. Canada’s Canadarm3 \footnote{\url{https://www.asc-csa.gc.ca/eng/canadarm3/about.asp}} reflects the same requirement on the robotics side: inspection and maintenance tasks must be performed reliably under time delay, with local sensing and conservative autonomy supporting routine operations.

On the lunar surface, sustained habitation concepts similarly motivate automation and local decision support. The core issue is not novelty but workload and safety: power and thermal management, life-support supervision, and resource scheduling must remain stable when crews are absent or when they are occupied with time-critical tasks. The most credible pattern is conservative autonomy paired with clear escalation. Routine adjustments can be performed locally within verified envelopes; ambiguous situations trigger safe modes and structured alerts that allow crews or Earth-based teams to make final decisions. Recent research and technology development programmes increasingly treat this style of operations management as a first-class capability for habitats and surface assets.

Lunar volatile prospecting further illustrates how programmes evolve while the underlying technical need remains stable. VIPER was conceived to characterise water ice at the lunar south pole, where terrain hazards and illumination extremes impose strict operational constraints. Although the programme has faced schedule and implementation changes, the mission class remains important: prospecting requires hazard-aware mobility, sensor-driven target selection, and tight integration between mapping products and sampling decisions. This is a recurring theme for settlement planning: missions must turn sparse observations into actionable resource maps while tolerating uncertainty in both environment and hardware performance.

Internationally, long-horizon plans such as the International Lunar Research Station\footnote{\url{https://www.cnsa.gov.cn/english/n6465652/n6465653/c6812150/content.html}} roadmap emphasise infrastructure, power, and communications as prerequisites for sustained operations. Across these architectures, autonomy is increasingly framed as a practical workforce multiplier: operating robotic systems, supervising infrastructure health, and coordinating heterogeneous assets across contact gaps. The shared requirement is not aggressive autonomy for its own sake, but stable operations and predictable performance with limited human time and limited opportunities for intervention.

Next-generation planetary missions and mobile explorers extend these constraints to higher latency and higher consequence. Mars rover missions continue to push autonomy in navigation, science operations sequencing, and safety-aware execution under limited downlink and planning cycles. The ExoMars Rosalind Franklin rover\footnote{\url{https://science.nasa.gov/mission/rosalind-franklin/}} concept, for example, combines long traverses with deep subsurface drilling and complex instrument sequences, which increases the need for robust localisation, safe site approach, and well-structured execution logic. Although the specific degree of onboard autonomy varies across missions, the trajectory is clear: mobility, hazard assessment, and measurement prioritisation are increasingly designed with onboard decision support in mind, because waiting for daily command cycles can become the limiting factor for science return.

Dragonfly\footnote{\url{https://science.nasa.gov/mission/dragonfly/}} to Titan represents an even stronger autonomy case because flight mobility requires rapid local decisions under extreme communication delay. A rotorcraft must manage navigation, hazard avoidance, safe landing selection, and energy-limited sorties in an environment where uncertain terrain and atmospheric effects cannot be fully pre-scripted. The mission therefore embodies a broader design principle for deep exploration: high-level objectives can be set by humans, but execution must be local, conservative, and robust to unmodelled variation.

Sample-return concepts further sharpen the operational requirements. Multi-vehicle campaigns require time-critical phases such as entry, descent and landing, ascent, rendezvous, and capture, all under tight margins. Surface retrieval and orbit operations must also be coordinated without relying on continuous ground teleoperation. Even when campaign architectures are revised over time, the underlying capability set remains: reliable onboard perception, precise relative navigation, and safe coordination across multiple vehicles and mission phases. Parallel programmes with similar goals reinforce the point that these are mission-critical capabilities rather than niche demonstrations.

In space telescopes and astrophysics missions, the key driver is scale. Survey observatories generate data volumes that demand automated triage and robust pipelines, especially for time-domain science where rapid identification of transient events improves follow-up success. As detectors grow larger and survey cadence increases, the bottleneck shifts toward screening, artefact rejection, prioritisation, and queue management rather than raw image acquisition alone. In practice, this work is often performed on the ground, but the system-level requirement is end-to-end: data products must be ranked and routed efficiently, and observing schedules must remain responsive to time-sensitive opportunities. As a result, automated analysis and scheduling tools increasingly become part of the mission design rather than post-hoc additions.

Robotic in-orbit servicing and space infrastructure highlight autonomy as a safety requirement in close-proximity operations. Rendezvous, proximity operations, and manipulation involve tight coupling between perception and control, uncertain target behaviour, and strict collision-avoidance constraints \cite{GuffantiGammelliEtAl2024}. Programmes such as OSAM-1 illustrate both the technical ambition and the programme volatility common in this domain; even when individual missions are cancelled or reshaped, the technology agenda persists. Commercial servicing demonstrations, including life-extension docking in GEO, show that operational servicing is feasible and provide stepping stones toward more dexterous interventions, such as component handling, repair, and assembly. Active debris removal adds another layer of difficulty: targets may be tumbling, non-cooperative, and poorly characterised, requiring reliable relative navigation, motion estimation, and conservative capture strategies.

Finally, as orbital regimes become more congested, the operational burden of conjunction assessment and collision avoidance continues to rise. Manual handling of large volumes of alerts does not scale to future constellations and mixed-use orbital environments. Decision support that can prioritise risk, support consistent manoeuvre selection, and maintain auditable records of actions becomes increasingly important. Across servicing, debris removal, assembly, and traffic management, the common thread is not ``smarter'' spacecraft in an abstract sense, but practical autonomy: perception, planning, fault handling, and conservative execution that together enable safe operations in high-stakes environments with limited human attention.

\subsubsection{The Emerging Field of Space AI}
Taken together, historical practice and near-future mission concepts suggest that Space AI is consolidating into a distinct field at the intersection of artificial intelligence and space science and engineering. Early work often appeared as isolated capabilities \cite{jonsson2007autonomy,chien2014space}: autonomy for rover driving, rule-based fault handling, or planning tools for ground operations. What is changing is the systems-level framing. AI is increasingly treated as a mission-architectural component that enables autonomy, resilience, and scalable operations when continuous human oversight is impractical.

Several drivers recur across otherwise diverse missions. First, communication constraints are fundamental. Long latency, intermittent contact, and limited bandwidth push decision-making toward the spacecraft, motivating onboard perception, planning, and local control rather than continuous teleoperation. Second, space operations are data-intensive. Telemetry, imagery, spectra, and health and system logs quickly exceed what can be inspected manually, so automated triage, anomaly identification, and prioritisation become necessary for both operations and science. Third, space systems operate in harsh and uncertain conditions, where failures are costly and repair is limited. This shifts emphasis from accuracy alone to mission assurance: verification and validation, fault tolerance, interpretable decision logic for safety cases, and reliable behaviour under distribution shift. Fourth, as missions become longer and more complex, the interaction between humans and onboard automation becomes central. The goal is not to remove humans from the loop, but to design collaboration that reduces cognitive load while preserving clear authority, transparency, and accountability.

The field is also becoming more visible institutionally. Space agencies increasingly articulate autonomy and onboard intelligence as strategic technology directions, and they support dedicated programmes for spacecraft autonomy, distributed missions, and AI-enabled science operations \cite{mesbahi2025introduction}. Cross-sector initiatives and applied research accelerators further indicate that AI in space is no longer viewed only as experimentation, but as mission-critical capability development. In parallel, academic communities at the boundaries of astronomy, robotics, and data-intensive science have grown more organised, with recurring venues focused on autonomy, space robotics, and computational astronomy.

Overall, Space AI is transitioning from scattered demonstrations into an enabling discipline for future exploration and infrastructure. The constraints that dominate space missions—long-duration autonomy, limited intervention, resource-constrained compute, and high consequence of failure—naturally elevate research problems in assured autonomy, system-level validation, robust learning and planning, distributed coordination, and human–machine interaction under uncertainty. In that sense, Space AI is not simply “AI applied to space”; it is a forcing function that reshapes how missions are designed and how trustworthy autonomous systems are developed.

\section{AI on Earth: Enabling Space Exploration}

Space missions depend on intelligent capabilities in flight, but they are enabled just as decisively by the systems that remain on Earth. Design and verification, mission planning, operations support, and data exploitation all take place largely on the ground. In practice, many mission risks are retired before launch through modelling and testing, and many mission outcomes are determined after launch through planning, monitoring, and analysis. This section reviews how data-driven methods and optimisation tools on Earth strengthen space exploration, focusing on ground support systems, launch and recovery, and education and public engagement.

\subsection{AI in Ground Support Systems}

\begin{figure*}
    \centering
    \includegraphics[width=\linewidth]{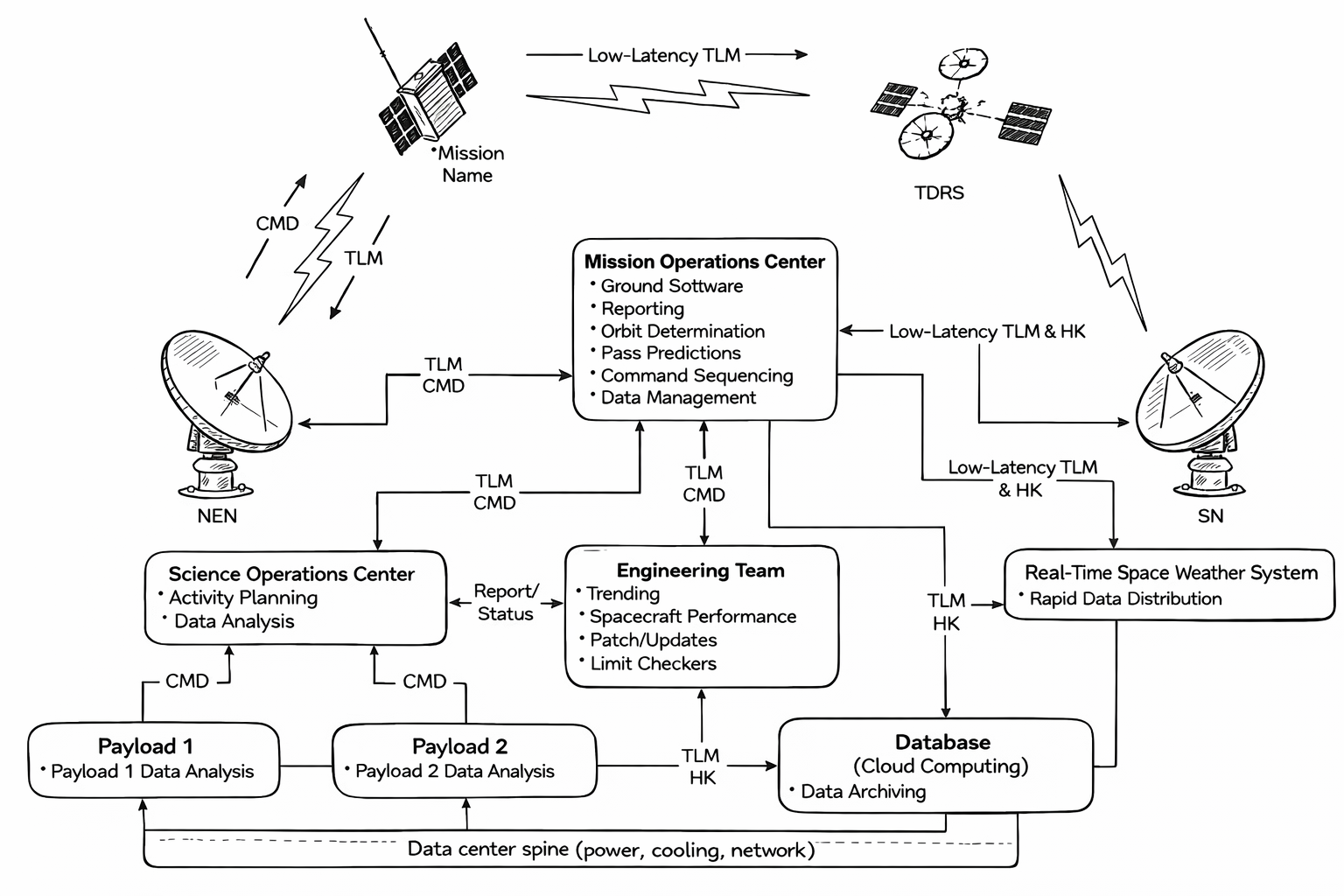}
    \caption{The framework of NASA ground support system for satellite.}
    \label{fig:ground}
\end{figure*}

Ground support systems cover the end-to-end pipeline that precedes and accompanies a mission: engineering design, high-fidelity simulation, planning and scheduling, and the analysis of telemetry and payload data (Seen in Figure \ref{fig:ground}\footnote{\url{https://www.nasa.gov/smallsat-institute/sst-soa/ground-data-systems-and-mission-operations/}}). Recent progress in learning-based modelling and optimisation has improved the speed and coverage of these workflows, allowing engineers to evaluate more design options, test more failure modes, and respond more effectively to operational constraints. The result is not only higher efficiency but also stronger assurance, as missions become more complex and less tolerant of late-stage surprises.

\subsubsection{Space Infrastructure Development and Simulation}
Digital engineering increasingly relies on high-fidelity simulation as the primary environment for design trade-offs, verification, and operational rehearsal. A digital twin can be understood as an executable virtual representation of a spacecraft, habitat, or launch system that is calibrated using design data, test results, and, when available, in-flight telemetry \footnote{\url{https://www.nasa.gov/jstar-digital-twins/}} \footnote{\url{https://www.esa.int/Enabling_Support/Space_Engineering_Technology/Large_Space_Simulator_gets_a_digital_twin}} \footnote{\url{https://sa.catapult.org.uk/blogs/digital-twin-showcase-opportunities-for-the-space-industry/}} \cite{9499045,YE2020107076,10115665}. While the roots of such practices trace back to mission simulators and ground testbeds used throughout the history of human spaceflight, modern digital twins place greater emphasis on continuous calibration, uncertainty quantification, and predictive capability.

Learning-based components are useful where first-principles models are incomplete, expensive to run, or difficult to parameterise. Examples include predicting component degradation under thermal cycling and radiation exposure, estimating remaining useful life for critical subsystems, and identifying early indicators of abnormal behaviour from high-dimensional telemetry. When embedded within simulation and verification pipelines, these models help prioritise test campaigns, narrow the space of plausible failure modes, and support earlier design fixes, which are typically far cheaper than late-stage mitigation.

Another area where optimisation has become central is structural and configuration design under mass, stiffness, and manufacturability constraints. Generative design and topology optimisation methods can explore large design spaces and propose candidate geometries that satisfy load and safety requirements while reducing structural mass \cite{zhu2016topology,peng2025topology,opgenoord2019design,di2018design,han2025generation}. In conceptual studies and early-phase engineering, such approaches have reported substantial mass reductions and improved load paths, enabling either increased payload capacity or larger design margins for shielding, redundancy, and thermal management. These methods are most valuable when treated as part of a disciplined engineering loop: candidate designs are generated, screened against constraints, validated with high-fidelity analysis, and then iterated with clear manufacturability assumptions. In this way, ground-based optimisation does not replace engineering judgement; it expands the set of options that can be examined within realistic schedules and budgets, improving both performance and confidence before hardware is committed.

\subsubsection{Intelligent Mission Planning and Scheduling}
Mission operations involve continual allocation of limited resources such as crew time, power, thermal capacity, onboard memory, and communications windows. Planning is therefore a constraint-heavy optimisation problem with frequent updates, because disturbances are inevitable: weather affects ground links, instruments behave differently than expected, and anomalies force replanning. Classical approaches remain essential, but they become difficult to scale when decision spaces grow large or when planning must be repeated under tight time constraints \cite{fukunaga1997aspen,barreiro2012europa,muscettola1998remote}.

Constraint-based scheduling and heuristic search are widely used to build feasible timelines under multiple interacting requirements, including for complex crewed environments such as the International Space Station \cite{marquez2010evolving,marquez2013supporting}. The practical value of modern decision-support tools lies in how quickly they can recompute plans when conditions change and how transparently they can expose trade-offs to operators. As missions expand to multi-asset architectures, such as satellite constellations and coordinated surface--orbital campaigns, the planning burden grows further: actions taken by one vehicle affect others through shared resources and coupled constraints \cite{frank2001planning,zilberstein2025decentralized}.

Learning-based optimisation can complement these pipelines by improving policies or heuristics in regimes where explicit modelling is difficult or where repeated simulation is feasible \cite{herrmann2023reinforcement,zhao2020two,marinelli2011lagrangian}. For example, policies trained in simulation can propose high-quality initial schedules, choose when to downlink or buffer data, and balance science objectives against energy and thermal constraints. Similarly, data-driven anomaly detection trained on historical telemetry and simulated fault scenarios can support contingency planning by flagging subtle precursors of failure and recommending candidate responses \cite{hayden2004livingstone}. The key contribution of these approaches is not merely automation; it is improved responsiveness and coverage. By enabling faster replanning and earlier detection of off-nominal trends, ground-based planning and monitoring systems raise mission utilisation and strengthen resilience, which in turn supports more ambitious exploration with manageable operational risk.

\begin{table*}[htbp]
\small
\setlength{\tabcolsep}{3pt}
\renewcommand{\arraystretch}{0.95}
\centering
\caption{Survey of AI Applications in Spacecraft Design Optimization, Part I.}
\label{tab:ai_spacecraft_design}
\resizebox{\textwidth}{!}{%
\begin{tabular}{p{3.0cm}p{2.75cm}p{4.85cm}p{4.85cm}p{1.45cm}}
\toprule
\textbf{Application Area} & \textbf{AI Methods Used} & \textbf{Description} & \textbf{Benefits Achieved} & \textbf{Ref.} \\
\midrule
Generative Component Design &
Generative models (GAN/VAE; diffusion emerging) &
Generates candidate geometries or layouts from prior designs; candidates are then screened/optimised using physics-based constraints and simulation &
Faster concept exploration; broader candidate set; potential to surface non-intuitive designs for further engineering validation &
\cite{goodfellow2020generative,kazemi2022multiphysics} \\

Topology Optimization &
CNN-assisted topology optimisation; neural-augmented solvers &
Accelerates material-layout search for structural/thermal components under load/thermal constraints, often by learning mappings from loads/boundaries to near-optimal layouts &
Lighter parts with comparable performance; reduced optimisation iterations relative to FEM-only loops; faster design turnaround &
\cite{banga20183d,budzyn2021topology} \\

Surrogate Modeling for Simulation Acceleration &
Gaussian processes; DNN/XGBoost surrogates; neural operators (e.g., DeepONet) &
Learns fast predictors of high-fidelity simulation outputs (FEM/thermal/CFD responses) to enable rapid evaluation inside optimisation and trade-study loops &
Substantial speed-ups in iterative design studies; enables larger design-space exploration under fixed compute budgets; supports sensitivity analysis and uncertainty-aware screening &
\cite{tanaka2023thermal} \\

Multi-Disciplinary Design Optimization (MDO) &
Evolutionary/heuristic search; surrogate-assisted optimisation; (BO/RL in simulation) &
Optimises coupled trade-offs across subsystems (mass, power, thermal margins, reliability) under mission-level requirements and interface constraints &
Improved compromise solutions across interacting subsystems; better coverage of the trade space; scalable exploration with constraints explicit &
\cite{boggero2018design,bartholomew1998role} \\

Thermal Management Design &
Surrogate models (ANN/GP); response-surface methods; sparse regressors &
Predicts temperature/heat-flow responses for candidate layouts (radiators, heat pipes, insulation) and supports optimisation under power/mass constraints &
Reduces overdesign; identifies efficient thermal paths; accelerates thermal trade studies and integration iterations &
\cite{mattos2013optimal} \\

Micrometeoroid Shielding Design &
Supervised learning (trees/SVM/DNN); probabilistic risk models &
Learns impact/failure likelihood from test data and simulations to support risk-informed shielding choices and placement decisions &
Mass-aware risk reduction; probabilistic failure assessment to guide shielding trade-offs and margin allocation & \cite{RYAN2023104727,RYAN2015522,9462134}
 \\

Structural Mass Reduction (Design Space Search) &
Genetic algorithms; autoencoders for design embeddings; surrogate-assisted search &
Searches structural configurations (panels/struts/trusses) under mission load envelopes, often using low-dimensional embeddings for efficient exploration &
Improved mass--margin trade-offs; faster iteration in early-phase structural configuration studies & \cite{4284021,ZHU202191}
 \\

Reliability Engineering \& Health Monitoring (Design Feedback) &
RUL prediction (LSTM/GRU/RF); fault classification (CNN/SVM); anomaly detection &
Uses ground test + flight data to identify failure precursors, inform redundancy allocation, monitoring strategy, and component selection during design &
Improved mission robustness and maintainability planning; reduced likelihood of mission-ending failures through risk-informed design decisions & \cite{SATYAHANUSH20227373}
 \\
\bottomrule
\end{tabular}%
}
\end{table*}

\begin{table*}[htbp]
\small
\setlength{\tabcolsep}{3pt}
\renewcommand{\arraystretch}{0.95}
\centering
\caption{Survey of AI Applications in Spacecraft Design Optimization, Part II (continued).}
\resizebox{\textwidth}{!}{%
\begin{tabular}{p{3.0cm}p{2.75cm}p{4.85cm}p{4.85cm}p{1.45cm}}
\toprule
\textbf{Application Area} & \textbf{AI Methods Used} & \textbf{Description} & \textbf{Benefits Achieved} & \textbf{Ref.} \\
\midrule
AI--Human Collaborative Design Tools &
Active learning; preference learning; human-in-the-loop optimisation &
AI proposes candidates or highlights trade-offs; engineers provide constraints/preferences and verify feasibility; loop repeats with targeted simulation/testing &
Higher engineering productivity; better exploration without sacrificing accountability; improved traceability of decisions and constraints & \cite{COUTINHO202413,WANG2024111950,BIANCHI2022100647}
 \\

Data Fusion for Design Insight &
Multi-modal learning (text+CAD+sim); graph learning &
Links requirements, CAD/MBSE artefacts, simulation/test logs, and outcomes to extract patterns of successful designs and recurring failure modes &
Design reuse and knowledge transfer; early identification of integration risks; systematic learning from prior missions and test campaigns &
 \\

Propulsion \& Engine Component Optimization &
Surrogate-assisted optimisation; physics-informed ML; Bayesian optimisation (where applicable) &
Accelerates nozzle/combustor/thermal--structural trade studies by reducing reliance on repeated high-cost CFD/FEA evaluations &
Faster propulsion design cycles; improved performance--mass trade-offs under operational constraints; better-informed prototyping decisions & \cite{NEELAKANDAN2025104439,MA2026111002,LI20171}
 \\

GNC Co-design with Vehicle Constraints &
Optimal control with learned models; RL in high-fidelity simulation &
Co-optimises guidance/control with actuation limits, aerodynamics, thermal/structural constraints, and mission objectives through simulation-driven policy search &
Improved closed-loop performance under constraints; reduced conservatism when validated against high-fidelity models; better robustness analysis in simulation & \cite{doi:10.1126/scirobotics.adi6421,BRANDONISIO20245741,OGHIM20251129}
 \\

Electrical Power System (EPS) Sizing \& Trade-offs &
Multi-objective optimisation; surrogate modeling; data-driven load modelling &
Joint sizing and trade studies across solar arrays, batteries, power electronics, thermal coupling, and mission profiles &
Reduced mass with preserved energy margins; improved robustness to load uncertainty; faster EPS design iteration and integration checks & \cite{asif2006multi,MOKHTAR2024100092}
 \\

RF/Communications Payload Layout Optimization &
Gradient-based optimisation; evolutionary search; learned surrogates &
Optimises antenna/array placement and configuration subject to structural/thermal integration constraints and interference considerations &
Better link performance with fewer redesign cycles; improved integration feasibility; reduced late-stage coupling surprises & \cite{yang2021antenna,koziel2024antenna,FAKOOR2017172}
 \\

Radiation Shielding \& Fault-Tolerant Avionics Design &
Risk-informed ML; probabilistic modelling; redundancy optimisation &
Estimates failure risk versus shielding/placement/redundancy using environment models plus test/heritage data to support mass-aware reliability design &
Improved reliability under radiation with explicit mass trade-offs; better-informed redundancy and placement decisions & \cite{JUN2024}
 \\

System Architecture and MBSE-informed Optimization &
MDAO; mixed-integer optimisation; surrogate-assisted architecture search &
Explores subsystem architecture options under requirements, interfaces, cost/mass/power/risk constraints, maintaining traceability to system models &
More consistent requirement-to-design traceability; stronger global trade studies; earlier detection of integration and feasibility risks & \cite{ZHANG2025104258}
 \\
\bottomrule
\end{tabular}%
}
\end{table*}

\subsubsection{AI-Assisted Data Analysis for Space Missions}
Modern space missions generate large volumes of heterogeneous data, including continuous engineering telemetry, payload measurements, and imagery. Ground-based data analysis is therefore a mission-critical capability: it supports timely operational decisions, improves fault response, and accelerates scientific interpretation. In practice, the most valuable tools are those that reduce operator burden while remaining interpretable and robust under changing operating conditions.

A long-standing operational use case is telemetry monitoring and anomaly detection. Spacecraft and stations stream thousands of parameters to mission control, and subtle deviations can be difficult to identify reliably with manual inspection alone. Data-driven monitoring systems can learn nominal operating envelopes and flag ``out-of-family'' behaviour for further investigation. NASA’s Inductive Monitoring System (IMS), for example, has been used as an anomaly-detection tool for ISS operations and has been trained for the Control Moment Gyroscope subsystem\footnote{\url{https://spinoff.nasa.gov/Spinoff2008/ct_8.html}}. The practical significance is early warning: when weak precursors are surfaced sooner, teams can schedule corrective actions before faults propagate, which is particularly valuable for long-duration missions where spare capacity and crew time are limited.

A second major use case is scientific data exploitation at scale. Earth observation missions and astronomy surveys routinely produce datasets too large for purely manual analysis. As a result, ground pipelines increasingly include automated triage and classification, enabling rapid mapping of hazards and environmental change, as well as fast screening of scientific targets for follow-up. In Earth science, a recent trend is the emergence of foundation models trained on broad, harmonised remote-sensing archives. The Prithvi family of geospatial foundation models, developed through NASA--IBM collaboration and released openly, is trained on NASA’s Harmonized Landsat and Sentinel-2 (HLS) data and supports downstream tasks such as flood mapping, burn-scar detection, and crop classification\footnote{\url{https://science.nasa.gov/science-research/ai-geospatial-model-earth/}; \url{https://huggingface.co/ibm-nasa-geospatial}}. In parallel, domain-adapted scientific language models are being released for search and knowledge access across NASA-relevant disciplines. The INDUS suite, for instance, is reported as openly available and targeted to Earth science, heliophysics, planetary science, astrophysics, and related areas\footnote{\url{https://science.nasa.gov/open-science/ai-language-model-science-research/}; \url{https://huggingface.co/papers/2405.10725}}. These models are less about replacing scientific reasoning than about improving retrieval, triage, and workflow efficiency across large corpora and high-throughput data products.

A third, emerging direction is distributed learning across fleets and sensor networks. When many assets observe related phenomena but bandwidth is constrained, it can be attractive to exchange compact model updates rather than raw data. Federated and distributed learning frameworks explore this idea by training models across multiple devices and aggregating updates to improve a shared model \cite{zhai2023fedleo}. Although operational adoption in space remains early, the concept is well aligned with constellations and multi-asset architectures where local inference is necessary and downlink is a scarce resource.

Overall, AI-assisted data analysis on Earth should be framed as an operational and scientific throughput layer. Its value is realised when it integrates cleanly with mission control practices, makes uncertainty visible, and improves responsiveness without obscuring accountability.

\subsubsection{AI in Spacecraft Design Optimization}
Spacecraft design involves tightly coupled trade-offs across structures, thermal control, power, avionics, and reliability. Many of these trade-offs are explored through repeated simulation and optimisation, making them natural targets for data-driven acceleration, particularly in early-phase design when the search space is large.

One useful role for learning-based methods is to provide fast surrogate models for expensive physics-based simulations. When trained on a representative set of high-fidelity analyses, surrogate predictors can approximate quantities such as thermal gradients, structural stresses, or performance margins with substantially lower computation cost. This enables rapid iteration over configuration choices, supports sensitivity analysis, and helps engineers identify promising regions of the design space before committing to detailed modelling. In multidisciplinary design optimisation workflows, these surrogates are most effective when paired with clear validation boundaries and conservative use of uncertainty estimates, so that speed does not come at the expense of overlooked failure modes.

Generative methods can also support early concept exploration. Beyond topology optimisation for macroscopic structures, researchers have explored using generative models to propose candidate geometries or configurations for specific components. For example, generative adversarial networks and related models can learn design priors from historical artefacts and then generate new candidates that are subsequently screened with physics simulation and constraint checks \cite{goodfellow2020generative}. In practice, such approaches are best viewed as idea-generation tools that broaden the set of candidates considered, while certification-critical decisions remain grounded in verified analysis.

A further benefit of data-driven methods lies in reliability and maintainability planning. Large test campaigns and multi-mission engineering datasets contain signals about failure precursors and component degradation. When integrated carefully, predictive diagnostics can inform design choices such as redundancy allocation, component selection, and monitoring strategies, improving long-term robustness. The key is to connect predictions to engineering actions: what should change in the design or operational concept when a risk indicator rises, and what evidence is required to justify that change.

In summary, AI-enabled optimisation is most valuable when it strengthens the design loop rather than bypassing it: accelerating exploration, improving coverage of trade-offs, and supporting reliability engineering with quantified confidence.

\subsection{AI in Launch and Recovery}
Launch and recovery are dominated by safety, timing, and reliability. On Earth, data-driven methods mainly contribute by improving manufacturing quality, supporting inspection and refurbishment, and strengthening decision support for launch readiness and window selection. In most operational systems, these tools augment classical engineering rather than replacing it, with the goal of increasing cadence without eroding assurance.

\subsubsection{AI-Driven Rocket Engineering and Manufacturing}

\begin{figure}
\centering
\includegraphics[width=\linewidth]{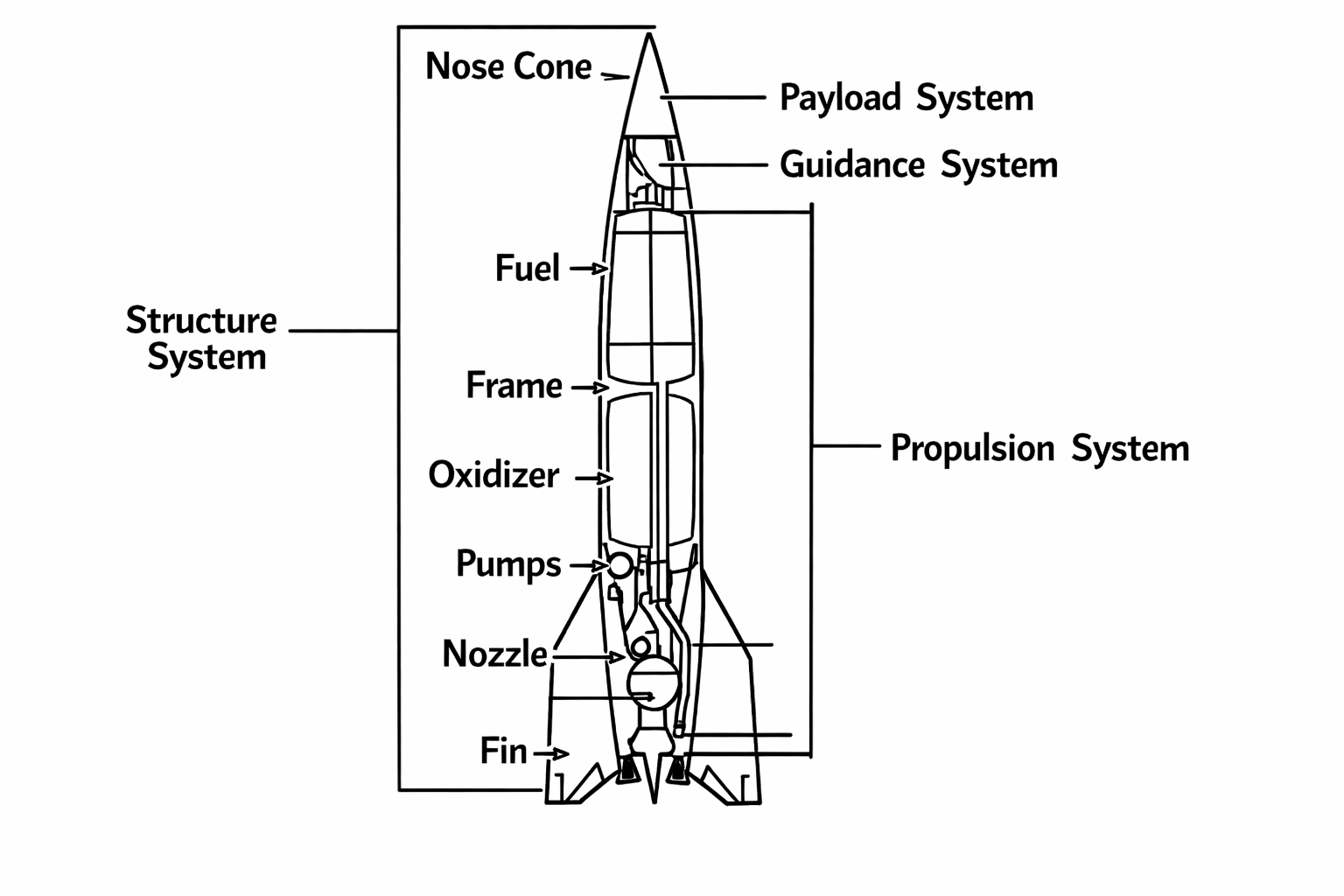}
\caption{The basic structure and parts of a rocket.}
\label{fig:rocket}
\end{figure}

Modern launch vehicles increasingly rely on advanced manufacturing processes, including additive manufacturing for selected components and high-throughput inspection for reuse. In these settings, the principal contributions of data-driven methods are process monitoring and quality control \cite{herzog2024process,zhang2025advancing}. Closed-loop manufacturing can use sensor feedback (e.g., thermal signatures, images, acoustic signals) to detect deviations during fabrication and to reduce defect rates. After recovery, automated inspection can compare high-resolution imagery and metrology data to baseline conditions to identify damage, abnormal wear, or regions requiring targeted non-destructive evaluation \cite{deng2025tdga}. The operational value is shorter turnaround: faster inspection and more consistent defect detection support reuse and improve fleet-wide reliability \cite{logiou2022benchmark,mishra2023machine}.

More broadly, design-for-manufacture and material screening increasingly benefit from computational search and data-driven prioritisation, especially when the space of candidate process parameters or compositions is large \cite{wei2021deep}. The most defensible framing is that these tools accelerate iteration and help focus expensive test resources where they matter most, rather than claiming that manufacturing has become ``autonomous'' in an end-to-end sense.

\subsubsection{Launch Window Optimization and Scheduling}
Launch decisions must balance orbital mechanics, range availability, vehicle readiness, and meteorological constraints. Data-driven forecasting and decision support are most useful when they reduce last-minute scrubs and help quantify risk trade-offs. Weather-driven constraints (e.g., winds aloft, lightning risk \cite{merceret2010history,willett2010rationales}) can be represented probabilistically, allowing teams to evaluate the likelihood of violating launch commit criteria across candidate windows. When combined with trajectory constraints and operational logistics, scheduling tools can compare feasible opportunities over a planning horizon and support more stable launch campaigns, particularly as launch cadence increases and range resources become more contested.

\subsubsection{Autonomous Landing and Recovery Systems}

\begin{figure*}
    \centering
    \includegraphics[width=\linewidth]{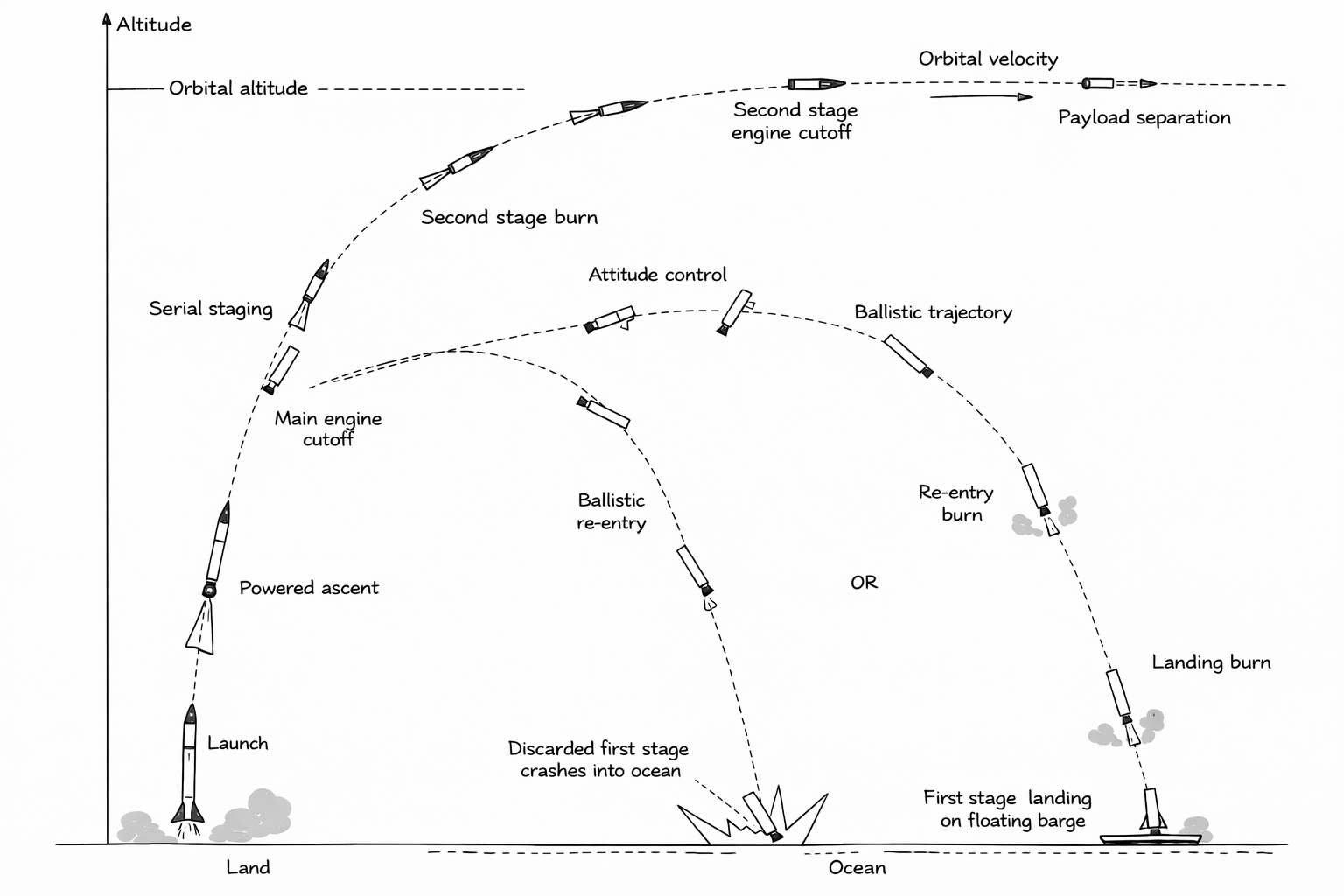}
    \caption{The illustration of rocket landing and recovery.}
    \label{fig:rocketlanding}
\end{figure*}

Precision landing and recovery are enabled primarily by robust guidance, navigation, and control, supported by real-time sensing \cite{blackmore2016autonomous}. In reusable systems, recovery requires continuous state estimation and closed-loop control under aerodynamic uncertainty and vehicle constraints \cite{blackmore2010minimum}. Data-driven components can contribute where perception or parameter estimation is needed, for example in interpreting sensor measurements, detecting off-nominal trends, or supporting rapid post-flight assessment. It is important, however, to distinguish learning-based perception or adaptation from the core safety-critical control logic, which is typically engineered with strong stability and assurance arguments \cite{sanchez2018real,izzo2024optimality}. The practical outcome of improved landing and recovery is lower cost per flight through reuse and higher operational tempo, which directly expands the feasible envelope of exploration programmes.

\subsection{AI for Education, Outreach, and Public Engagement}
Earth-based tools for training, education, and participation shape the long-term sustainability of space exploration by developing skills, broadening access, and maintaining public engagement. Here, the value of AI is primarily in scale and interactivity: it can make complex information accessible to more people, support repeated practice in realistic simulations, and help non-experts engage with real mission data.

\subsubsection{VR and AR Applications in Space Education}
Immersive simulation is now widely used for astronaut training and for communicating mission concepts. Mixed-reality systems can provide realistic rehearsal environments and enable trainees to practice procedures with repeatable variations \cite{zhivotovskiy2024reality}. A prominent public-facing example is ``Destination: Mars'', developed through collaboration between NASA and Microsoft using HoloLens technology and imagery from Mars missions\footnote{\url{https://science.nasa.gov/resource/mixed-reality-tech-brings-mars-to-earth/}; \url{https://www.jpl.nasa.gov/news/a-mixed-reality-trip-to-mars/}}. In training settings, similar simulation platforms can introduce controlled perturbations and failure scenarios, allowing trainees to practice response strategies without the cost and constraints of physical mockups \cite{garcia2020training,byrne2019treadmill,wang2022augmented}.

\subsubsection{AI-Driven Educational and Interactive Platforms}
Interactive educational platforms increasingly include conversational interfaces, structured retrieval from curated content, and adaptive learning pathways. In the space context, technology demonstrations have also tested how voice-based interfaces could operate under spacecraft communication constraints. The Callisto technology demonstration, flown on Orion during Artemis I, integrated voice interaction and collaboration tools to evaluate how such interfaces might support future missions\footnote{\url{https://www.nasa.gov/missions/callisto-technology-demonstration-to-fly-aboard-orion-for-artemis-i/}}. For education and outreach, the underlying point is not that a spacecraft needs a ``chatbot'', but that natural-language interfaces can lower barriers to accessing procedures, documentation, and mission information for trainees and the public, provided the underlying content is validated and curated.

\subsubsection{Citizen Science and Crowdsourcing Initiatives}
Citizen science has become an important complement to professional analysis in domains where data volumes are large and where human pattern recognition remains useful, especially for rare or unexpected phenomena. Contemporary practice often combines human input with automated models: volunteers label or verify subsets of data, and the resulting labels improve model performance, which then scales analysis to new datasets. Alongside volunteer platforms, open-data hackathons also broaden participation. NASA’s International Space Apps Challenge is a long-running global hackathon that encourages participants to build tools and analyses using NASA and partner open data\footnote{\url{https://spaceappschallenge.org/}}. These mechanisms expand the effective workforce for space-data analysis and help create a pipeline of tools, datasets, and educational resources that benefit both space science and Earth applications.

\section{AI in Orbit: Advancing Space Technology}

\begin{figure}
\centering
\includegraphics[width=\linewidth]{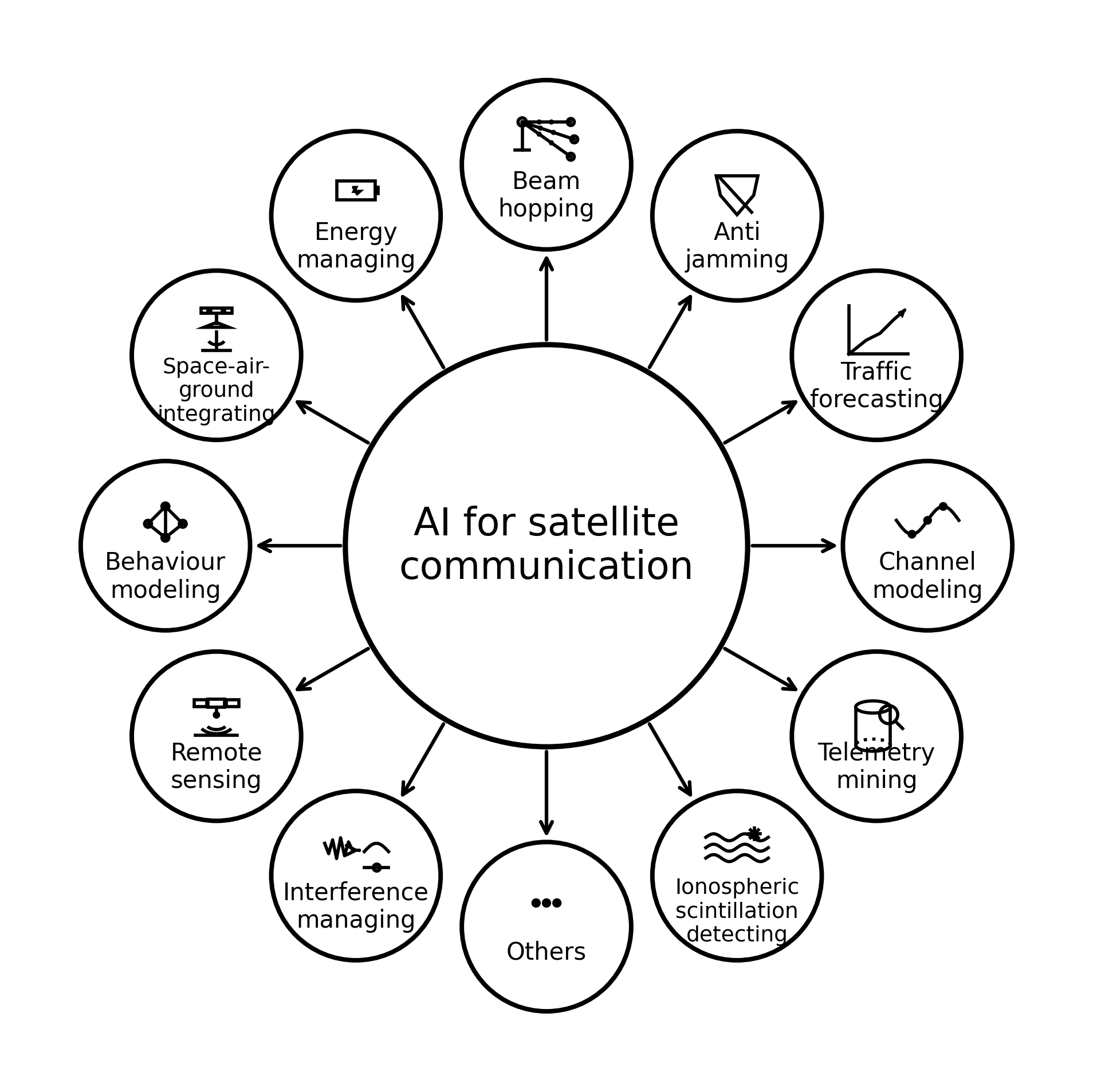}
\caption{AI for satellite communications examples.}
\label{fig:satellitecommunication}
\end{figure}

Orbit is where space systems operate at scale: large constellations, long-lived platforms, and increasingly complex interactions among satellites, stations, and ground networks. The operational environment is dynamic and congested, with limited opportunities for human intervention and strict constraints on power, compute, and communication (seen in Figure \ref{fig:satellitecommunication}). In this context, data-driven automation and decision support are becoming practical enablers of safer and more efficient orbital operations. This section reviews how autonomy, communications intelligence, and onboard decision-making are reshaping orbital space technology.

\subsection{Autonomous Satellites and Space Stations}

\begin{figure}
\centering
\includegraphics[width=\linewidth]{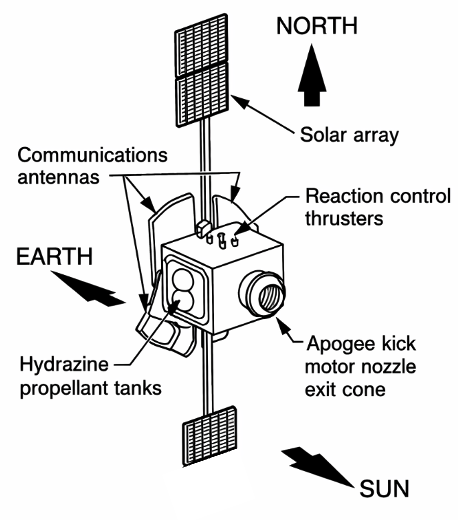}
\caption{The basic structure and parts of a satellite.}
\label{fig:satellite}
\end{figure}

\subsubsection{Autonomous Satellite Deployment and Operations}
Modern satellite missions, especially those involving constellations, require coordinated deployment, orbit acquisition, and routine operations across many assets \cite{williams1996model,hayden2004livingstone}. While ground control remains essential, parts of these workflows increasingly rely on onboard autonomy to reduce operational load and to respond within short time scales. During post-launch commissioning and orbit-raising, onboard decision logic can help execute manoeuvres, manage propulsion and attitude constraints, and compensate for dispersions introduced by the launch vehicle or environmental perturbations such as atmospheric drag in low Earth orbit \cite{11068406,el2025intelligent}. For constellation missions, this includes maintaining appropriate spacing and phasing while adhering to power and thermal limits.

Once on station, autonomy contributes most directly through health management and operations continuity. Onboard monitoring can track telemetry trends, manage power distribution, thermal regulation, and pointing, and trigger safe-mode procedures when anomalies are detected. In parallel, mission planning can be partially automated to prioritise imaging opportunities or communications tasks when constraints change (e.g., cloud cover, eclipses, or unexpected payload events) \cite{dt-icra-2024,narayanan2025tle,zilberstein-icaps-2024}. The practical value is responsiveness: when communication opportunities are limited, the spacecraft can maintain safe operations and preserve mission objectives without waiting for detailed ground instructions \cite{thangavel2024artificial}. In aggregate, this reduces routine workload for operators and supports higher mission tempo, particularly for large fleets \cite{frank2001planning}.

\subsubsection{Intelligent Communication Systems Optimization}

\begin{figure}
    \centering
    \includegraphics[width=\linewidth]{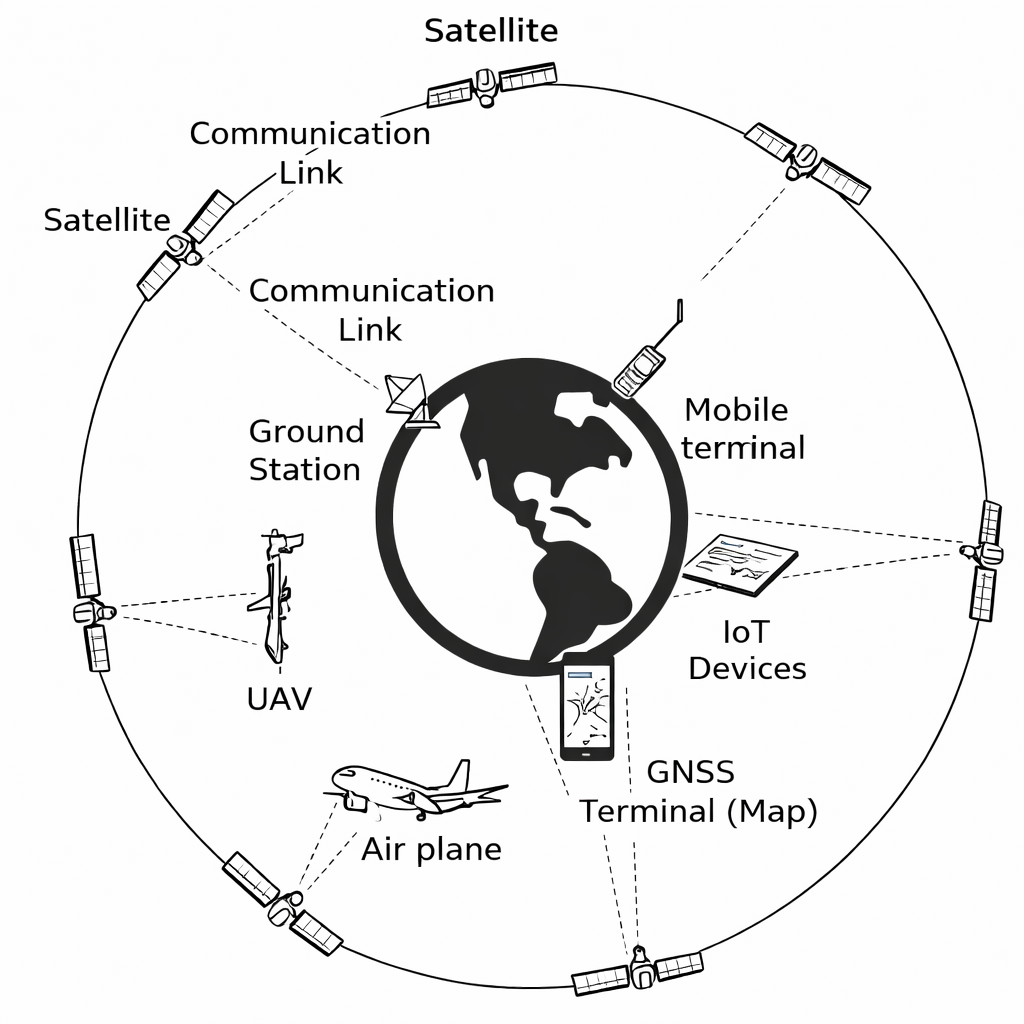}
    \caption{The Framework for satellite communication system.}
    \label{fig:communication}
\end{figure}

Orbital communications must manage time-varying link geometry, interference, limited spectrum, and rapidly changing demand (Seen in Figure \ref{fig:communication}). AI-enabled network optimisation is therefore increasingly used to support adaptive resource allocation and robust connectivity. At the physical and link layers, cognitive-radio style approaches can sense spectrum conditions and adjust frequencies, coding, and modulation to sustain performance when interference or propagation conditions change \cite{heo2023mimo,ya2022large}. At the network layer, constellation operators must schedule downlinks, route traffic through inter-satellite links, and allocate bandwidth across competing services, all under strict power and visibility constraints. Data-driven traffic prediction and automated scheduling can improve utilisation of scarce contact windows and reduce latency by selecting efficient routes and prioritising high-value data.

A critical consideration is that space communications are also exposed to adversarial and fault scenarios, including jamming, spoofing, and misconfiguration. For operational resilience, anomaly detection can be applied to traffic patterns and link metrics to flag abnormal behaviour, while adaptive waveform selection and routing policies can help mitigate degradation \cite{li2026rfi}. Similar principles apply to navigation and timing: detecting spoofed or inconsistent signals and cross-checking against alternative sensors can reduce the risk of corrupted state estimation. The broader point is not that learning replaces established communications engineering, but that adaptive decision support can improve robustness, especially in dense orbital environments where both benign interference and intentional disruption are plausible.

Overall, intelligent communications and autonomous operations are converging: as satellites become more capable of deciding what to collect, when to transmit, and how to route information, orbital systems move toward higher throughput and higher reliability with less dependence on continuous ground control.

\begin{table*}[htbp]
\centering
\caption{Example Studies on AI Applications in Space- and Ground-based Telescope Operations}
\label{tab:ai_telescope_operations}
\resizebox{\textwidth}{!}{
\begin{tabular}{p{5cm}p{1cm}p{1.5cm}p{1.5cm}p{1cm}p{5cm}p{0.8cm}}
\toprule
\textbf{Title of Study} & \textbf{Year} & \textbf{AI Technique(s)} & \textbf{Application Area} & \textbf{Telescope Type} & \textbf{Summary of Contribution} & \textbf{Ref.} \\
\midrule
Scheduling with neural networks—The case of the Hubble Space Telescope & 1992 & Neural Networks & Mission Planning & Space & Neural network-based scheduling for HST observations, handling complex constraints automatically. & \cite{johnston1992scheduling} \\
Galaxy types in the Sloan Digital Sky Survey using supervised artificial neural networks & 2004 & Neural Networks & Data Analysis & Ground & Automated classification of galaxy morphology in large-scale imaging surveys. & \cite{ball2004galaxy} \\
Artificial intelligence for the EChO mission planning tool & 2015 & Genetic Algorithms & Mission Planning & Space & Genetic algorithm-based planning for maximizing scientific returns of exoplanet observations. & \cite{garcia2015artificial} \\
Rotation-invariant convolutional neural networks for galaxy morphology prediction & 2015 & CNN & Data Analysis & Ground & CNN-based galaxy morphology classification leveraging rotational symmetry in imaging data. & \cite{dieleman2015rotation} \\
Identifying exoplanets with deep learning & 2018 & CNN & Data Analysis & Space & CNN-based exoplanet detection in Kepler mission data, identifying previously unnoticed planets. & \cite{shallue2018identifying} \\
A framework for telescope schedulers: Applications to the LSST & 2019 & Heuristic Algorithms & Operations, Mission Planning & Ground & Dynamic scheduling framework to optimize LSST observation plans, adapting to conditions and priorities. & \cite{naghib2019framework} \\
Adaptive optics control using model-based reinforcement learning & 2021 & Reinforcement Learning & Operations & Ground & Model-based RL control for adaptive optics, compensating for atmospheric turbulence effectively. & \cite{nousiainen2021adaptive} \\
Astronomaly: Personalised active anomaly detection in astronomical data & 2021 & Active Learning & Data Analysis & Ground & Interactive anomaly detection pipeline for discovering novel astronomical objects in large datasets. & \cite{lochner2021astronomaly} \\
The ROAD to discovery: Machine-learning-driven anomaly detection in radio astronomy & 2023 & Self-supervised Learning & Operations & Ground & Real-time self-supervised anomaly detection framework for LOFAR telescope data. & \cite{mesarcik2023road} \\
Machine learning-driven anomaly detection for Euclid Space Telescope operations & 2024 & Gradient Boosted Trees & Operations & Space & XGBoost-based telemetry anomaly detection and forecasting system for spacecraft health monitoring. & \cite{gomez2024machine} \\
\bottomrule
\end{tabular}
}
\end{table*}

\subsubsection{AI-Enhanced Space Telescope Operations}
Space telescopes are scarce and expensive assets, and their scientific value depends heavily on how efficiently observation time is planned and executed. Scheduling is a constrained optimisation problem governed by visibility windows, slewing overheads, calibration needs, thermal constraints, and instrument availability. Decision-support tools can help generate feasible schedules that maximise scientific return, and can also support rapid replanning when time-critical opportunities arise, such as transient alerts from other facilities. In practice, these capabilities matter most for survey-style missions and time-domain astronomy, where the ability to react quickly can determine whether a transient is captured at all \cite{kudryavtsev2025neural}.

Operational performance also depends on maintaining stable pointing and instrument health over long periods. Data-driven monitoring can assist in tracking trends in guidance sensors, reaction wheels/gyros, and power/thermal behaviour, supporting earlier detection of off-nominal patterns and faster diagnosis. Where appropriate, automated procedures can support routine recalibration and help preserve image quality by identifying drift or degradation signals early, reducing the burden on ground teams while keeping the observatory within safe operating margins.

Finally, the downstream data pipeline increasingly relies on automated triage and classification. For missions that produce high-cadence, large-volume survey data, it is often impractical to treat all data products equally \cite{lee2025classification,ortiz2025multi,djorgovski2022applications}. Automated screening can prioritise subsets for rapid downlink and follow-up analysis, while routine artefact rejection and anomaly flagging can improve the reliability of scientific products. The goal is not to replace scientific interpretation, but to ensure that the observatory’s limited time and downlink capacity are directed toward the most informative observations.

\subsubsection{AI-Driven Data Processing in Space}
Onboard data processing is becoming a practical necessity as sensors outpace downlink capacity. Instead of transmitting all raw measurements to Earth, spacecraft can increasingly perform early-stage processing in flight and downlink only the most relevant data products \cite{bazarov2022sensitivity}. This includes compression and quality assessment, event detection, and prioritisation of imagery or spectra under strict power and compute budgets. The operational value is immediate: bandwidth is conserved, time-to-insight is reduced, and missions can respond faster to emerging events.

\begin{figure}
    \centering
    \includegraphics[width=\linewidth]{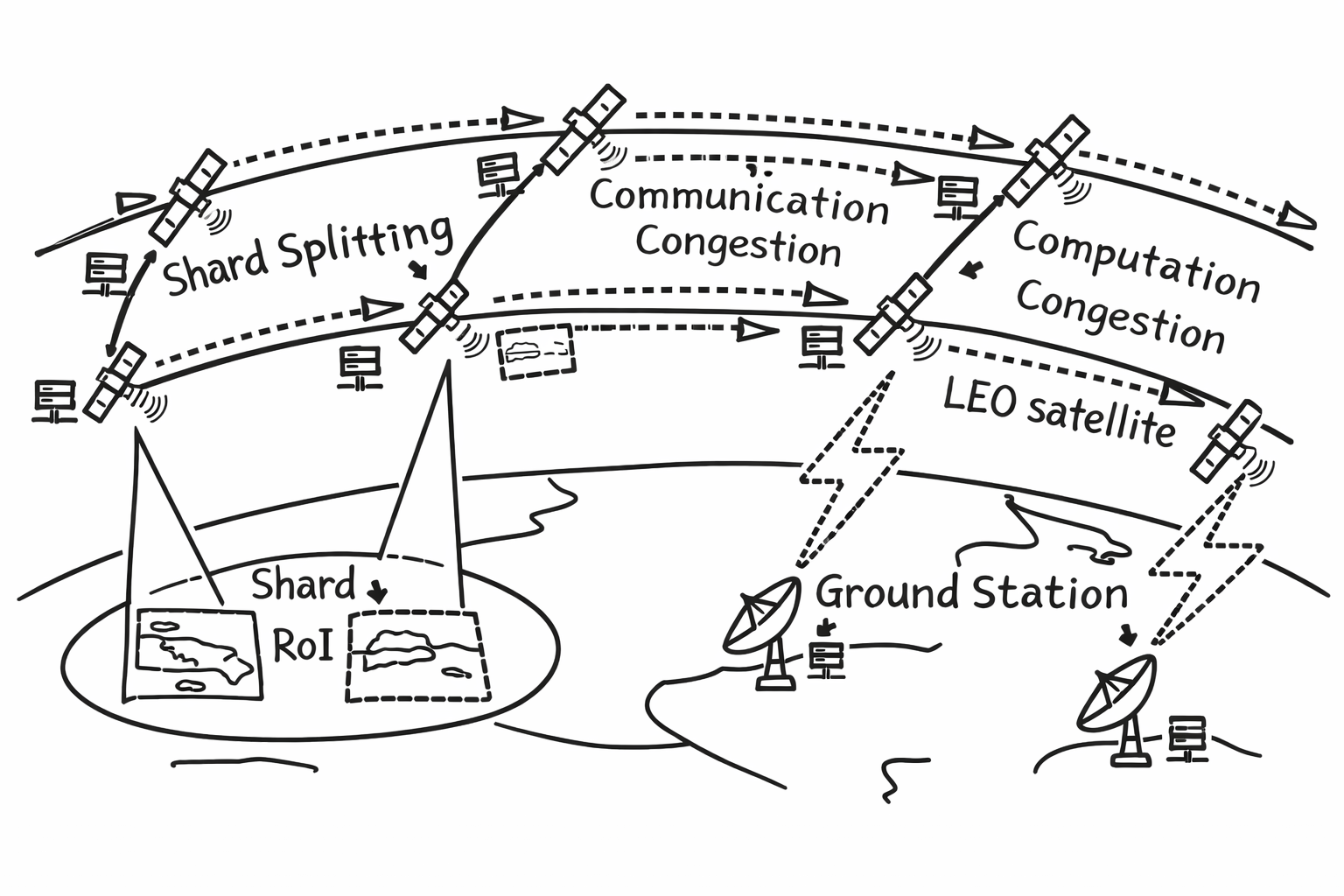}
    \caption{Example multi-satellite cooperative observation, routing and computation process.}
    \label{fig:seco}
\end{figure}

For Earth observation, onboard screening can identify scenes that contain time-critical phenomena such as fires, floods, or unusual changes, enabling selective downlink and more responsive monitoring. For planetary and deep-space missions, onboard prioritisation is even more important because contact opportunities are sparse and latency is high. Rovers and orbiters can use in-flight processing to rank candidate targets, select representative samples, and reduce the risk that the most informative measurements are delayed or missed. In constellation settings, partial onboard processing also supports distributed sensing, where each satellite extracts compact summaries and shares cues with the fleet, improving collective coverage without saturating crosslinks or downlink windows \cite{zhai2024seco,wang2018multi}. Example framework of multi-satellite cooperative observation, routing and computation process is illustrated in Figure \ref{fig:seco}.

These trends are enabled by progress in flight-qualified computing and by careful co-design of algorithms with hardware constraints \cite{10115594}. The most useful systems are those that make their confidence visible and fail safely: when uncertainty is high, they preserve raw data, request additional observations, or defer to conservative modes rather than over-committing to a single interpretation.

\subsubsection{Space-Based Data Centers and In-Space Cloud Infrastructure}
\begin{figure*}
    \centering
    \includegraphics[width=\linewidth]{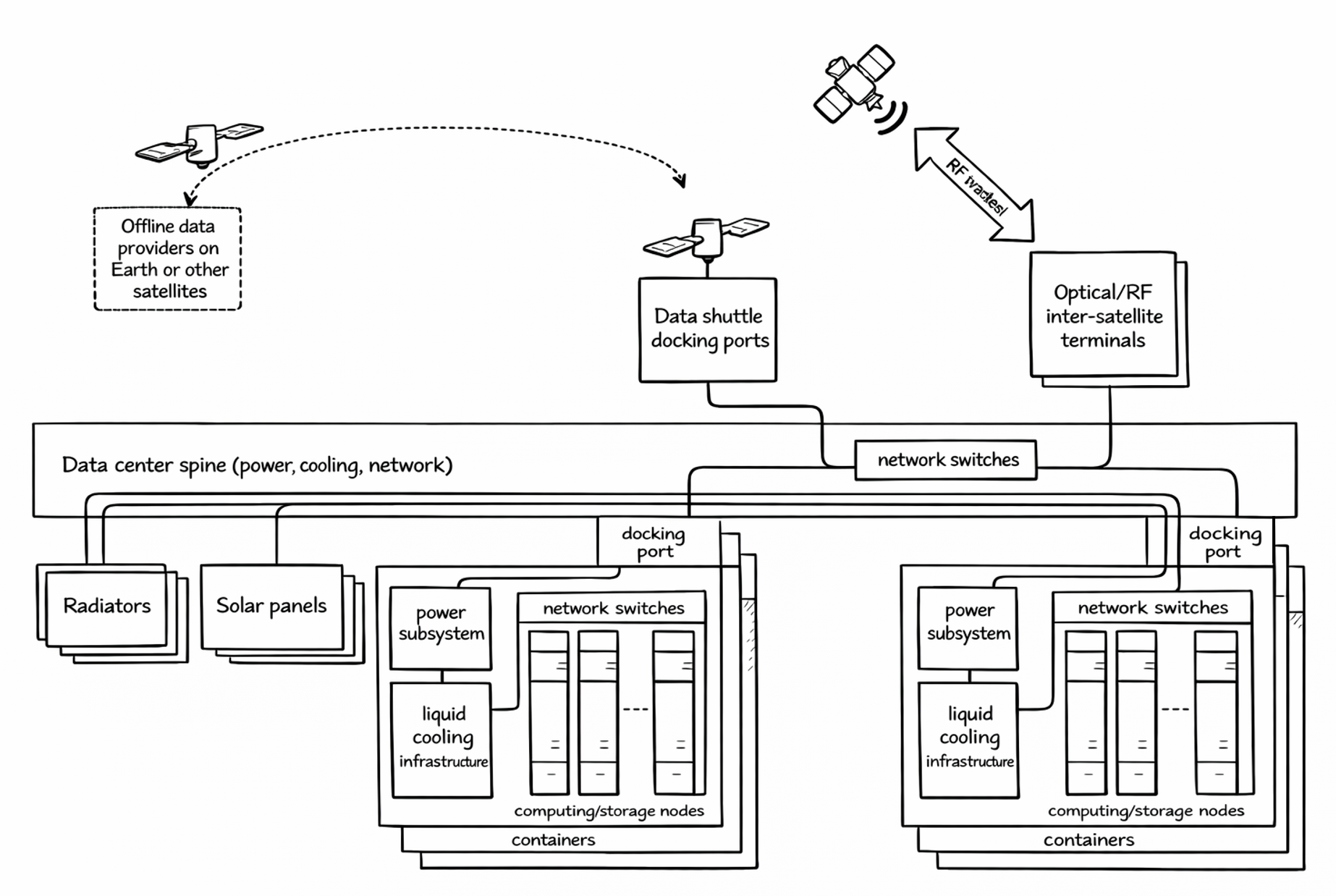}
    \caption{Orbital data center network architecture.}
    \label{fig:datacentre}
\end{figure*}

As onboard processing grows, a complementary concept is emerging: placing scalable compute and storage capacity in space to form an ``in-space cloud'' or orbital data center layer. The motivation is similar to terrestrial edge--cloud systems \cite{meoni2024ops}. Onboard compute is power- and volume-limited, while downlink is constrained by spectrum, contact time, and latency. Intermediate space-based compute nodes could support aggregation, caching, and higher-throughput analytics closer to the data source, particularly for constellations producing continuous high-volume streams. Industry initiatives are beginning to position ``space-based cloud'' and orbital data-center services as a future infrastructure layer (e.g., Starcloud\footnote{\url{https://www.starcloud.com/}} and proposed orbital data center network architecture in Figure \ref{fig:datacentre}).

From a Space AI perspective, the key research questions are systems-level rather than purely algorithmic. In-space compute infrastructure must operate under compounded environmental stressors, including radiation exposure and physical degradation from micrometeoroid and orbital debris impacts. Experimental total ionizing dose (TID) radiation testing of commercial embedded GPU platforms, such as the NVIDIA Jetson Nano\footnote{\url{https://developer.nvidia.com/embedded/jetson-nano}}, demonstrates that COTS AI accelerators can continue operating under moderate radiation exposure, but with non-negligible performance degradation and reliability risks that must be addressed through radiation-aware design, redundancy, and fault-tolerant system architectures \cite{slater2020total}. In parallel, studies of micrometeoroid and debris micro-impacts on space-exposed solar arrays highlight how long-term surface damage and material erosion can progressively reduce power generation capability, directly constraining the sustained operation of power-hungry onboard and in-orbit compute nodes \cite{berthoud1997meteroid}. 

Together, these results underscore that orbital data centers and in-space cloud infrastructure are governed by coupled constraints on compute reliability, power availability, and environmental resilience. Workloads must therefore be scheduled under tight power and thermal budgets; resilience must explicitly account for radiation-induced faults and gradual hardware degradation; and computation must be coordinated with network routing and downlink planning to ensure graceful degradation rather than catastrophic failure. Security and governance further compound these challenges, as orbital compute introduces new attack surfaces and raises questions of data sovereignty, accountability, and compliance across jurisdictions \cite{zengshan2025comprehensive}. Although the practical maturity of space-based data centers remains uneven and mission-dependent, these constraints reinforce a broader point: Space AI increasingly concerns end-to-end data and compute architecture design—spanning onboard inference, in-orbit aggregation, and ground pipelines—rather than isolated onboard models.

\begin{table*}[htbp]
\small
\centering
\setlength{\tabcolsep}{4pt}
\renewcommand{\arraystretch}{1.05}
\caption{Fault Detection and Autonomous Recovery in Orbital Platforms}
\label{tab:fault-detection}
\begin{tabular}{p{1.6cm}p{3.4cm}p{2.3cm}p{3.2cm}p{3.0cm}p{3.0cm}}
\toprule
\textbf{Subsystem} & \textbf{Fault Types} & \textbf{Detection (ML/Model/Stat)} & \textbf{Detection Keywords} & \textbf{Recovery Action} & \textbf{Mission Example} \\
\midrule
Attitude Control & Gyro drift; reaction wheel degradation/failure; star tracker loss & ML; model-based; statistical & Kalman filtering; residuals; thresholds & Switch to backup sensor/actuator; re-tune controller; momentum management & Kepler reaction wheel failures \\
Power (EPS) & Solar array degradation/failure; battery capacity fade & ML; model-based; statistical & Thresholding; PCA; clustering & Load shedding; reconfiguration; safe mode & Cassini power anomalies (RTG / bus undervoltage) \\
Thermal Control & Cooling loop leak; overheating; thermal sensor drift/failure & Model-based; statistical & Trend analysis; limit checks; residuals & Re-orient for thermal balance; activate redundancy; reduce loads & ISS ammonia leak events \\
Communications & Antenna pointing loss; transponder anomalies; signal dropouts & Statistical; model-based & Link margin monitoring; watchdogs; packet loss & Switch to backup chain; re-point; safe mode & Voyager 2 communications anomalies \\
Data Handling & CPU reset/crash; memory upsets (SEUs); software fault & Model-based; statistical & Watchdogs; ECC; exception logs & Reboot; failover; memory scrubbing & Hubble onboard computer anomalies \\
Propulsion & Thruster stuck-on/off; valve issues; propellant leak & Model-based; statistical & Pressure/flow residuals; performance monitoring & Switch to backup thrusters; adjust burn plan; isolate leak & Akatsuki main engine anomaly \\
Mechanisms & Solar array deployment issues; antenna deployment anomalies & Model-based; statistical & Limit switches; motor current spikes; timing checks & Retry sequence; alternate release path; reconfigure operations & Galileo high-gain antenna deployment failure \\
Life Support & Cabin leak; CO$_2$ scrubbing anomalies; O$_2$ generation faults & Statistical; model-based & Pressure decay; gas sensor trends; alarm logic & Isolate module; switch to backup unit; conserve consumables & ISS minor atmosphere leak responses \\
\bottomrule
\end{tabular}
\end{table*}

\subsubsection{Fault Detection, Diagnosis, and Autonomous Recovery}
Fault management is a core requirement for long-duration orbital missions, where limited contact windows and high operational complexity make purely manual monitoring increasingly difficult. Historically, spacecraft fault detection and diagnosis relied on rule-based limits, predefined fault trees, and extensive ground analysis. These approaches remain widely used, but they can be brittle when operating conditions drift, when multiple subsystems interact, or when the number of monitored variables grows \cite{newhouse2011fault,tipaldi2015survey,morgan2005fault}.

Data-driven monitoring can complement traditional methods by modelling nominal behaviour from telemetry and flagging deviations that merit attention \cite{hundman2018detecting,cuellar2024explainable,iverson2009general}. Depending on the mission and available training data, this may include supervised classifiers trained on labelled fault cases, or unsupervised methods that learn normal operating regimes and detect out-of-family behaviour. Such tools are particularly useful for early warning, where weak precursors appear as subtle trend changes rather than clear threshold violations. Model-based approaches provide a different advantage: by encoding expected causal relationships among subsystems, they help generate consistent explanations for observed symptoms and support structured diagnosis when multiple hypotheses are plausible.

Detection and diagnosis are only useful if they translate into safe responses. Autonomous recovery in orbit is typically constrained by conservative safety logic: isolate the suspected component, reconfigure to a known-safe operating mode, and preserve essential functions until ground teams can verify the situation. Examples include switching to redundant units, adjusting attitude-control allocation when a reaction wheel degrades, or shedding non-critical loads to protect power margins. In well-defined fault classes, onboard procedures can execute these steps without waiting for immediate ground intervention, improving continuity of operations during communication gaps.

Learning-based decision policies can be explored in simulation to propose recovery strategies, but operational adoption must be bounded by verification, fail-safe fallbacks, and clear authority limits \cite{seto1998simplex,schierman2020runtime,ravaioli2022universal}. A practical framing is therefore ``assured autonomy'': onboard systems handle frequent, well-characterised anomalies quickly and conservatively, while complex or ambiguous cases are escalated to ground operators with richer context. Done well, this combination reduces downtime, improves mission robustness, and extends spacecraft service life by responding earlier and more consistently to developing faults.

\subsubsection{AI for Satellite Remote Sensing and Earth Observation}

Earth observation has become one of the most data-intensive branches of space activity. Modern optical and radar satellites generate continuous streams of imagery and derived products, while users increasingly demand timely, reliable information rather than raw pixels. AI contributes at two levels. On the ground, automated pipelines support large-scale mapping and monitoring tasks such as land-cover classification, change detection, disaster assessment, and infrastructure analytics \cite{10.1145/2168752.2168763}. In orbit, onboard screening and prioritisation can reduce downlink pressure and shorten time-to-insight by transmitting only the most informative scenes or compressed summaries, which is particularly valuable for responsive applications where delays reduce utility.

In practice, the most robust uses of AI in remote sensing combine learning-based perception with explicit uncertainty handling and domain constraints. Atmospheric effects, seasonal variation, sensor drift, and domain shift between regions can degrade performance if models are deployed without careful calibration. As a result, operational systems increasingly incorporate quality flags, confidence estimates, and human-in-the-loop verification for high-stakes outputs. For multi-sensor settings, AI also supports data fusion across modalities such as optical, SAR, and multispectral imagery, enabling more reliable inference under cloud cover, low light, or complex terrain \cite{8738825,rs13122364,CHOI2024153}. As constellations expand, these methods scale from single-scene interpretation to continuous spatiotemporal monitoring, where the key challenge becomes consistency over time and across platforms.

Ultimately, AI-enhanced Earth observation is most valuable when framed as a decision-support layer for society: enabling faster situational awareness during hazards, more frequent environmental accounting, and more efficient use of limited sensing and downlink resources. These benefits also feed back into space operations by informing space weather response, tracking orbital-risk drivers such as atmospheric density changes, and supporting broader sustainability and governance objectives for the orbital environment.

\begin{table*}[htbp]
\footnotesize
\setlength{\tabcolsep}{3pt}
\centering
\caption{Open-Access Remote Sensing Benchmark Datasets for AI Applications (Part I)}
\label{tab:remote_sensing_datasets}
\begin{tabular}{p{2.4cm}p{1.2cm}p{2.7cm}p{2.3cm}p{2.5cm}p{1.5cm}p{2.0cm}p{2.8cm}}
\toprule
\textbf{Dataset Name} & \textbf{Image Type} & \textbf{Coverage/Source} & \textbf{Size} & \textbf{Tasks Supported} & \textbf{Resolution} & \textbf{License} & \textbf{Website/Link} \\
\midrule
AID Scene Dataset \cite{xia2017aid} & Aerial & Global (Google Earth) & 10,000 images in 30 classes & Scene classification & $\sim$0.5–2 m & Free for research & \url{https://captain-whu.github.io/AID/} \\
BigEarthNet \cite{sumbul2019bigearthnet} & Satellite & Europe (Sentinel-2) & 590,326 patches (120×120 px) & Multi-label land cover classification & 10 m & CDLA Permissive 1.0 & \url{https://bigearth.net} \\
COWC \cite{mundhenk2016large} & Aerial & 6 global locations & 32,716 cars labeled & Object detection, counting & 0.15 m & Public domain & \url{https://gdo152.llnl.gov/cowc/} \\
DeepGlobe 2018 \cite{demir2018deepglobe} & Satellite & Multiple (WorldView-3) & 3 tracks, 25k+ images & Segmentation, detection & 0.5 m & Free for research & \url{https://deepglobe.org/} \\
DIOR \cite{li2020object} & Satellite & Diverse global (Google Earth) & 23,463 images, 20 classes & Object detection & $\sim$0.5–1 m & CC BY 4.0 & \url{https://gcheng-nwpu.github.io/} \\
DOTA v1.0/v2.0 \cite{9560031,Ding_2019_CVPR} & Aerial & Global (Gaofen-2, JL-1) & 5,215 images, 18 classes & Oriented object detection & 0.3–1 m & CC0 Public & \url{https://captain-whu.github.io/DOTA/} \\
EuroSAT \cite{helber2019eurosat} & Satellite & Europe (Sentinel-2) & 27,000 images, 10 classes & Land use classification & 10 m & MIT License & \url{https://github.com/phelber/EuroSAT} \\
fMoW \cite{christie2018functional} & Satellite & Global & 1M+ images, 63 categories & Scene classification & $\sim$0.3 m & Research-only & \url{https://fmow.labs.gov} \\
HRSCD \cite{daudt2019multitask} & Aerial & France (Urban Atlas) & 291 registered pairs & Semantic change detection & 0.5 m & Open data & \url{https://rcdaudt.github.io/hrscd/} \\
Inria Aerial Labeling & Aerial & 10 cities (USA, Austria) & 360 images (5000×5000 px) & Building segmentation & 0.3 m & Public domain & \url{https://project.inria.fr/aerialimagelabeling/} \\
LandCover.ai \cite{boguszewski2021landcover} & Aerial & Poland & 41 orthophotos & Semantic segmentation & 0.25 m & CC BY-NC-SA 4.0 & \url{http://landcover.ai} \\
LEVIR-CD \cite{chen2020spatial} & Satellite & Texas, USA & 637 pairs (1024×1024 px) & Change detection (buildings) & 0.5 m & Free for research & \url{https://justchenhao.github.io/LEVIR/} \\
\bottomrule
\end{tabular}
\end{table*}

\begin{table*}[htbp]
\ContinuedFloat
\footnotesize
\setlength{\tabcolsep}{3pt}
\centering
\caption{Open-Access Remote Sensing Benchmark Datasets for AI Applications (Part II)}
\begin{tabular}{p{2.4cm}p{1.2cm}p{2.7cm}p{2.3cm}p{2.5cm}p{1.5cm}p{2.0cm}p{2.8cm}}
\toprule
\textbf{Dataset Name} & \textbf{Image Type} & \textbf{Coverage/Source} & \textbf{Size} & \textbf{Tasks Supported} & \textbf{Resolution} & \textbf{License} & \textbf{Website/Link} \\
\midrule
LoveDA \cite{wang2021loveda} & Satellite & China (3 cities) & 5,987 images & Segmentation, domain adaptation & 0.3 m & CC BY-NC-SA 4.0 & \url{https://github.com/Junjue-Wang/LoveDA} \\
OpenEarthMap \cite{xia2023openearthmap} & Both & Global (44 countries) & 5,000 images, 2.2M polygons & Land cover segmentation & 0.25–0.5 m & CC BY 4.0 & \url{https://open-earth-map.org} \\
Onera OSCD \cite{daudt2018urban} & Satellite & 24 cities (Sentinel-2) & 24 pairs (multispectral) & Change detection (binary) & 10 m & CC BY 4.0 & \url{https://rcdaudt.github.io/osi-challenge/} \\
PatternNet \cite{zhou2018patternnet} & Satellite & USA (Google Earth) & 30,400 images, 38 classes & Scene classification & 0.3 m & CC BY 4.0 & \url{https://sites.google.com/view/zhouwx/dataset} \\
RESISC45 \cite{cheng2017remote} & Satellite & Global & 31,500 images, 45 classes & Scene classification & 0.2–30 m & Free for research & \url{http://www.escience.cn/people/JunweiHan/NWPU-RESISC45.html} \\
SEN1-2 \cite{schmitt2018sen1} & Satellite & Global (Sentinel-1/2) & 282,384 image pairs & SAR-optical fusion, matching & 10 m & CC BY 4.0 & \url{https://mediatum.ub.tum.de/1436631} \\
SpaceNet (1–7) \cite{van2021multi} & Satellite & Global & 67,000+ km$^2$, 11M buildings & Detection, segmentation, change detection & 0.3–0.5 m & CC BY-SA 4.0 & \url{https://spacenet.ai/datasets/} \\
UC Merced \cite{neumann2019domain} & Aerial & USA & 2,100 images, 21 classes & Scene classification & 0.3 m & Public domain & \url{http://weegee.vision.ucmerced.edu/datasets/landuse.html} \\
UAVid \cite{lyu2020uavid} & Aerial (drone) & UK (Southampton) & 420 frames & Semantic segmentation & $\sim$0.05 m & CC BY-NC-SA 4.0 & \url{https://uavid.nl/} \\
VisDrone \cite{zhu2021detection} & Aerial (drone) & China (urban scenes) & 10,209 images + 260k video frames & Object detection, tracking & Varying (cm-level) & CC BY-NC-SA 3.0 & \url{http://aiskyeye.com/} \\
xBD (xView2) \cite{xview2XView2} & Satellite & Global (15 countries) & 22,068 image pairs, 850k annotations & Change detection, damage mapping & 0.3 m & CC BY-NC-SA 4.0 & \url{https://xview2.org/} \\
xView \cite{gupta2019xbd}& Satellite & Global (DIUx, NGA) & 1,413 images, 1M+ objects & Object detection & 0.3 m & CC BY-NC-SA 4.0 & \url{http://xviewdataset.org/} \\
\bottomrule
\end{tabular}
\end{table*}

\begin{table*}[htbp]
\small
\centering
\setlength{\tabcolsep}{4pt}
\renewcommand{\arraystretch}{1.05}
\caption{NLP Applications in Satellite Remote Sensing and Earth Observation}
\label{tab:nlp_rs_summary}
\begin{tabular}{p{3.6cm}p{4.3cm}p{2.2cm}p{6.2cm}}
\toprule
\textbf{RS Task Category} & \textbf{Representative Method/Model} & \textbf{Ref.} & \textbf{Technical Highlights} \\
\midrule
Description Generation (Captioning) & RSICD CNN-RNN & \cite{hoxha2020new}  & First large-scale caption dataset (RSICD); CNN-RNN baseline for diverse land-use scenes.\\
& Geospatial Relation Captioning & \cite{chen2019geospatial} & Attention mechanism for capturing spatial relations in captions. \\
& MLAT Transformer & \cite{liu2022remotet} & Transformer with multi-scale features, outperforming CNN-RNN in benchmarks. \\
& Count-aware Captioning & \cite{ni2024incorporating} & Integrates object counting with numerical adjustment in captions for precision. \\
& CRSR Retrieval-Refinement & \cite{li2024cross} & Uses retrieved semantic cues for enhanced caption generation. \\
\midrule
Visual Question Answering (VQA) & RSVQA CNN-RNN Baseline & \cite{lobry2020rsvqa} & Established RSVQA dataset; CNN for visual, RNN for textual input. \\
& Bi-Modal Transformer VQA & \cite{bazi2024rs} & Dual-stream Transformer for improved image-question fusion. \\
& Curriculum Learning VQA & \cite{yuan2022exploring} & Trains VQA with progressively complex queries; improves accuracy and robustness. \\
& Mask-based Dual-Stream Network & \cite{li2024enhancing} & Uses mask-guided attention to enhance image-region relevance. \\
\midrule
Image Generation from Text & StrucGAN & \cite{zhao2021text} & Multi-stage GAN; generates spatially coherent RS images from text. \\
& CRS-Diff (Diffusion Model) & \cite{tang2024crs} & Diffusion-based controllable image synthesis from textual attributes. \\
& Txt2Img-MHN & \cite{xu2023txt2img} & Modern Hopfield Network for semantic alignment in generated images. \\
\midrule
Cross-Modal Retrieval & Regional Cross-Attention Retrieval & \cite{zheng2022cross} & Aligns local image regions with words; improves retrieval accuracy. \\
& Attention Correction \& Filtering & \cite{yang2025remote} & Iterative refinement of embeddings, reducing modality mismatch. \\
& Multilingual Transformer Retrieval & \cite{al2022multilanguage} & Enables multilingual text-to-image retrieval leveraging semantic relations. \\
\midrule
Land Use/Land Cover Classification & TACOSS & \cite{zermatten2025learning} & Open-vocabulary classification via text embeddings; easy schema transfers. \\
& Zero-Shot Scene Classification & \cite{al2023vision} & NLP embeddings (CLIP-based) enable unseen land-use class classification. \\
\midrule
Change Detection with Text & Dual-Branch Change Captioning (LEVIR-CC) & \cite{liu2022remote} & Dual-transformer model for bi-temporal change captioning; LEVIR-CC dataset introduced. \\
\midrule
Information Extraction from RS Documents & NLQ to GeoSPARQL QA Interface & \cite{potnis2020towards} & Translates natural language queries into GeoSPARQL for EO databases. \\
\midrule
Semantic Understanding \& Language Grounding & EarthGPT & \cite{zhang2024earthgpt} & Domain-specific multimodal LLM integrating diverse RS tasks and sensors. \\
\bottomrule
\end{tabular}
\end{table*}

\subsubsection{Intelligent Space Robotics for Maintenance and Assembly}
Robotic servicing and in-orbit assembly are increasingly viewed as enabling capabilities for a sustainable orbital economy, where spacecraft are maintained, upgraded, and constructed rather than discarded \cite{nanjangud2018robotics,vaquero-eels-scirob-2024}. Compared with ground robotics, the orbital setting introduces distinctive challenges: targets may be uncooperative and tumbling, contact dynamics are unforgiving, illumination varies sharply, and communication delays or limited bandwidth can constrain continuous teleoperation. These realities motivate higher autonomy in perception, planning, and control, while still preserving strict safety constraints.

\begin{figure*}
\centering  
\includegraphics[width=\linewidth]{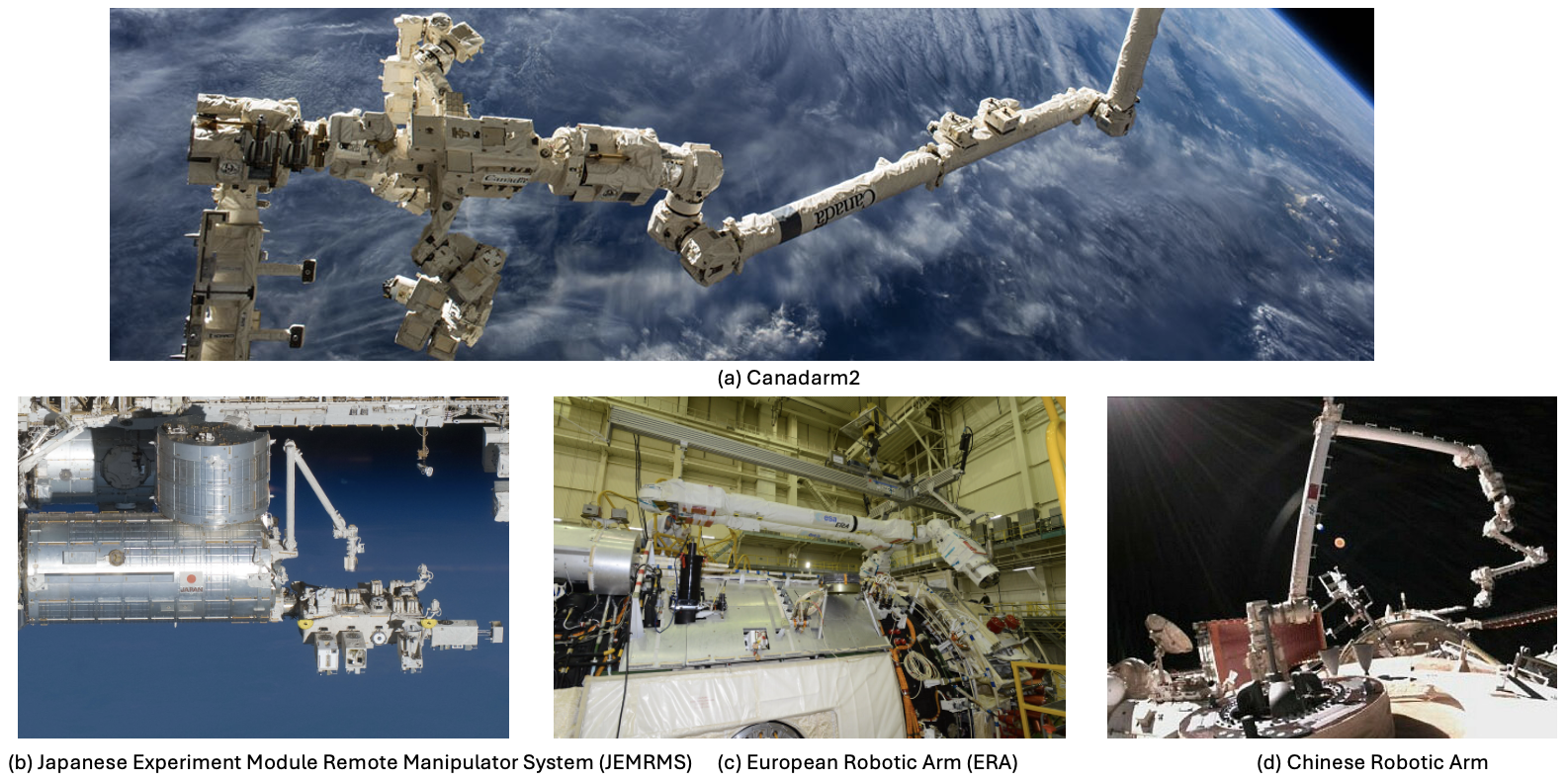}  
\caption{The Images of Robotic Arm on Space Station.}
\label{fig:robotarm}  
\end{figure*}

For maintenance and servicing, robotic manipulators on servicing vehicles or stations can support inspection, approach, capture, and intervention tasks \cite{montanaro2023multi,PARK20245726,10610138}. Vision-based pose estimation and interface recognition help identify grasp points, docking fixtures, and tool interaction sites on a client satellite \cite{zhuang2025off} (major space-station robotic arms are summarised in Table~\ref{tab:robotic-arms} and Figure~\ref{fig:robotarm}) \cite{CASSINIS20235061,ahmed2024speed}. During close-proximity operations, autonomy is most valuable for fast, local decisions: stabilising relative motion, compensating for small misalignments, and maintaining safe separation margins. Once a stable capture is achieved, robotic procedures can support tasks such as module replacement, connector manipulation, or propellant transfer. The practical impact is longer spacecraft lifetime and reduced reliance on risky human extravehicular activity, particularly for routine or repetitive servicing tasks.

In-orbit assembly extends these ideas from ``repair'' to ``construction''. Building large-aperture telescopes, modular platforms, and other structures that are impractical to launch fully integrated requires coordinated manipulation, alignment, and verification in microgravity. Autonomy supports this pipeline by converting high-level assembly goals into executable sequences, selecting feasible grasp and placement actions under geometric and dynamic constraints, and using sensor feedback to refine alignment and tolerance control during mating operations \cite{rodriguez2021autonomous,zimpfer2005autonomous,herz2014onboard,liu2025goal}. When multiple robots or arms are involved, coordination becomes a central challenge: collision avoidance, task allocation, and contingency handling must be managed across agents while maintaining overall structural accuracy and safety. In this setting, AI contributes most when embedded within a conservative engineering envelope, providing efficient planning and perception while deferring to verified safety logic for contact-critical steps.

Overall, intelligent orbital robotics shifts the feasibility frontier for space infrastructure. By enabling maintenance, upgrade, and incremental assembly, robotics reduces lifecycle cost, supports modularity and reuse, and helps scale orbital capabilities without requiring every complex structure to be launched as a single monolithic payload.

\subsubsection{Human-AI Collaboration in Orbital Environments}
\begin{figure*}
\centering  
\includegraphics[width=\linewidth]{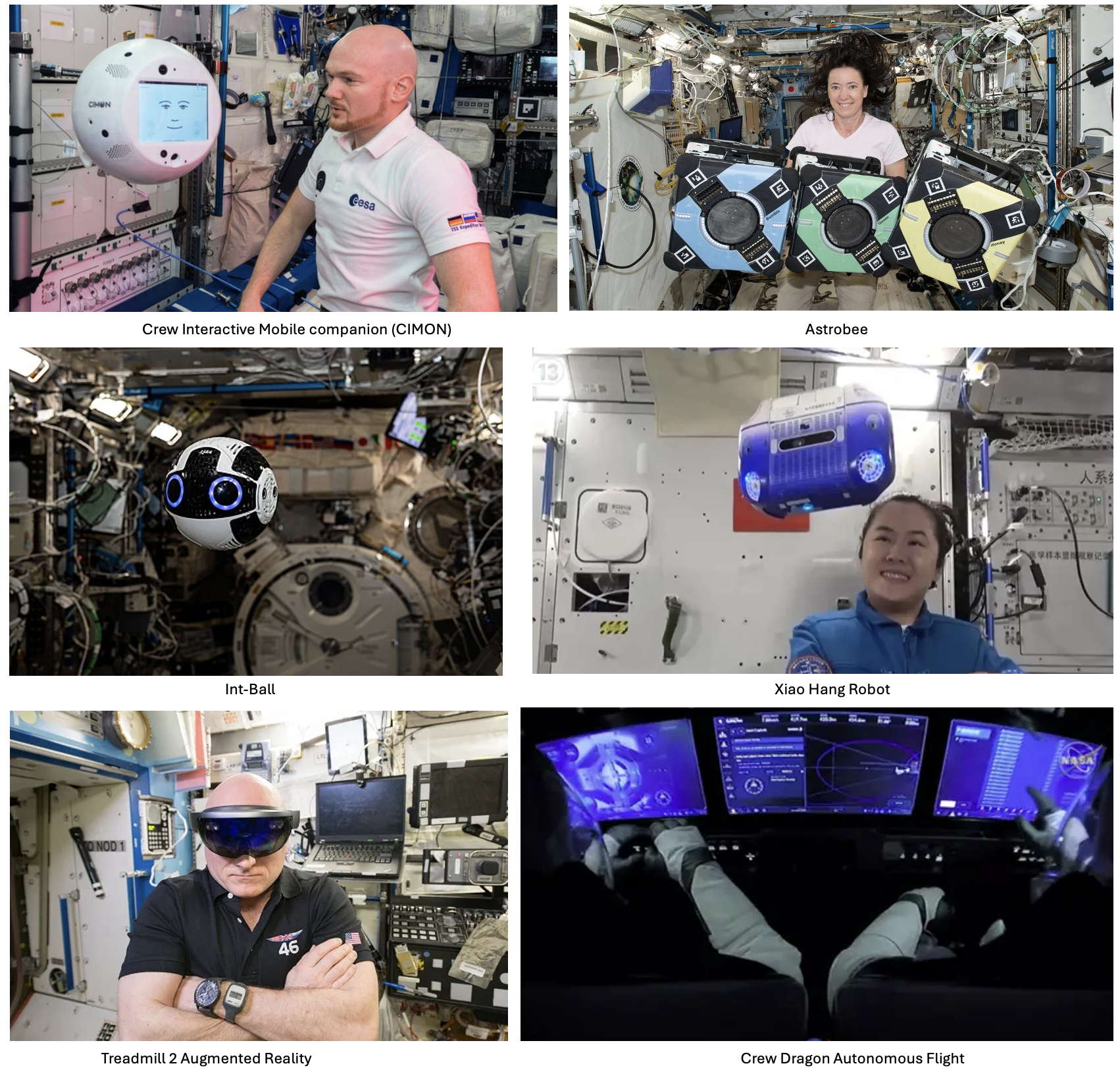}  
\caption{The Example Images of Human–AI Collaboration Systems in Orbital Environments.}
\label{fig:robothuman}  
\end{figure*}

As orbital systems become more complex and crews are expected to operate with greater autonomy, the interface between humans and onboard decision-support becomes a defining factor for safety and performance. In practice, the most valuable ``AI'' in orbit is often not a single monolithic agent, but a collection of tools that help crews manage information, maintain situational awareness, and execute procedures consistently \cite{hirano2024int,schmitz2020towards}. Examples include continuous monitoring of life-support and vehicle health, summarisation of telemetry and logs, and procedure guidance that reduces the chance of omissions under time pressure (Examples seen in Figure \ref{fig:robothuman}). When well integrated, these tools reduce routine cognitive load, enabling astronauts to allocate attention to mission-level decisions and unexpected events \cite{ROMETSCH20222145}.

Effective collaboration depends less on conversational fluency than on predictable behaviour and clear authority boundaries \cite{parasuraman2000model,o2022human}. Crews must be able to understand what the system is doing, why it is raising an alert, and what confidence it assigns to its own recommendations \cite{luo2022evaluating}. Transparency can be achieved through simple design choices: exposing the evidence behind an alert (which sensors changed and how), showing plausible alternative explanations, and making the consequences of suggested actions explicit. In safety-critical contexts, the operational pattern is typically conservative: onboard tools propose options and automate well-characterised routines, while astronauts retain final authority over actions with irreversible consequences. This division of labour supports trust calibration, reduces automation surprise, and keeps accountability aligned with human responsibility.

Looking forward, the importance of human--AI teamwork increases as missions shift toward longer duration and reduced ground support \cite{chen2021toward}. When communication is delayed or intermittent, crews benefit from reliable onboard decision support for fault triage, resource planning, and schedule replanning. The goal is not to replace human judgement, but to ensure that crews can make good decisions with limited time, incomplete information, and high operational stakes.

\begin{table*}[htbp]
\centering
\caption{Comparative Performance of Major Space Station Robotic Arms}
\label{tab:robotic-arms}
\resizebox{\textwidth}{!}{%
\begin{tabular}{lccccccccccc}
\toprule
\textbf{Robotic Arm} & \textbf{Length (m)} & \textbf{Mass (kg)} & \textbf{Payload (kg)} & \textbf{DoF} & \textbf{Stop Dist. (m)} & \textbf{Modes} & \textbf{Pos. Acc. (mm)} & \textbf{Att. Acc. (°)} & \textbf{Lin. Vel. (m/s)} & \textbf{Ang. Vel. (deg/s)} & \textbf{Stiffness (Nm/rad)} \\
\midrule
Canadarm2 (Canada) & 17.6 & 1497 & 116000 & 7 & 0.6 & Manual/Auto & 50 & 1.0 & 0.37 & 5.0 & $9 \times 10^{5}$ \\
JEMRMS (Japan) & 10.0 & 780 & 7000 & 6 & 0.12 & Manual/Limited Auto & 15--30 & 0.37 & 0.05--0.10 & N/A & N/A \\
ERA (Europe) & 11.3 & 630 & 8000 & 7 & 0.15 & Auto/Tele-op & 5 & N/A & 0.10 & N/A & N/A \\
Chinese Arm (China) & 10.2 & 738 & 25000 & 7 & N/A & Manual/Semi-Auto & 45 & N/A & N/A & N/A & N/A \\
\bottomrule
\end{tabular}%
}
\end{table*}

\begin{table*}[htbp]
\small
\setlength{\tabcolsep}{3pt}
\centering
\caption{Examples of Human–AI Collaboration Systems in Orbital Environments}
\label{tab:human_ai_orbit}
\begin{tabular}{p{2.8cm} p{2.5cm} p{4.8cm} p{2.3cm} p{2.5cm} p{1.8cm}}
\toprule
\textbf{System Name} & \textbf{Developer/Agency} & \textbf{Description of Human–AI Interaction} & \textbf{Interface Type} & \textbf{Collaboration Role} & \textbf{Platform} \\
\midrule
CIMON-2 & Airbus/DLR/IBM & Free-flying AI assistant interacting naturally with crew; voice-controlled, conversational support, displays visual instructions. & Voice, natural language, visual display & Procedure guidance, information retrieval, crew support & ISS (Columbus) \\[0.5em]

Astrobee & NASA Ames & Autonomous robots assisting with routine chores, inventory checks, experiment documentation, navigational assistance in cabin. & Touchscreen, autonomous navigation, remote control & Routine task automation, crew time optimization & ISS (U.S. Segment) \\[0.5em]

Int-Ball2 & JAXA & Free-floating spherical drone autonomously photographing crew operations, reducing manual documentation workload. & Autonomous, remote operator control & Autonomous documentation, situational awareness & ISS (JEM “Kibo”) \\[0.5em]

Xiao Hang Robot & China Manned Space Agency (CMSA) & Free-floating robotic assistant with multimodal interaction, captures imagery/videos, supports crew activities and inspections. & Voice, multimodal (possibly gesture/eye tracking) & Crew support, photography, routine task assistance & Tiangong Station \\[0.5em]

T2 AR Maintenance Assistant & NASA/Microsoft & Augmented Reality headset providing real-time, step-by-step visual and auditory guidance during maintenance tasks. & AR headset (HoloLens), voice, gesture, gaze input & Maintenance guidance, error detection, decision support & ISS (NASA, ESA, JAXA crews) \\[0.5em]

Crew Dragon Autonomous Flight & SpaceX & Semi-autonomous spacecraft managing critical flight phases (docking, re-entry); astronauts monitor via touchscreen and intervene if needed. & Touchscreen, automated alerts, manual override & Autonomous spacecraft control, anomaly alerting, human override capability & Crew Dragon (ISS missions) \\
\bottomrule
\end{tabular}
\end{table*}

\subsection{AI-Based Space Resource and Environment Management}

\subsubsection{Energy and Resource Management Systems}
Resource management is a central constraint for orbital platforms and habitats. Power, thermal capacity, consumables, and maintenance opportunities are all limited, and the system must remain stable under disturbances such as eclipses, payload duty-cycle changes, or unexpected faults. For this reason, modern mission concepts increasingly emphasise automated resource scheduling and predictive monitoring as part of the flight software and ground support stack.

Power management illustrates the broader principle \cite{lollar1987automated}. Platforms typically must allocate generation and storage across competing loads while maintaining safety margins for critical subsystems. Automated controllers can prioritise essential functions (e.g., thermal regulation, communications, and life-support in crewed contexts), schedule high-demand activities into favourable periods, and shed non-critical loads when margins tighten. Data-driven forecasting can complement this by predicting near-term demand and generation profiles from historical patterns and mission timelines, improving the quality of scheduling decisions and reducing unnecessary conservatism \cite{RINGER1991365}.

A similar logic applies to consumables and closed-loop systems. Water and air revitalisation, waste processing, and other regenerative subsystems are coupled and can exhibit complex dynamics. Predictive models can estimate consumption rates under different activity profiles and environmental conditions, supporting planning and early warning for off-nominal trends. For longer missions, the engineering objective is stable operation with minimal resupply, which places a premium on monitoring, fault diagnosis, and control policies that can maintain performance despite gradual drift and component aging.

Bioregenerative life-support concepts further highlight the need for careful monitoring and control \cite{gratius2024digital,lasseur2010melissa}. ESA’s MELiSSA programme \footnote{\url{https://www.esa.int/Enabling_Support/Space_Engineering_Technology/Melissa}}, for example, investigates a multi-compartment loop that uses biological and physico-chemical processes to recycle waste streams into oxygen, water, and biomass. Such systems require tight regulation of variables such as nutrient balance, microbial growth, and gas exchange. Here, automation and data-driven optimisation can support stability by detecting early signs of imbalance and adjusting operating points before the loop destabilises. Across these examples, the common aim is mission assurance: using predictive monitoring and automated control to keep resource usage within safe envelopes, increase efficiency, and maintain habitability over extended durations.

\begin{table*}[htbp]
\small
\centering
\setlength{\tabcolsep}{3pt}
\renewcommand{\arraystretch}{1.05}
\caption{Operational Space Debris Tracking Technologies (Ground-based \& Space-based), Part I: Ground-Based Systems.}
\label{tab:debris_tracking_1}
\begin{tabular}{p{3.2cm} p{2.2cm} p{1.7cm} p{3.2cm} p{4.9cm}}
\toprule
\textbf{System} & \textbf{Sensor Type} & \textbf{Orbit Domain} & \textbf{Operational Conditions} & \textbf{AI Methods Used} \\
\midrule
\multicolumn{5}{l}{\textbf{Ground-Based Systems:}} \\
\addlinespace[0.4ex]
\textit{Space Fence} (USSF, Kwajalein) &
Radar (S-band phased array) &
LEO, MEO, GEO &
All-weather, 24/7 continuous scan; detects objects as small as &
Automated track initiation and orbit updates (advanced signal processing; proprietary algorithms, incl.\ Kalman filtering) \\

\textit{AN/FPS-85} (USAF Eglin) &
Radar (UHF phased array) &
LEO (extends to MEO) &
All-weather; wide-area sector scans; regular catalog maintenance &
Standard tracking filters (Kalman filter-based processing; no public use of ML) \\

\textit{LeoLabs Radar Network} (commercial) &
Radars (phased-array, S-band and UHF at multi-sites) &
LEO &
All-weather; multi-site coverage with frequent revisits (sub-10 cm object tracking) &
AI-driven data analytics for event detection and object characterization (e.g., anomaly detection, maneuver identification) \\

\textit{TIRA} (Fraunhofer, Germany) &
Radar (L-band tracking/imaging) &
LEO (up to \(\sim\)3000 km) &
All-weather; dish antenna tracks one object at a time (high-resolution imaging) &
Deep learning-based detection (YOLO CNN tested to enhance small-debris detection); classical orbit determination via Kalman filters \\

\textit{GESTRA} (Germany) &
Radar (phased-array surveillance) &
LEO (up to \(\sim\)3000 km) &
All-weather; electronically scans wide sky region; experimental continuous operation &
Automated detection and tracking (software-defined radar); no specific AI reported (designed for future ML integration) \\

\textit{GRAVES} (France) &
Radar (bistatic UHF fence) &
LEO (low inclinations) &
All-weather; continuous horizontal fan-beam transmission; objects detected on crossing beam &
Conventional signal processing and orbit correlation (no known AI; uses radar Doppler/track filtering) \\

\textit{Okno} (Nurek, Tajikistan) &
Optical telescopes (multi-aperture station) &
GEO, HEO (up to 40{,}000 km); also MEO &
Night-only, clear skies (passive detection of sun-lit objects); automated wide-field scanning &
Automated image analysis and orbit computation (no published AI; relies on brightness/trajectory detection algorithms) \\

\textit{Chinese PMO Telescopes} (Purple Mountain Obs.) &
Optical telescopes (network of sites) &
LEO, MEO, GEO &
Night-only, clear weather; multiple observatories (7+ sites) for full orbital regime coverage &
Research on deep learning (e.g.\ YOLOv8-based debris detection) shows improved optical detection efficiency (operational system algorithms not publicly disclosed) \\

\textit{Chinese SSA Radars} (PLA network) &
Radars (large phased-array, multi-band) &
LEO (primarily) &
All-weather; at least six fixed phased arrays (missile-warning-class) used for surveillance; near continuous tracking &
Predominantly classical radar processing (tracking filters); any AI/ML usage not public (focus on catalog maintenance via automated software) \\

\textit{GEODSS} (USSF, global sites) &
Optical telescopes (1~m aperture, electro-optical) &
Deep space (GEO/HEO mainly) &
Night-only operations; clear skies required; \(\sim\)3 sites cover GEO belt (tasked search \& revisit) &
Automated streak detection with image processing; standard orbit determination (no specific AI reported in ops, though ML-based streak detectors are under evaluation) \\

\textit{Space Surveillance Telescope (SST)} (Australia/US) &
Optical telescope (3.5~m wide-FOV) &
Deep space (GEO, cislunar) &
Night-only, clear weather; wide field survey scans; fast revisit for faint objects &
Advanced image processing pipeline for autonomous detection; considering ML enhancements for dim object extraction (current system uses classical algorithms; details classified) \\

\textit{Bisei Spaceguard Center} (JAXA, Japan) &
Optical telescopes (0.5--1~m) &
LEO, GEO &
Night-only, clear skies; scheduled observations of GEO and LEO targets &
Automated detection and tracking software (classical); Japanese research exploring AI (e.g.\ neural nets for streak detection) not yet in routine operation \\
\bottomrule
\end{tabular}
\end{table*}

\begin{table*}[htbp]
\small
\centering
\setlength{\tabcolsep}{3pt}
\renewcommand{\arraystretch}{1.05}
\caption{Operational Space Debris Tracking Technologies (Ground-based \& Space-based), Part II: Space-Based Systems (continued).}
\label{tab:debris_tracking_2}
\begin{tabular}{p{3.2cm} p{2.2cm} p{1.7cm} p{3.2cm} p{4.9cm}}
\toprule
\textbf{System} & \textbf{Sensor Type} & \textbf{Orbit Domain} & \textbf{Operational Conditions} & \textbf{AI Methods Used} \\
\midrule
\multicolumn{5}{l}{\textbf{Space-Based Systems:}} \\
\addlinespace[0.4ex]
\textit{SBSS / ORS-5 ``SensorSat''} (USA) &
Optical telescope (on spacecraft) &
LEO-based sensor, surveying GEO/HEO &
Operates above atmosphere; observes targets against space background (mostly night-side of Earth for sun-lit GEO objects) &
Onboard automation for detecting resident space objects (conventional image thresholding \& tracking; future models consider neural-net vision for onboard processing) \\

\textit{GSSAP constellation} (USSF, 4 sats) &
Optical sensors (on maneuverable satellites) &
Near-GEO orbit (monitoring GEO) &
Space-based, no weather constraints; flexible proximity operations around GEO; frequent revisits of high-value objects &
Autonomous navigation and imaging algorithms (exact methods classified); likely uses AI-aided target recognition and refined Kalman filtering for close approaches \\

\textit{Sapphire} (Canada) &
Optical telescope (small satellite) &
Sun-synchronous LEO (\(\sim\)700 km); monitors MEO/GEO &
Space-based imaging of objects in GEO; primary ops in Earth’s shadow with target in sunlight &
Ground-processing uses automated image analysis and orbit update (standard Kalman filters; no public info on AI usage in this mission) \\

\textit{GEO Inspector sats} (e.g.\ China \textit{SJ-17}) &
Optical sensors (on-orbit inspection craft) &
GEO &
On-orbit close inspection of satellites/debris; can perform rendezvous; not weather-limited &
AI-based vision and relative navigation systems suspected (e.g.\ pattern recognition for target approach), though specifics are not publicly released \\
\bottomrule
\end{tabular}
\end{table*}

\begin{figure}
\centering  
\includegraphics[width=\linewidth]{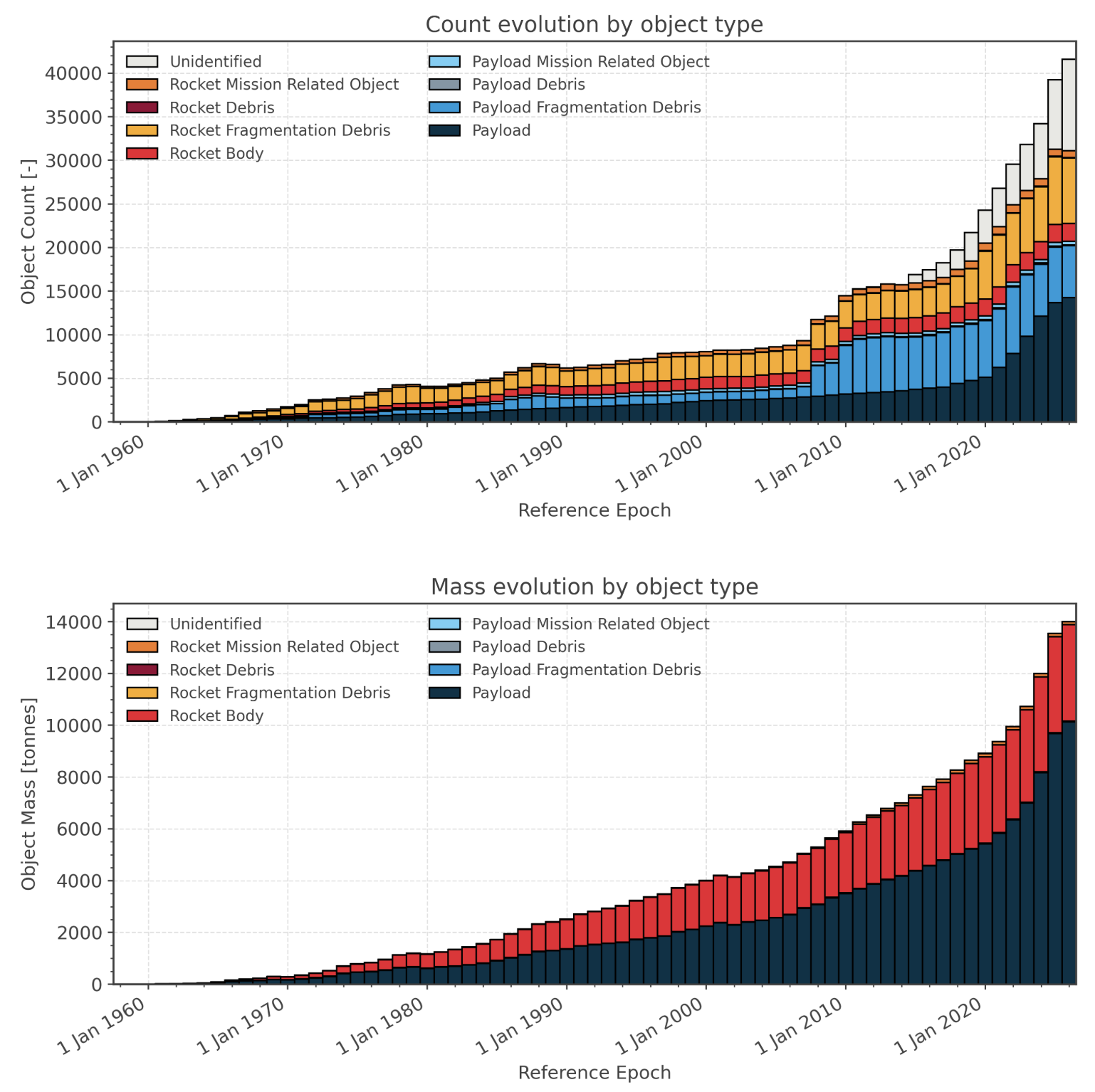}  
\caption{The Evolution of Number and Mass of objects in Earth Orbit \cite{esa2021esa}.}
\label{fig:objectearth}  
\end{figure}

\subsubsection{Space Debris Tracking, Removal, and Mitigation}

Space debris has become a defining constraint for safe and sustainable operations in orbit, especially in low Earth orbit (LEO). NASA's Orbital Debris Program Office (ODPO) provides illustrative visualisations of catalogued objects in LEO and GEO (Figure~\ref{fig:debris})\footnote{\url{https://orbitaldebris.jsc.nasa.gov/photo-gallery/}}, and reports that the catalogued population is dominated by debris rather than active spacecraft. As the number of objects grows, the main operational challenge is no longer detection alone, but timely tracking, reliable risk assessment, and scalable decision-making for avoidance and remediation \cite{kazemi2024orbit}.

\begin{figure}
\centering
\includegraphics[width=0.5\linewidth]{figure/debris.png}
\caption{(a) LEO stands for low Earth orbit and is the region of space within 2,000 km of the Earth's surface. It is the most concentrated area for orbital debris. (b) The GEO images are images generated from a distant oblique vantage point to provide a good view of the object population in the geosynchronous region (\(\sim\)35,785 km altitude). (c) The GEO Polar images are generated from a vantage point above the north pole, showing the concentrations of objects in LEO and in the geosynchronous region.}
\label{fig:debris}
\end{figure}

\begin{figure}
\centering
\includegraphics[width=\linewidth]{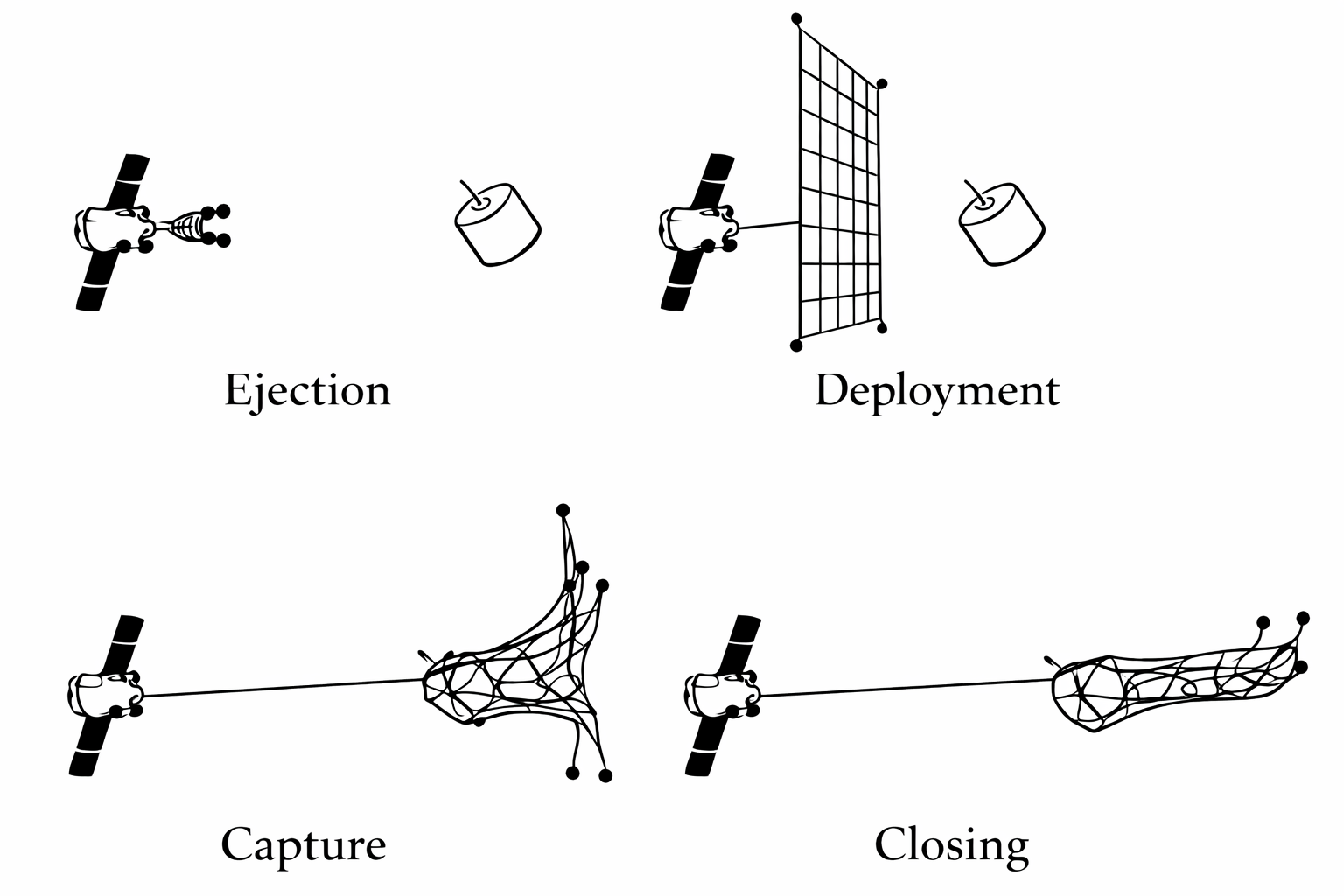}
\caption{Satellite net capture process diagram \cite{aslanov2023survey}.}
\label{fig:debris2}
\end{figure}

A first role for data-driven methods is to improve the throughput and robustness of detection and catalog maintenance. Radar and optical sensors produce large streams of measurements where false alarms, missed detections, and intermittent observations are unavoidable. Learning-based classifiers can assist with filtering and association by distinguishing likely objects from sensor artefacts and background noise, and by supporting track linking when observations are sparse or heterogeneous (e.g., mixing optical imagery and radar returns) \cite{aldahoul2022rgb}. In engineering terms, the goal is not to replace orbit determination, but to reduce operator burden and improve the reliability of downstream estimation by supplying cleaner candidate measurements and better data association hypotheses.

A second role is conjunction assessment and collision-risk prioritisation. Physics-based propagation remains the backbone of risk modelling, but operational pipelines must cope with large volumes of alerts and uncertainty in state estimates. Data-driven components can help characterise systematic prediction errors, triage low-consequence encounters, and highlight cases where uncertainty growth or poor geometry makes the risk assessment fragile. The most valuable outputs are therefore risk-ranked candidate events accompanied by uncertainty indicators that help operators decide when an avoidance manoeuvre is justified, rather than blanket automation that may increase manoeuvre burden or introduce unnecessary fuel penalties.

Finally, active debris removal introduces autonomy requirements that are qualitatively different from passive tracking. Capturing non-cooperative and potentially tumbling targets requires reliable relative navigation, real-time perception, and conservative contact strategies \cite{napolano2025cnn}. Here, onboard vision and learning-based pose estimation can support terminal guidance, while planning and control must remain bounded by strict safety constraints to avoid unintended collisions or unstable contact dynamics. At the mission level, optimisation tools can support sequencing and route planning for multi-target concepts, but practical designs must account for fuel margins, capture uncertainty, and the risk of failed attempts.

Overall, AI contributes to orbital debris management most effectively when embedded within a conservative, uncertainty-aware workflow: improving measurement triage and association for tracking, supporting scalable risk prioritisation for conjunction management, and enabling safer perception and decision support for removal missions. These capabilities complement, rather than replace, established astrodynamics and safety engineering, and they become increasingly important as orbital environments move toward higher density and tighter operational margins.

\subsubsection{Life Support System Automation and Optimization}
Life support is a safety-critical subsystem that must maintain a stable habitat under tightly coupled constraints on power, mass, consumables, and maintenance opportunities \cite{dewberry1990space,gratius2024digital}. As missions extend in duration and operate with longer communication delays, greater onboard automation becomes essential not only for efficiency but also for early warning and resilience. In practice, the most credible role of AI is decision support embedded within conservative control architectures: forecasting demand, detecting off-nominal trends, and recommending actions that keep the system within certified operating envelopes.

Atmosphere management illustrates the need for adaptive operation. Oxygen supply, carbon dioxide removal, humidity control, and ventilation must respond to changing crew activity and cabin conditions \cite{coker2016predictive}. Automated controllers can adjust setpoints and operating modes based on real-time sensor inputs, while predictive models estimate near-term metabolic load and help prevent undesirable excursions, such as local CO$_2$ build-up during high exertion. Similar principles apply to CO$_2$ scrubbing and regeneration scheduling, where consumption varies over the daily timeline and where optimal cycling must balance removal efficiency, power usage, and component wear \cite{beard2020characterization}. Data-driven monitoring can complement these controllers by flagging gradual performance degradation and by supporting maintenance planning before margin is exhausted.

Long-duration habitation also requires careful management of water loops and microbial risk \cite{dewberry1991intelligent}. Closed-loop recycling systems can drift over time due to biofilm formation, filter saturation, or changes in chemistry \cite{stahl2021real}. Early detection is therefore as important as control. Data-driven anomaly detection can fuse signals such as conductivity, turbidity, pressure drops, and trace chemical indicators to surface potential contamination events earlier than simple threshold rules. When coupled to safe response procedures, this enables timely corrective actions, from increased purification cycles to targeted disinfection, without imposing constant manual monitoring on the crew.

Finally, life support cannot be separated from crew health and space medicine. Physiological monitoring, workload, sleep, radiation exposure, and nutrition all interact with habitat conditions and operational planning. Onboard decision support can help crews and flight surgeons interpret trends, prioritise interventions, and manage limited medical resources, particularly when immediate ground consultation is not available \cite{ibrahim2025generative}. The operational aim is robust habitability: keeping the habitat stable, detecting failures early, and maintaining crew health with minimal cognitive burden and clear accountability.

\subsubsection{Space Manufacturing with Autonomous System}
In-space manufacturing and repair reduce dependence on resupply and enable infrastructure that is impractical to launch fully assembled \cite{arney2021orbit}. The core challenge is assurance: manufacturing processes must remain stable and predictable under microgravity, thermal variation, and limited opportunities for hands-on inspection. Automated monitoring and control therefore become central, with perception and decision support layered on top of conservative process constraints.

Additive manufacturing in orbit provides a representative example \cite{prater20193d,hoffmann2023space}. Printing large structural elements or functional components requires tight control of process parameters and continuous quality checks. Sensor-driven control can detect deviations such as layer misalignment, nozzle issues, or thermal distortion and can pause, adjust parameters, or trigger corrective routines \cite{cai2023review,abouelnour2022situ,sani2024artificial}. Vision-based inspection and metrology can then verify geometry and surface integrity, allowing defects to be detected early when they are still recoverable. This transforms quality assurance from an offline activity into a continuous, in-process capability, which is critical when failed parts are costly and replacement is not immediate.

Autonomous manufacturing also links naturally to on-orbit maintenance and resilience. As robotic assembly becomes more common, inspection, repair, and fabrication form a single operational loop: identify damage or wear, manufacture or retrieve a replacement part, and perform installation with verified alignment. Over longer time horizons, this loop may incorporate self-sensing and self-repairing materials, where embedded sensors detect structural changes and automated procedures initiate protective or restorative actions. Here again, the engineering objective is not speculative autonomy for its own sake, but predictable operation under uncertainty: detecting true damage reliably, choosing safe interventions, and preserving structural integrity without excessive crew time or manual procedures.

Overall, autonomous manufacturing and repair extend the logistics horizon of space systems. By enabling fabrication, inspection, and maintenance in situ, they support modularity, reuse, and scalable infrastructure, which are all prerequisites for sustained activity in orbit and beyond.

\subsection{Intelligent Space Traffic Management}

\subsubsection{Satellite Collision Avoidance and Predictive Algorithms}

\begin{figure}
\centering
\includegraphics[width=\linewidth]{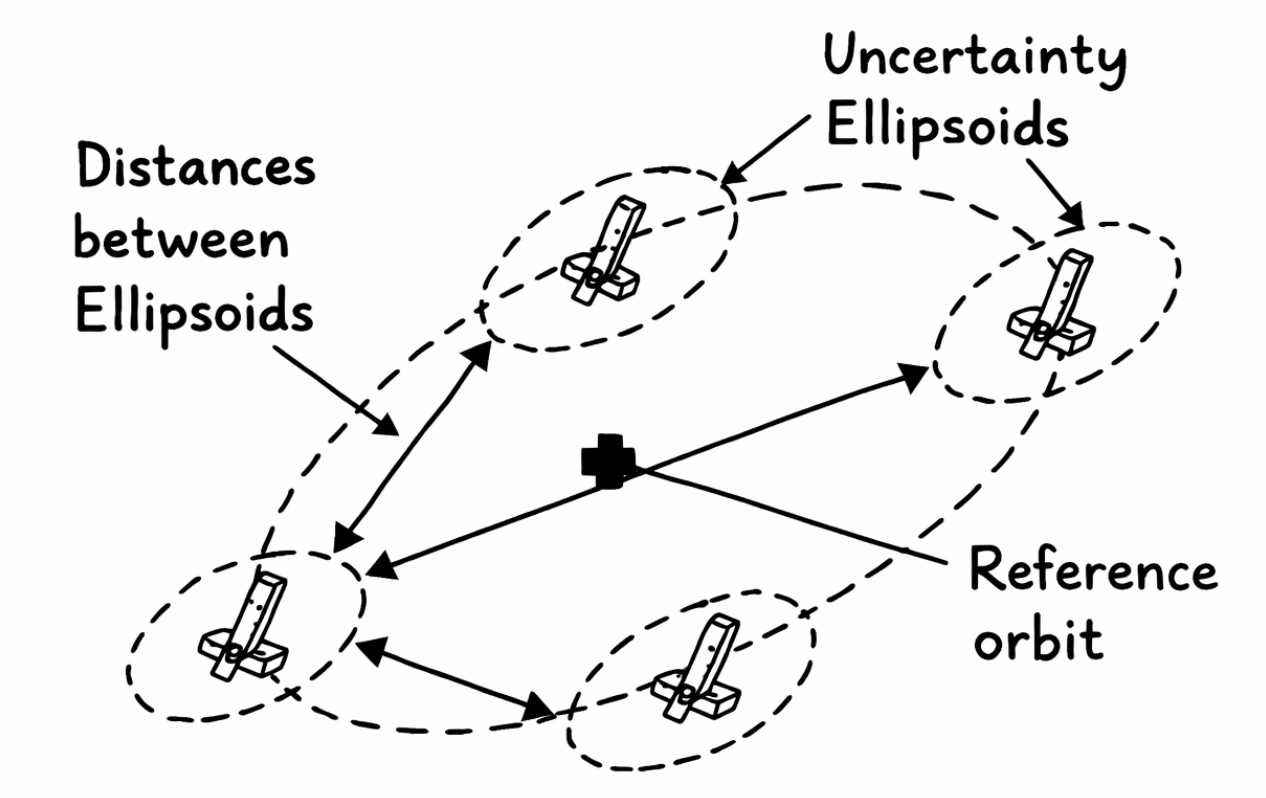}
\caption{The collision avoidance approach using ellipsoids.}
\label{fig:collision}
\end{figure}

The rapid growth of active spacecraft and the persistent debris environment have made conjunction management a day-to-day operational burden. Collision avoidance begins with tracking and orbit determination, but the decision bottleneck often lies in uncertainty: limited observations, modelling error, and covariance growth can produce large volumes of alerts with widely varying credibility \cite{stoll2013operational,slater2006collision,bonnal2020just}. In this setting, data-driven methods are most useful when they improve triage and reduce false urgency. For example, statistical and machine-learning models can help characterise systematic prediction error under different regimes (altitude, solar activity, manoeuvre behaviour), refine uncertainty estimates, and prioritise events where the risk assessment is both high and reliable enough to justify action.

A second emerging direction is privacy-preserving collaboration across operators. Many of the most informative signals for risk assessment are proprietary, including precise ephemerides, manoeuvre history, and internal uncertainty models. Federated or distributed learning has therefore been proposed as a way to improve shared predictors without requiring raw data exchange \cite{fang2022olive}. The practical aim is modest but important: better calibrated risk ranking and fewer missed high-risk events, achieved through learning from a broader population of behaviours while respecting data boundaries.

Finally, autonomy is increasingly relevant to execution. When contact windows are limited or time-to-conjunction is short, an onboard system may need to recommend or initiate conservative responses within pre-authorised envelopes. Here, the priority is assured behaviour rather than aggressive optimisation. A safe pattern is to automate well-defined steps (validation checks, candidate manoeuvre generation, feasibility screening) while keeping human operators in the loop for final approval when time allows, and ensuring robust fallbacks when ambiguity remains.

\subsubsection{Traffic Optimization and Coordination in Low Earth Orbit (LEO)}
Beyond individual conjunctions, congestion in LEO is a system-level problem shaped by constellation geometry, common altitude shells, station-keeping practices, and the accumulation of non-manoeuvrable objects \cite{lei2024spatial,hu2022traffic,wei2024towards}. Traffic optimisation therefore involves coordination across many spacecraft over long time horizons, not only last-minute avoidance. Modelling tools can represent the orbital environment as a time-evolving interaction network and evaluate how local decisions propagate into global risk. Within such frameworks, optimisation can support deconfliction strategies such as phasing policies, altitude-band allocation, or coordination rules that reduce repeated close approaches.

Scenario simulation is particularly valuable for policy and planning. By running large ensembles of ``what-if'' analyses that vary deployment rates, disposal compliance, tracking performance, and manoeuvre norms, operators and regulators can assess how different practices affect long-term collision risk and operational workload. In this role, AI is best understood as an accelerator for simulation and decision support, helping evaluate large design spaces and exposing trade-offs rather than replacing governance.

\subsubsection{AI for Regulatory Compliance and Safety in Orbit}
As commercial activity expands, compliance and accountability become as important as technical avoidance \cite{ancona2025technical,oltrogge2018we}. Many regulatory obligations are operational in nature: adherence to licensed frequency use, disposal timelines, coordination responsibilities, and reporting requirements. Automated compliance monitoring can reduce burden by continuously checking operational states against declared constraints and producing structured logs and reports \cite{reed2020blockchain}. This is especially relevant as fleets scale, where manual compliance tracking becomes error-prone and costly.

Safety in orbit also depends on shared norms and auditable behaviour. For autonomous or semi-autonomous decision-making, it is often useful to treat policies as ``constrained autonomy'' with explicit safety cases: what the system is allowed to do, what triggers escalation to humans, and what evidence is recorded for post-event review. Secure logging and provenance mechanisms (e.g., cryptographic signing of manoeuvre intentions and execution records) can strengthen transparency without requiring all parties to share raw telemetry \cite{suh2025encrypted,berry2018evolution}. Distributed-ledger approaches are sometimes discussed in this context, but their operational value depends on careful integration with real-time constraints and on clear governance of data access and liability. In practice, the near-term priority is simpler: consistent, machine-checkable reporting; interoperable data formats; and audit trails that make responsibility and compliance verifiable at scale.

\section{AI in Deep Space: Expanding Human Horizons}

Deep-space missions operate under long communication delays, harsh environments, and limited opportunities for real-time support from Earth. Autonomy is therefore not a convenience but a practical requirement. AI contributes most when it increases operational independence while preserving safety margins: improving navigation in uncertain terrain, prioritising scientific measurements under constrained contact, and enabling coordinated behaviour across multiple assets. This chapter reviews deep-space autonomy from two complementary angles: navigation and exploration, and AI-enabled scientific discovery. The emphasis is on how missions translate perception into reliable action and how they turn limited observation opportunities into maximal scientific return.

\subsection{Autonomous Navigation and Exploration}
Robots exploring planetary surfaces and small bodies must navigate without GPS and often with infrequent ground guidance. Autonomy depends on reliable local perception and mapping, safe path planning under uncertain terrain properties, and resource-aware decision-making when energy and time are limited. Modern systems combine classical robotics methods such as SLAM and optimisation-based planning with learned perception modules that improve scene understanding under varying illumination and terrain appearance.

\subsubsection{Path Planning, Obstacle Avoidance, and Terrain Analysis}
Surface navigation on the Moon, Mars, and asteroids is dominated by risk: rocks and cliffs threaten immobilisation; soft regolith and slopes can induce slippage or entrapment; and localisation errors accumulate when terrain lacks stable landmarks. Early Mars rovers demonstrated basic autonomy by periodically stopping to assess hazards and then proceeding cautiously. Even with such precautions, terrain interactions remained mission-limiting, motivating more capable perception and traversability reasoning.

A key development is learned terrain classification \cite{10687827,chen2022tcn}. Systems such as JPL's Soil Property and Object Classification (SPOC)\footnote{\url{https://nasa-jpl.github.io/SPOC/}} use convolutional models to label terrain classes in imagery and estimate the likelihood of hazardous surfaces such as soft sand. These outputs do not replace physics-based reasoning, but they provide a practical ``risk map'' that can be fused with stereo geometry, wheel odometry, and inertial sensing. In turn, planners can incorporate learned cost models to bias routes away from likely hazards while still making progress toward mission targets. When combined with local 3D mapping and localisation, this supports continuous replanning as new obstacles enter view, rather than relying on sparse waypoint updates from Earth.

Recent missions demonstrate the operational impact of this integration. Perseverance's navigation architecture extends earlier AutoNav concepts by processing imagery more continuously and enabling longer autonomous segments per planning cycle, reducing idle time and improving driving efficiency in cluttered terrain. In parallel, European rover designs such as ExoMars (Rosalind Franklin) have emphasised autonomous vision-based navigation and slip-aware mobility strategies, reflecting the same engineering lesson: reliable traversability assessment and conservative replanning are prerequisites for sustained surface exploration.

Local planning is inseparable from localisation and map-building. Visual odometry and SLAM track features across images to estimate motion and maintain a consistent map, but performance can degrade under lighting changes, repeated textures, and feature-poor regions. Learned feature extraction and adaptive mapping strategies can improve robustness by recognising more stable landmarks and by supporting re-localisation when drift grows. The practical goal is not millimetre accuracy, but sufficient consistency that planners can trust the robot's pose estimate when negotiating hazards and executing science operations.

\subsubsection{Autonomous Long-Distance Traversal and Exploration}
Long-distance traversal extends local autonomy from metres to mission-scale mobility \cite{chen2025robust}. Beyond avoiding immediate hazards, the robot must manage energy, thermal constraints, time-to-communication windows, and contingency handling across multi-step plans. Route selection becomes a decision problem under uncertainty: a shorter path may be riskier or more energy-intensive; a safer path may delay science objectives. Autonomy helps by evaluating alternatives, updating plans as conditions change, and ensuring that progress remains compatible with resource margins.

Energy-aware planning is particularly important for solar-powered assets and for operations in extreme illumination conditions. Lunar polar exploration provides a clear example because permanent shadow and intermittent sunlight create time-varying constraints that must be respected to avoid power depletion. Planning systems can combine terrain models, illumination forecasts, and activity timelines to schedule safe excursions and ensure return to favourable conditions. Although specific programmes evolve over time, this class of capability is representative of what deep-space autonomy must deliver: decisions that are explicitly conditioned on environmental dynamics rather than assuming static operating envelopes.

A second frontier is multi-agent exploration, where multiple robots or spacecraft cooperate to achieve sensing and science goals that exceed the capability of any single asset. Coordination requires task allocation, communication-aware planning, and safety-aware deconfliction. NASA's Cooperative Autonomous Distributed Robotic Exploration (CADRE) project\footnote{\url{https://www.jpl.nasa.gov/missions/cadre/}} illustrates this direction by aiming to demonstrate a team of small rovers that coordinate through a mesh network, share observations, and divide tasks while operating with limited continuous human oversight. The scientific motivation is distributed sensing, such as multi-point measurements and rapid area coverage, while the engineering motivation is robustness through redundancy and shared situational awareness.

Multi-agent ideas also arise in deep-space proximity operations and small-body missions. ESA's Hera mission\footnote{\url{https://www.esa.int/Space_Safety/Hera}}, which will study the Didymos--Dimorphos system, includes accompanying CubeSats that enable complementary measurements from multiple vantage points. Such architectures depend on autonomous relative navigation, timing, and coordination policies that keep assets safe while maximising joint measurement value. Even when assets are not peer-to-peer, coordination between heterogeneous platforms is increasingly relevant. The Mars 2020 mission\footnote{\url{https://science.nasa.gov/mission/mars-2020-perseverance/}} demonstrated how an aerial scout can support ground navigation: Ingenuity performs autonomous flights using onboard visual navigation and can provide terrain context that informs rover-level planning. This pattern generalises to future missions where orbiters, landers, and mobile agents cooperate across modalities and scales.

Overall, deep-space navigation autonomy is moving from cautious, stop-and-go hazard checks toward sustained, resource-aware mobility and coordinated exploration. The outcome is not merely faster driving; it is a qualitative increase in what missions can attempt under latency and uncertainty, enabling broader coverage, richer measurements, and more resilient execution in environments where help from Earth is delayed.

\subsection{AI-Driven Scientific Research and Discovery}
Beyond navigation, deep-space autonomy increasingly extends into the scientific loop itself. Many missions can no longer rely on a simple ``collect everything and decide later'' model: downlink is limited, latency is high, and transient opportunities may be missed by the time data reaches Earth. Practical autonomy therefore targets three capabilities. First, instruments and robots execute measurement sequences with limited supervision, adjusting settings or sampling strategy when intermediate results indicate higher value. Second, onboard processing performs rapid triage, compression, and event detection to prioritise what should be transmitted or followed up. Third, missions use decision support to select and prioritise targets under strict constraints on time, energy, and contact windows. Together, these elements shift spacecraft from passive data collectors toward systems that can allocate attention and resources in response to what they observe.

\subsubsection{Autonomous Scientific Experimentation}
Fully autonomous hypothesis generation remains beyond current operational practice, but modern missions increasingly support closed-loop experimentation in narrower, well-defined forms. The common pattern is conservative autonomy: onboard software selects targets or adapts measurement parameters within pre-authorised bounds, using locally available sensing to make decisions when the ground team is out of contact.

A representative example is autonomous target selection for rover instruments. The AEGIS system on Mars rovers enables automated selection of candidate rock targets for follow-on measurements based on onboard imagery and user-defined priorities \cite{francis2017aegis}. Rather than waiting for daily command cycles, the rover can use otherwise idle time after a drive to acquire additional measurements, improving utilisation of limited operational windows. The key point is not that the rover replaces geologists, but that it executes a safe, bounded extension of the science plan when ground interaction is unavailable.

Perseverance extends this idea to adaptive measurement at the instrument level. PIXL\footnote{\url{https://science.nasa.gov/resource/pixl-instrument-on-nasas-perseverance-studies-bills-bay/}} supports ``adaptive sampling'', where intermediate spectra guide subsequent measurement allocation, allowing the instrument to concentrate effort on locally informative regions of a target rather than uniformly scanning with a fixed pattern. This reduces wasted integration time and helps reach actionable conclusions within a single activity block. Importantly, the autonomy here is tightly coupled to the mechanics of measurement: fine positioning, stability control, and parameter selection determine whether high-resolution mapping is successful, and onboard logic can correct for small disturbances that cannot be teleoperated in real time.

Autonomous experimentation is also evident in sample acquisition under extreme latency. OSIRIS-REx\footnote{\url{https://science.nasa.gov/mission/osiris-rex/}} provides a clear case: the Touch-and-Go sampling sequence at Bennu required autonomous guidance and hazard-aware navigation because real-time teleoperation was impossible. Vision-based Natural Feature Tracking (NFT) enabled robust localisation against a rocky, cluttered surface and supported execution of the sampling manoeuvre within tight safety margins \cite{lorenz2017lessons}. Similar autonomy patterns appear in other small-body missions, where the combination of microgravity dynamics and uncertain terrain makes pre-scripted, open-loop procedures risky \cite{9076194}.

Finally, orbital laboratories provide a pathway toward more general closed-loop experimentation. On the ISS and related platforms, experiments increasingly incorporate automation for imaging, monitoring, and procedure execution. While most decision-making remains ground-directed, these environments are natural testbeds for ``self-driving'' experimental loops in which measurement frequency, experimental conditions, or sample handling are adjusted based on observed results. The operational relevance is clear for deep space: when contact is delayed, systems that can run safe closed-loop protocols autonomously can increase science throughput and reduce missed opportunities.

\subsubsection{Real-Time and Adaptive Data Analysis}
Onboard analysis is becoming essential because sensor output often exceeds downlink capacity and because time-critical events require rapid response. The purpose of onboard processing is therefore triage and action: identify what is worth transmitting, what should trigger follow-up observations, and what can be safely compressed or deferred \cite{8742190}. Classic automated planning/scheduling systems (e.g., ASPEN \cite{fukunaga1997aspen}) have been used to generate and repair constraint-rich activity plans for spacecraft operations. EUROPA \cite{barreiro2012europa} is another platform providing reusable planning architecture rather than a mission-specific solution.

Instrument-level adaptive analysis provides one template, as in PIXL’s real-time interpretation of spectra to guide additional measurements. Similar patterns occur across modalities. Imaging systems can flag scenes with rare or time-critical content, spectrometers can perform onboard feature detection to identify candidate compounds or anomalous signatures, and environmental sensors can trigger opportunistic measurements when unusual conditions are detected. The engineering requirement is reliability under uncertainty: the system must expose confidence, avoid over-committing to brittle classifications, and preserve raw data when ambiguity is high.

Adaptive observation can also be framed as event-driven replanning. The Autonomous Sciencecraft Experiment on EO-1\footnote{\url{https://science.nasa.gov/mission/eo-1/}} demonstrated a closed loop in which onboard analysis identified events of interest and retasked the observation plan accordingly \cite{chien2004eo,tran2005autonomous}. Although EO-1 operated in Earth orbit, the underlying principle generalises directly to deep-space science: a spacecraft that can recognise a plume, impact event, or transient atmospheric change and then reallocate instrument time can capture phenomena that would otherwise be missed due to planning latency. In deep-space settings, this requires careful integration with resource constraints and safety rules, since replanning competes with power, thermal limits, and communication priorities.

A related capability is anomaly and novelty detection in large data streams. Rather than classifying predefined classes, these methods highlight outliers that merit attention, such as unusual textures, colours, or spectral patterns. When coupled to conservative follow-up actions, novelty detection can guide additional measurements and prioritise downlink for potentially high-value discoveries. This bridges the gap between systematic survey operations and truly discovery-driven exploration.

Finally, there is growing interest in distributed learning and knowledge transfer across missions and assets. Full federated learning across deep-space platforms is not yet an operational norm, but the motivation is clear: multiple instruments and vehicles observe related phenomena under different conditions, and sharing compact model updates could enable collective improvement without transferring raw data. Even today, sequential transfer across missions is common, where models trained on heritage datasets or prior missions are adapted for new environments. As multi-asset exploration expands, this direction points toward architectures in which orbiters, landers, and mobile robots not only share data products but also share and update the models that interpret them.

In summary, AI-driven science in deep space is best understood as a set of bounded capabilities that increase scientific throughput under latency and bandwidth constraints. By enabling conservative closed-loop experimentation, fast triage, and event-driven replanning, missions can allocate limited time and resources toward the most informative observations, improving discovery potential while maintaining operational safety.

\subsubsection{Scientific Target Identification and Prioritization}
A recurring constraint in exploration is that the number of candidate targets far exceeds available time, energy, and downlink \cite{smith2005rover}. Target selection is therefore a resource-allocation problem: which scene, sample, or phenomenon deserves additional measurements now, and what can be deferred or ignored without compromising mission objectives. AI supports this process through three complementary capabilities: novelty detection to surface unexpected signals, utility-based prioritisation to balance competing objectives, and cross-instrument or cross-mission data fusion to connect evidence across scales.

Novelty and anomaly detection are particularly valuable when missions operate in survey mode or collect long streams of similar measurements. Rather than specifying every interesting pattern in advance, novelty detection flags observations that differ from recent context or from learned nominal distributions, enabling opportunistic follow-up. On rovers, this idea aligns with autonomous target selection workflows in which image context is used to propose candidate rock targets for additional spectroscopy when the ground team is unavailable \cite{francis2017aegis,chien2004eo,tran2005autonomous}. In orbiter and telescope settings, analogous triage is used to surface transient or rare signatures that warrant higher-rate sampling or rapid downlink, especially when contact windows are limited \cite{bowkett2025autonomous}.

Prioritisation can also be framed as decision-making under constraints. Given a limited set of actions (e.g., imaging, drilling, repositioning, or allocating instrument time), planners must balance expected scientific value against cost in time, risk, power, and thermal margins. Probabilistic planning and decision-theoretic formulations, including POMDP-style approaches, provide a structured way to encode uncertainty and exploration--exploitation trade-offs. Reinforcement learning has been explored in simulation as a way to learn policies that maximise science utility over long horizons, but operational deployment must remain bounded by verification and safety rules. In practice, these methods are most immediately useful as decision aids that evaluate alternative observation sequences and expose trade-offs, rather than as fully autonomous ``science directors.''

A third dimension is evidence integration across instruments and missions. Many science questions require linking orbital context to in situ measurements, or fusing data across modalities such as imagery, spectroscopy, radar, and environmental sensors. Machine learning can assist by discovering correlations across heterogeneous datasets and producing ranked hypotheses about where evidence is most consistent. The outcome is practical: better targeting of scarce in situ investigations and more informative follow-up strategies. The same principle applies to planetary defense, where integrating observations across multiple sensors and epochs can improve orbit estimation and guide follow-up scheduling \cite{irureta2025deep,wolfe2025near,caldas2024machine}.

Overall, target identification and prioritisation is best seen as an extension of mission autonomy into the scientific loop. When implemented conservatively, it increases scientific yield by directing limited resources toward the most informative measurements, while keeping humans in control of objectives, interpretation, and high-consequence choices.

In summary, deep-space AI contributes most when it closes small, well-defined loops: identifying candidates, allocating attention, and triggering follow-up within verified safety envelopes. These capabilities complement human science teams by increasing responsiveness under latency and bandwidth constraints and by ensuring that high-value opportunities are less likely to be missed.

\subsection{Geological Feature Recognition and Resource Detection}
Identifying habitable environments and potential in-situ resources requires geological interpretation across scales, from orbital mapping to surface and subsurface sensing \cite{wagstaff2015landmark}. AI contributes by extracting structured information from high-dimensional remote-sensing products, supporting probabilistic resource mapping, and prioritising follow-up observations that reduce uncertainty. These capabilities are relevant both to scientific questions (e.g., past water activity, sedimentary history) and to future utilisation scenarios (e.g., water ice for ISRU).

\subsubsection{Geological Feature Recognition}
Planetary remote sensing instruments such as hyperspectral imagers, radar sounders, and seismometers produce data products that are rich but difficult to interpret at scale. Hyperspectral imaging provides mineralogical fingerprints across many spectral bands, and machine-learning pipelines can assist with dimensionality reduction, denoising, and classification of surface materials when trained and validated against laboratory spectra and heritage datasets \cite{wang2025advanced}. These tools are particularly useful for mapping hydrated minerals or alteration products that serve as indicators of past water-related processes, while also exposing uncertainty when signatures are weak or confounded by illumination and viewing geometry.

Subsurface sensing adds another layer of complexity. Ground-penetrating radar and related instruments can reveal stratigraphy, buried interfaces, and subsurface heterogeneity, but interpretation often depends on assumptions about material properties and noise. Data-driven methods can support segmentation of radargrams, detection of interfaces, and fusion with rover localisation and surface context to build more coherent stratigraphic models. For example, Perseverance’s RIMFAX\footnote{\url{https://science.nasa.gov/blog/searching-for-buried-treasure-on-mars-with-rimfax/}} instrument provides radar observations that can be processed into stratigraphic interpretations; learning-based approaches may assist with feature extraction and pattern recognition, particularly when combined with geological priors and simulation-informed training data.

Resource probability modelling can further translate geological indicators into actionable maps. By combining variables such as temperature, topography, illumination, radar/spectral signatures, and heritage observations, statistical or machine-learning models can produce ranked likelihood surfaces for volatiles such as water ice or other target resources. These products are most useful when presented as probabilistic estimates with uncertainty, enabling mission designers to balance expected value against landing and traversal risk, and to plan follow-up measurements that reduce ambiguity.

\subsubsection{Near-Earth Object Detection and Planetary Defense}
Planetary defense relies on early detection, orbit determination, and risk assessment for near-Earth objects (NEOs) \footnote{\url{https://science.nasa.gov/planetary-defense/}}. Modern surveys produce large image streams where faint, fast-moving objects can be difficult to identify reliably. AI-assisted pipelines can support candidate detection, artefact rejection, and triage for follow-up observations, complementing classical image differencing and orbit fitting \cite{duev2019deepstreaks,denneau2013pan,irureta2025method,acciarini2025eclipsenets}. The operational value is throughput and prioritisation: focusing limited telescope time on candidates most likely to be real objects with non-negligible risk.

Risk assessment ultimately depends on probabilistic orbit determination under uncertainty and on long-horizon propagation with perturbations \cite{milani2005nonlinear,chamberlin2001sentry,tommei2021impact,solin2018monitoring}. Decision-support tools can help maintain and update risk lists as new observations arrive, and can assist in prioritising follow-up based on uncertainty reduction potential rather than on point estimates alone \cite{national2010defending,virtanen2006time}. In mission design and mitigation planning, simulation frameworks can explore deflection strategies and observation campaign design; learning-based optimisation has been studied in this context to accelerate scenario evaluation and to explore large parameter spaces, but such outputs must remain grounded in validated dynamics models \cite{wang2022intelligent,yuan2023integrated,zohdi2022machine,vasile2008optimal}.

Overall, AI contributes to planetary defense by scaling detection and triage, improving the efficiency of follow-up scheduling, and accelerating scenario analysis for risk reduction, while core astrodynamics remains the foundation for orbit estimation and impact probability assessment.

\subsection{Human-AI Collaboration in Deep Space}
Deep-space missions place crews and systems under conditions that differ qualitatively from near-Earth operations: communication is delayed or intermittent, maintenance opportunities are limited, and the cost of error is high. Under these constraints, the central question is not whether AI is ``useful,'' but how to design human--AI teaming that improves safety and mission performance while preserving clear authority, accountability, and predictable behaviour. This section focuses on two aspects: onboard assistance and shared decision-making for crews, and remote operation frameworks where autonomy compensates for latency.

\subsubsection{Astronaut Assistance Systems and Shared Decision-Making}
On long-duration missions, crews face continuous operational demands: monitoring habitat status, managing schedules, executing procedures, and maintaining awareness of health and environmental risk. Decision-support systems can reduce routine cognitive load by consolidating information, maintaining checklists, and supporting retrieval of procedures and configuration states through hands-free interfaces \cite{foutter2024space}. When integrated with onboard sensors, monitoring can provide early warning for off-nominal trends in life-support, power, thermal conditions, and radiation exposure, helping crews respond before margins are exhausted.

For maintenance and troubleshooting, context-aware guidance can improve consistency. Mixed-reality interfaces, for example, can overlay step-by-step procedures and safety checks on equipment, reducing the likelihood of omission errors when tasks are performed under fatigue or time pressure \cite{he2025gaze}. In deep space, the value of such tools is amplified because external expertise cannot be consulted in real time; onboard guidance must therefore be robust to ambiguity and explicit about confidence and assumptions.

Shared decision-making requires conservative design. In safety-critical situations, an AI system may propose actions such as shelter protocols during solar activity or adjustments to environmental control settings, but the system must make its reasoning inspectable. Practical transparency can be achieved by exposing the evidence behind alerts, enumerating plausible alternatives, and stating the expected consequences of each option. Authority boundaries should be explicit: well-characterised routine actions may be automated within predefined envelopes, while irreversible or ethically sensitive decisions remain under human control. This structure supports calibrated trust, reduces automation surprise, and keeps accountability aligned with crew responsibility.

\subsubsection{Remote Operation and Support Systems for Deep Space Missions}
Deep-space exploration also relies on remote operation of robotic assets under latency. Because real-time teleoperation is often infeasible, effective control shifts from low-level commanding toward goal-based supervision \cite{bresina2005activity,bresina2005mixed}. Operators specify objectives and constraints, and onboard autonomy handles local perception, navigation, and execution, reporting results and state updates when communication permits. Predictive interfaces can further support supervision by propagating expected state forward in time, allowing controllers to reason about the likely conditions at command receipt and to plan around uncertainty \cite{hunter1992time,bejczy1990phantom}.

Autonomy at the remote end must be matched by strong support on the ground. Even when contact is delayed, telemetry and event logs provide critical signals for diagnosis, trend analysis, and maintenance planning \cite{iverson2009general,hundman2018detecting}. Data-driven analytics can help identify weak precursors of failure, highlight anomalies that merit immediate attention, and reduce the workload of monitoring many subsystems across multiple vehicles. Mission control environments can then present a unified operational picture that prioritises what matters, rather than overwhelming teams with raw streams.

Taken together, these approaches enable latency-tolerant operations: crews and autonomous systems handle routine execution locally, while Earth-based teams provide strategic oversight, verification, and longer-horizon planning. This division of labour is central to scaling deep-space activity safely as missions become longer, more complex, and more distributed.

\begin{figure*}
\centering  
\includegraphics[width=\linewidth]{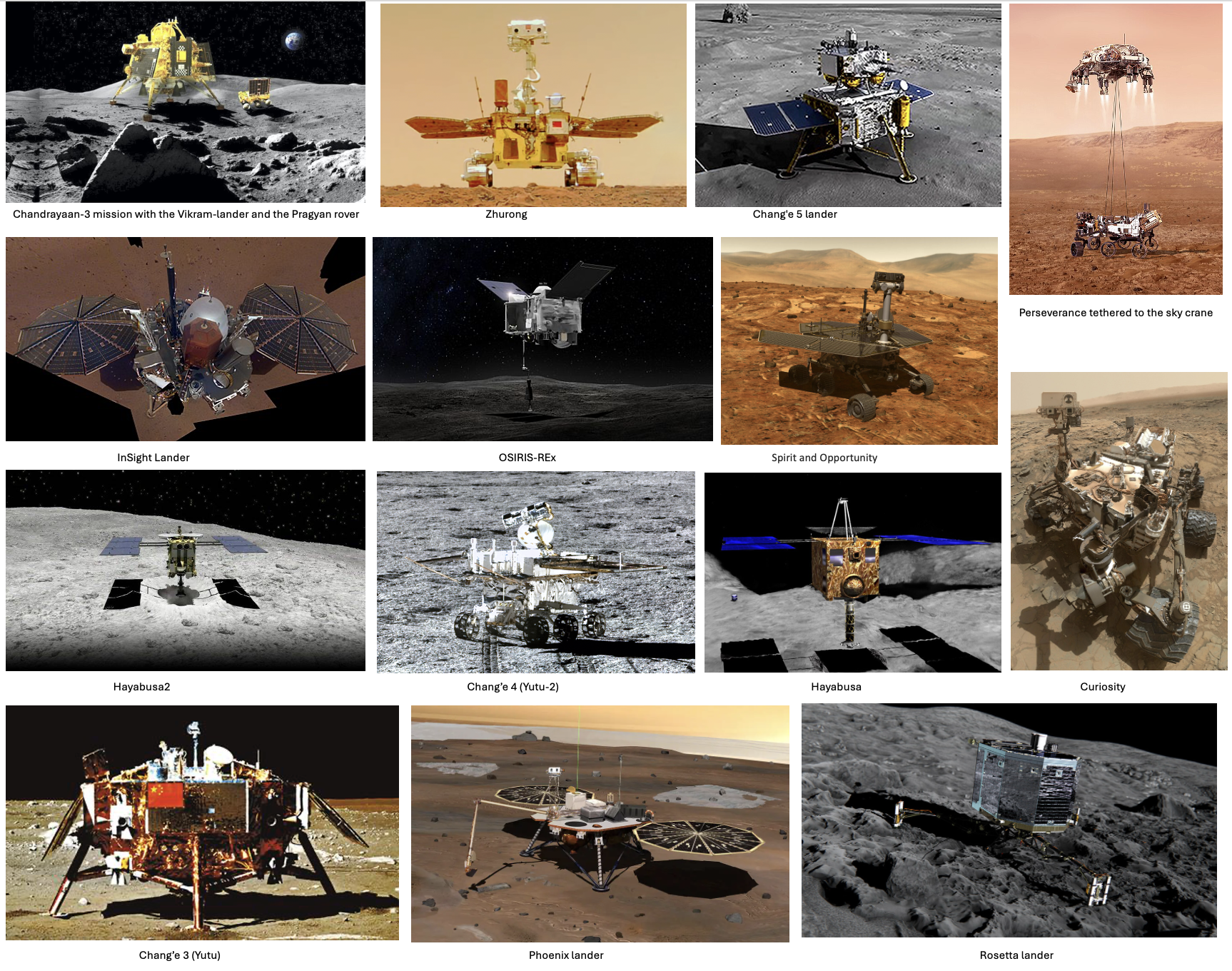}  
\caption{Images of Robotic Space Missions.}
\label{fig:robotspacerover}  
\end{figure*}

\section{AI for Multi-Planetary Life: Building the Future}

\subsection{AI-Enabled Extraterrestrial Settlement}
Establishing a sustained off-world presence begins with understanding local environments and resources. Before habitats, logistics, and industry can scale, missions must identify water, volatiles, and useful minerals, quantify their accessibility, and reduce uncertainty about where and how extraction can be performed safely. AI contributes most when it strengthens this chain from sensing to action: enabling autonomous surveying, fusing multi-modal measurements into coherent 3D models, and supporting decision-making under uncertainty. These capabilities form the technical basis for settlement planning, from landing-site selection to the design of in-situ resource utilisation (ISRU) systems.

\subsubsection{Autonomous Planetary Surveying and Mapping}
Robotic explorers already provide the primary means of surveying planetary surfaces and subsurfaces \cite{fong2008field}. As mission scales grow and operational timelines tighten, autonomy becomes essential for efficient prospecting. Local navigation and hazard-aware planning allow rovers and landers to cover more terrain per command cycle, while onboard or near-real-time processing can triage measurements and prioritise follow-up at locations that show stronger evidence of volatiles or relevant mineral signatures.

Lunar polar prospecting has frequently been used as a motivating example because illumination and terrain hazards impose strict constraints on energy and safety. The VIPER rover\footnote{\url{https://science.nasa.gov/mission/viper/}}, originally conceived to characterise lunar south-pole volatiles, illustrates the broader requirement: a rover must combine hazard detection, route planning, and instrument-driven decision support to sample subsurface regolith under changing illumination and limited contact. Regardless of specific programme timelines, this class of mission clarifies what autonomy must deliver for settlement: safe mobility, efficient prospecting, and credible uncertainty estimates for resource claims.

Surveying becomes actionable when disparate measurements are integrated into spatial models. A central requirement is constructing 3D geological representations from sparse, noisy data. Subsurface sampling and drilling instruments (e.g., lunar volatile sampling concepts such as ESA’s PROSPECT\footnote{\url{https://www.esa.int/Science_Exploration/Human_and_Robotic_Exploration/Prospect_searching_for_water_at_the_lunar_poles}}) produce discrete measurements that gain value when fused with orbital context, including thermal maps, topography, radar or neutron indicators, and imaging \cite{BOAZMAN2024116240}. AI-assisted data fusion can help interpolate between sparse drill points, identify consistent patterns across modalities, and produce layered 3D models that describe likely resource depth, concentration, and heterogeneity. The practical benefit is decision support: these models guide where to drill next, how to size extraction systems, and how to place infrastructure to reduce operational risk.

Predictive modelling extends this capability beyond the surveyed footprint. Many resources are patchy and sensitive to micro-environmental conditions. For instance, lunar polar water ice depends on local temperature history and illumination geometry; Martian subsurface volatiles depend on geology, thermal gradients, and shielding. Data-driven predictors can combine proxies such as slope, thermal stability, albedo, and remote-sensing signatures to estimate where resources are likely to be richer or more stable. For mission planning, the important output is not a single deterministic map but a probabilistic surface with uncertainty, allowing planners to weigh potential yield against landing, traversal, and operational risk. As new measurements arrive, models can be updated to reduce uncertainty and to support adaptive exploration strategies.

\subsubsection{Autonomous In-Situ Resource Utilization (ISRU)}
Resource identification is only the start; settlement requires converting local material into usable commodities such as oxygen, water, fuel, and construction feedstock. ISRU systems must operate under strict power budgets, variable feedstock composition, and limited maintenance. Here, AI is best framed as part of an assured control and operations layer: monitoring process health, adapting operating points within safe envelopes, and coordinating multiple coupled subsystems in closed-loop operation.

Process optimisation is one area of direct impact. MOXIE on Perseverance demonstrated oxygen production from Martian CO$_2$ under operational constraints, highlighting the importance of robust automation for thermal, pressure, and flow control \cite{doi:10.1126/sciadv.abp8636}. In future systems, adaptive controllers could tune operating parameters in response to feedstock variability, component aging, or changing environmental conditions, while maintaining safety constraints and minimising wear. Similar principles apply to regolith processing, where oxygen extraction, sintering, or metal recovery must be stabilised despite heterogeneity in grain size and composition.

Closed-loop operation is equally important at the habitat scale. Sustainable bases rely on coordinated material flows in which outputs and by-products of one process become inputs to another. ISRU therefore interacts with manufacturing, logistics, and life support. Automated monitoring and scheduling can help keep the overall loop stable: timing excavation and processing, balancing storage and demand, and triggering conservative responses when anomalies occur. The objective is not maximal autonomy, but predictable continuity of production with transparent margins and fallbacks.

Finally, off-world operations must tolerate uncertainty. Terrain conditions, dust contamination, thermal swings, and mechanical wear can disrupt excavation and processing. Adaptive autonomy can improve resilience by detecting when assumptions no longer hold, selecting alternative operating modes, and scheduling recovery actions (e.g., cleaning cycles, reduced-duty operation, or safe shutdown) before faults propagate. In multi-robot concepts, local learning may be complemented by shared updates across a fleet, enabling improvements in excavation or handling strategies without centralising raw data. The common requirement across these settings is conservative adaptability: systems respond to change, but do so within verified bounds that protect crew safety and preserve critical assets.

Together, autonomous surveying, 3D modelling, predictive resource maps, and ISRU optimisation form a coherent pipeline for settlement planning. They reduce uncertainty about where resources are, how accessible they are, and how reliably they can be converted into life-supporting commodities and infrastructure feedstock.

\begin{table*}[htbp]
\centering
\caption{Robotic Space Missions with Surface and Autonomous Robots}
\label{tab:robotic-missions}
\resizebox{\textwidth}{!}{%
\begin{tabular}{lllllllllll}
\toprule
\textbf{Launch Year} & \textbf{Mission Name} & \textbf{Country/Agency} & \textbf{Target} & \textbf{Platform Type} & \textbf{Mobility} & \textbf{Robotic Arm} & \textbf{Sampler} & \textbf{Drill} & \textbf{Sample Return} & \textbf{AI/Autonomy Features} \\
\midrule
1966 & Luna 9 & USSR & Moon & Lander & N & N & N & N & N & Automated soft landing \\
1967 & Surveyor 3 & USA (NASA) & Moon & Lander & N & Y & Y & N & N & First robotic arm on Moon \\
1970 & Luna 16 & USSR & Moon & Lander & N & Y & Y & Y & Y & Autonomous lunar soil return \\
1971 & Lunokhod 1 & USSR & Moon & Rover & Y & N & N & N & N & Teleoperated rover \\
1975 & Viking 1 \& 2 & USA (NASA) & Mars & Lander & N & Y & Y & N & N & Autonomous sampling and analysis \\
1981 & Venera 13 \& 14 & USSR & Venus & Lander & N & Y & Y & N & N & Autonomous drilling and sampling \\
1996 & Pathfinder (Sojourner) & USA (NASA) & Mars & Rover & Y & Y & N & N & N & Autonomous obstacle avoidance \\
2003 & MER (Spirit \& Opportunity) & USA (NASA) & Mars & Rover & Y & Y & N & Y & N & Autonomous navigation, drilling \\
2003 & Hayabusa & Japan (JAXA) & Asteroid Itokawa & Orbiter/Sampler & N & N & Y & N & Y & Autonomous sampling \\
2004 & Rosetta (Philae) & ESA & Comet 67P & Lander & N & N & Y & Y & N & Autonomous comet landing, drilling \\
2008 & Phoenix & USA (NASA) & Mars & Lander & N & Y & Y & N & N & Autonomous trenching and sampling \\
2011 & Curiosity & USA (NASA) & Mars & Rover & Y & Y & Y & Y & N & Advanced AutoNav, autonomous sampling \\
2013 & Chang’e 3 (Yutu) & China (CNSA) & Moon & Rover & Y & Y & N & N & N & Semi-autonomous navigation \\
2014 & Chang’e 4 (Yutu-2) & China (CNSA) & Moon & Rover & Y & Y & N & N & N & Autonomous hazard avoidance \\
2014 & Hayabusa2 & Japan (JAXA) & Asteroid Ryugu & Sampler/Rovers & Y & N & Y & N & Y & Autonomous touch-and-go sampling \\
2016 & OSIRIS-REx & USA (NASA) & Asteroid Bennu & Orbiter/Sampler & N & Y & Y & N & Y & Autonomous sample collection \\
2018 & InSight & USA (NASA) & Mars & Lander & N & Y & N & Y & N & Automated drill operations \\
2020 & Chang’e 5 & China (CNSA) & Moon & Lander & N & Y & Y & Y & Y & Autonomous lunar sample return \\
2020 & Perseverance & USA (NASA) & Mars & Rover/Drone & Y & Y & Y & Y & N & Advanced AutoNav, adaptive sampling \\
2020 & Tianwen-1 (Zhurong) & China (CNSA) & Mars & Rover & Y & N & N & N & N & Autonomous route planning \\
2023 & Chandrayaan-3 & India (ISRO) & Moon & Rover & Y & N & N & N & N & Autonomous descent and navigation \\
\bottomrule
\end{tabular}%
}
\end{table*}

\subsubsection{Planetary Robotics for Exploration and Resource Utilization}
Planetary robotics is central to both exploration and early-stage resource utilisation because it can operate where continuous human presence is infeasible. The fundamental constraint is latency: as distance increases, commands arrive too late for fine-grained control, and missions must rely on onboard perception, planning, and fault tolerance \cite{bresina2005mixed}. As a result, autonomy is increasingly treated as a mission requirement rather than an optional enhancement, particularly for mobile platforms that must navigate uncertain terrain, manage limited power, and execute instrument sequences safely.

Within the Solar System, next-generation missions illustrate this shift. Rotorcraft concepts such as Dragonfly require robust onboard navigation, hazard avoidance, and conservative decision-making to execute complex mobility under communication delay. The emphasis is typically not on ``open-ended intelligence,'' but on reliable autonomy within well-specified envelopes: safe flight or driving, local replanning, and predictable handling of anomalies. Similar principles apply to resource-oriented robotics, where excavation, sampling, and material handling must operate under variable feedstock properties and mechanical uncertainty \cite{9197500,verma-perseverance-science-robotics-2023,Europa-lander-scirob-2025}. Here, autonomy supports stable throughput by adapting operating parameters, scheduling actions under power and thermal constraints, and triggering safe recovery behaviours when performance drifts.

Looking forward, long-duration missions motivate tighter integration between autonomy and maintainability. Self-diagnostics, redundancy management, and limited forms of in situ repair are often discussed as ways to extend operational lifetime when spare parts and human intervention are unavailable. This direction intersects with advances in flight-qualified computing, where newer space processors and accelerators aim to support higher onboard throughput for perception, planning, and monitoring. The practical implication is broader autonomy headroom: more frequent local decisions, richer onboard triage of scientific data, and more capable fault management without relying on continuous ground support.

At the far end of the spectrum, interstellar mission concepts such as Breakthrough Starshot highlight why autonomy matters in principle. Proposals envision gram-scale probes travelling at relativistic fractions of the speed of light, where real-time supervision is impossible and where onboard systems would need to manage navigation, data prioritisation, and extremely constrained communications. These remain conceptual, but they usefully sharpen the research agenda: how to design autonomy that is robust over long durations, conservative under uncertainty, and efficient enough to operate under severe mass, power, and compute limits. In this sense, planetary robotics and deep-space autonomy sit on a continuum, ranging from near-term resource prospecting to long-horizon exploratory concepts.

\subsubsection{AI in Sustainable Base Construction and Infrastructure Development}
Building durable infrastructure off Earth requires construction, verification, and maintenance under environmental stresses that are difficult to replicate on Earth. Communication delay, limited crew time, and high consequence of failure motivate a construction paradigm dominated by robotic labour, conservative automation, and continuous monitoring. AI contributes most when it supports predictable execution: planning and coordinating tasks, verifying build quality, and detecting early signs of degradation so that small problems do not become mission-threatening failures \cite{leach2012robotic,azami2024comprehensive}.

A key direction is coordinated robotic construction \cite{thomas2005multirobot,govindaraj2020multi,cloud2021towards}. Rather than relying on large crews, base concepts often assume heterogeneous teams of machines for excavation, material processing, and fabrication. For example, lunar construction programmes and industry collaborations explore additive manufacturing of berms, landing pads, and structural elements using local regolith-derived feedstock \cite{thangavelautham2017autonomous}. In such settings, planning and scheduling tools allocate tasks, manage shared resources, and enforce safety constraints to prevent conflicts among robots. The engineering challenge is not merely coordination, but reliability under uncertainty: variable material properties, dust contamination, thermal cycling, and intermittent faults must be handled with conservative fallbacks and clear recovery procedures.

Design optimisation also plays a role, especially when structures must satisfy competing requirements such as pressurisation loads, radiation shielding, thermal insulation, and mass efficiency \cite{caiazzo2025structures}. Generative and optimisation-based design tools can explore architectures tailored to local materials and construction constraints, and can be coupled to high-fidelity simulation to verify safety margins. The value is early-stage trade exploration: identifying designs that are feasible to build with local feedstock while meeting structural and habitability requirements, and exposing sensitivity to uncertain material properties.

Maintenance is the third pillar. Over long time horizons, infrastructure will degrade due to micrometeoroid impacts, dust abrasion, thermal cycling, and radiation. Continuous monitoring using embedded sensors and periodic robotic inspection can support early detection of leaks, cracks, and anomalous strain patterns \cite{richards2013nasa,ahrens2025case}. When combined with conservative response policies, automated systems can isolate affected regions, schedule repair actions, and document events for later review. Self-healing materials and coatings are frequently discussed as complementary technologies, but their practical value depends on verification and on reliable triggering conditions. A pragmatic goal is therefore resilience by design and by operation: detect small defects early, repair or mitigate them with minimal crew involvement, and preserve habitability and safety margins across extended intervals between human visits \cite{sharma2025space}.

Together, robotic construction, optimisation-driven design, and monitoring-enabled maintenance provide a pathway toward infrastructure that is scalable, repeatable, and sustainable. This is essential not only for habitats, but also for power systems, landing and logistics zones, and the broader industrial base required for long-term off-world presence.

\subsubsection{Intelligent Life-Support and Ecological Management Systems}
Sustained habitation beyond Earth requires life-support systems that provide breathable air, clean water, food production, and safe waste recycling under severe constraints on mass, power, maintenance, and crew time. In long-duration missions, the challenge is not only to operate individual subsystems but to keep the overall habitat stable as a coupled system, where disturbances propagate across air, water, biomass, and waste loops. AI contributes most effectively when it strengthens monitoring, forecasting, and control in this coupled setting, enabling early warning, robust operating policies, and resource-efficient regulation within verified safety envelopes.

Closed-loop or partially closed-loop ecological concepts illustrate the complexity. ESA’s MELiSSA programme\footnote{\url{https://www.esa.int/Enabling_Support/Space_Engineering_Technology/Melissa}}, for example, investigates a multi-compartment loop in which biological and physico-chemical processes convert waste streams into oxygen, water, and biomass. Such architectures are intrinsically dynamic and interdependent: microbial activity affects nutrient availability; plant or algal growth affects gas exchange; and changes in crew activity alter load profiles. In this context, predictive monitoring and control are essential. Data-driven models can help estimate latent states (e.g., process health, microbial growth trends) from sensor streams, forecast near-term stability margins, and recommend conservative adjustments to operating points such as illumination schedules, flow rates, or recycling timing. The goal is homeostasis: keeping the loop within safe bounds despite variation in crew schedule, feedstock composition, and gradual component drift.

Food and oxygen production further motivate automation. Plant-growth systems in space, demonstrated in reduced form on the ISS (e.g., Veggie\footnote{\url{https://science.nasa.gov/mission/veggie/}}), must control light intensity and spectrum, temperature, humidity, nutrient delivery, and water use with high reliability. Imaging and environmental sensing can support early detection of plant stress, nutrient deficiency, or disease-like patterns, while optimisation can tune growth protocols for yield per unit power and water. The key requirement for settlement-scale agriculture is not merely better growth, but stable operations with minimal crew intervention, since crew time becomes the scarcest resource in a developing habitat. Automated supervision can therefore shift plant care from manual routine to exception handling: the system runs normal cultivation reliably and elevates only the cases that require human judgement.

Microbiome monitoring is a third pillar that links crew health and system reliability. Closed habitats inevitably develop distinctive microbial ecologies in air, water, and surfaces; some microbes are harmless or beneficial, while others create health risks or degrade hardware through biofilms and corrosion. Traditional culture-based monitoring is too slow for operational control, motivating sensor-driven and sequencing-enabled approaches. Miniaturised DNA sequencing and rapid biosensing have been demonstrated in microgravity settings, but their outputs are complex and noisy. AI can assist by interpreting multi-sensor signatures, distinguishing benign variation from meaningful contamination signals, and triggering conservative response actions such as intensified filtration, disinfection cycles, or targeted cleaning. Here, cautious decision logic is essential: false alarms can waste consumables and crew time, while missed detections can threaten both health and system integrity.

Across these components, the central design principle is assured autonomy. Intelligent life-support should expose confidence and uncertainty, degrade gracefully under faults, and preserve clear operator authority for high-consequence actions. When designed in this way, AI-enabled ecological management becomes a practical enabler of multi-planetary living: stabilising coupled resource loops, scaling food and oxygen production, and maintaining microbial safety over missions measured in months to years.

\subsection{Vision for Multi-Planetary AI Networks and Societies}
As space activity extends from single missions to sustained presence across multiple bodies, AI shifts from being a component of individual spacecraft to a capability that shapes networks, operations, and governance. Multi-planetary systems face three persistent constraints: long and variable latency, intermittent connectivity, and limited opportunities for repair or centralised supervision. Under these conditions, ``intelligence'' is most valuable when it supports resilient coordination at scale: routing and prioritising information across heterogeneous nodes, forecasting and managing environmental risks such as radiation, and maintaining long-term cultural and operational continuity.

\subsubsection{Autonomous Extraterrestrial Communication Networks}
A multi-planetary civilisation requires communications infrastructure that tolerates delay and disruption. Delay/Disruption-Tolerant Networking (DTN) provides a protocol foundation for intermittent links and long light-times, enabling store-and-forward communication and custody transfer across nodes. On top of DTN, autonomy can support routing and traffic management when link availability and demand are time-varying. The practical objectives are straightforward: preserve delivery guarantees for critical messages, allocate scarce bandwidth to high-priority data, and recover gracefully when links drop or nodes fail. Learning-based components can be explored as decision aids for routing and scheduling, for example by predicting contact opportunities, estimating link quality, and selecting forwarding policies that improve throughput while respecting operational constraints.

Lunar networking concepts such as LunaNet\footnote{\url{https://www.nasa.gov/humans-in-space/lunanet-empowering-artemis-with-communications-and-navigation-interoperability/}} reflect this architectural shift from point-to-point links toward an interoperable network spanning orbiters, landers, rovers, and relay nodes. In such architectures, autonomy is less about ``smart radios'' and more about system-level coordination: deciding what to transmit, when to buffer, how to route through alternative relays, and how to balance navigation services, telemetry, and science downlink. Similar principles extend to Mars-scale relay systems, where contact opportunities and energy budgets vary over time and where network behaviour must remain predictable under safety constraints.

Security becomes more challenging as networks expand and autonomy increases. Quantum-secure communications and quantum key distribution are often discussed as future ingredients for protecting critical links. Initiatives such as EuroQCI\footnote{\url{https://digital-strategy.ec.europa.eu/en/policies/european-quantum-communication-infrastructure-euroqci}} (largely Earth-focused) illustrate the direction of combining terrestrial and satellite segments for quantum-secure infrastructure. For space, the key point is not immediate deployment but the systems engineering problem: managing heterogeneous cryptographic modes, error rates, and availability under operational constraints. Here, automated control and scheduling can support robustness by switching strategies as conditions change, while governance must define what levels of assurance are required for different mission phases and services.

\subsubsection{AI for Space Weather Prediction and Risk Mitigation}
Space weather is a multi-planetary risk because it couples directly to astronaut safety, spacecraft reliability, and communications performance. Solar flares, coronal mass ejections, and energetic particle events can disrupt electronics and pose acute radiation hazards, while galactic cosmic rays create chronic exposure risk \cite{xue2025space}. Forecasting and mitigation therefore require both prediction and operational decision-making: recognising when conditions are changing, quantifying uncertainty, and triggering protective actions within tight time constraints.

AI is most valuable when it complements physics-based models rather than attempting to replace them \cite{price2025probabilistic,riggi2025solar}. Data-driven predictors can improve warning quality for specific indices or event classes by learning relationships between upstream solar wind measurements and downstream geomagnetic response, and by identifying precursors in imagery or time series that may be difficult to encode analytically. For operations, the practical output is calibrated risk: not only ``what is likely to happen,'' but also how confident the system is and what actions are justified under the mission’s safety case.

Mitigation depends on timely interpretation of onboard measurements and conservative response policies. Radiation monitors, dosimetry, and environmental sensors support early warning and exposure tracking, while decision-support tools can help determine when to enter sheltered configurations, how long to remain in protective modes, and how to schedule high-risk activities such as extravehicular operations. Over longer horizons, forecasting ``space climate'' across solar cycles supports planning for shielding, mission timelines, and maintenance schedules. Across these time scales, the central requirement is actionable uncertainty: predictions must be tied to operational thresholds and to safe fallbacks when models disagree or data is incomplete.

\subsubsection{AI for Space Archaeology, Heritage Preservation, and Cultural Sustainability}
Multi-planetary activity will create heritage sites and cultural records that merit preservation, both for historical value and for governance legitimacy. Heritage preservation in space includes protecting artefacts and sites of human exploration (e.g., landing sites and equipment remains) and ensuring long-term accessibility of cultural archives. The operational challenge is that enforcement and monitoring cannot rely on continuous human presence, and new missions may pose unintentional risks through plume impingement, surface disturbance, or uncoordinated traversal.

AI can support heritage protection through monitoring and change detection. High-resolution orbital imaging can be compared against baselines to identify disturbances near protected areas, while trajectory and operations logs can support compliance verification against agreed buffer zones. This is best framed as a verification pipeline rather than a punitive mechanism: detect changes, attribute likely causes with uncertainty, and provide auditable records that support dispute resolution and improved operational standards. Policy instruments such as NASA’s guidance on preserving Apollo sites and national legislation that operationalises such guidelines create the governance context in which automated monitoring becomes meaningful.

Cultural sustainability also includes durable off-world archiving. Projects such as Lunar Codex\footnote{\url{https://www.lunarcodex.com/}} illustrate an emerging practice of storing curated cultural artefacts beyond Earth. The technical issues are less about ``AI creativity'' and more about robust curation, indexing, compression, and error-resilient storage. AI can assist by organising large multimedia collections, supporting redundancy and provenance metadata, and enabling future retrieval and interpretation through structured knowledge representations.

Finally, autonomous pattern recognition can also support the scientific dimension of ``archaeology'' in a broad sense, including the identification of rare geological contexts and potential biosignature candidates. Here, the same methods used for target prioritisation and anomaly detection in planetary science can help allocate attention to the most informative observations. It is useful, however, to distinguish this scientific search from heritage protection: one concerns evidence of past natural processes or life, while the other concerns preservation of human artefacts and sites. Both benefit from autonomy, but they involve different governance and ethical constraints.

Overall, multi-planetary networks and societies will depend on AI most where scale and latency make centralised supervision impractical: resilient communications, risk-aware operations under space weather, and durable mechanisms for preserving both cultural records and the physical traces of exploration.

\subsubsection{Human-AI Collaboration and Social Systems in Multi-Planetary Communities}
As outposts evolve into long-duration communities, human--AI teaming becomes part of the settlement’s basic infrastructure rather than a mission add-on \cite{ulusoy2025investment}. Multi-planetary living amplifies three constraints simultaneously: delayed or intermittent contact with Earth, limited specialist staffing on site, and sustained exposure to isolation and environmental risk. Under these conditions, AI is most valuable when it supports predictable operations, preserves human agency, and strengthens social resilience without turning the habitat into an opaque automated system.

At the habitat level, autonomy is a safety requirement. Life support, power, thermal control, and fault management must continue through communication gaps, which motivates local monitoring, diagnosis, and conservative recovery policies. The most credible model is ``assured autonomy'': digital twins and onboard decision support help interpret sensor streams, forecast margins, and execute pre-authorised mitigation actions, while complex or ambiguous situations are escalated to crew judgement and, when possible, to Earth-based experts. The aim is not maximal automation, but continuity of safe operation and faster response when seconds or minutes matter. In this sense, AI becomes an embedded mission-control capability that lives with the crew.

At the individual level, assistance systems reduce cognitive overhead and procedure burden. Orbital demonstrations such as CIMON\footnote{\url{https://www.nasa.gov/image-article/robot-colleagues-meet-aboard-international-space-station/}} and voice-interface experiments like Callisto\footnote{\url{https://www.nasa.gov/missions/callisto-technology-demonstration-to-fly-aboard-orion-for-artemis-i/}} illustrate a practical direction: hands-free retrieval of procedures and configuration information, context-aware guidance during maintenance, and lightweight summarisation of system status. For deep-space crews, the design emphasis shifts from conversational novelty to reliability, privacy, and graceful degradation. Assistants should expose confidence and evidence behind recommendations, handle routine queries locally without dependence on external connectivity, and preserve clear override pathways. When implemented well, these tools increase crew autonomy without undermining accountability.

Social systems are the third pillar. Long-duration confinement creates predictable stressors: fatigue, mood decline, interpersonal friction, and the gradual erosion of cohesion. AI can contribute here in a bounded, ethically constrained manner. Continuous health and wellbeing monitoring may help detect early warning signs of sleep disruption, stress accumulation, or reduced functional capacity, but the handling of such data must be privacy-preserving and governed by clear norms to prevent surveillance from eroding trust within the community. Intervention should remain supportive rather than coercive, for example through optional coaching, workload recommendations, or facilitation of communication practices, and not through opaque ``scores'' or mandatory behavioural control. Companion robotics and interactive agents may provide additional psychological support in specific roles, but their benefits must be evaluated carefully to avoid dependency, manipulation, or unintended social substitution.

Finally, multi-planetary societies will require governance-by-design for human--AI collaboration. This includes explicit authority boundaries (what can be automated, what requires human confirmation), auditability of consequential actions, resilient fallback modes, and training that prepares crews to work effectively with decision-support tools without over-reliance. When these elements are aligned, human--AI collaboration can become a stabilising force: sustaining safe operations, expanding functional capability despite small crew sizes, and supporting psychological and social wellbeing under extreme isolation. In this way, AI becomes not only an operational assistant but part of the social infrastructure that enables communities to remain coherent, capable, and humane far from Earth.

\section{Conclusions and Future Outlook}

\subsection{Current Status and Trends of the Space Industry}

The space industry is increasingly interwoven with artificial intelligence, reshaping how missions are designed, operated, and scaled. Across both government agencies and NewSpace companies, AI is moving from ground support into flight-critical functions, enabling higher levels of autonomy in launch, on-orbit operations, and planetary exploration.

SpaceX exemplifies this shift through closed-loop guidance, navigation, and control (GNC) and onboard autonomy across key flight phases, including automated booster landings. In orbit, large constellations such as Starlink increasingly rely on automated conjunction assessment and maneuver planning to support collision avoidance at scale. NASA has likewise deployed autonomy in exploration, most visibly through Mars rover navigation and onboard decision-making. For instance, Perseverance uses an AI-enabled AutoNav capability to traverse terrain with reduced reliance on Earth-based command cycles, reflecting the broader trend toward spacecraft that can make time-critical decisions in situ.

Beyond flagship autonomy demonstrations, AI is rapidly becoming standard across engineering workflows and manufacturing. Companies such as Blue Origin emphasise autonomy and robust GNC for fully uncrewed flight operations, while newer entrants (e.g., Relativity Space) integrate machine learning and automation into manufacturing pipelines for process optimisation and quality assurance. Collectively, these developments indicate that AI is not only improving mission operations, but also accelerating the cadence and affordability of space system development.

At the same time, orbital environments are becoming more congested, increasing collision risk and driving demand for scalable space traffic management. AI is increasingly used to fuse tracking data, predict conjunctions, and automate maneuver decisions, particularly for multi-satellite constellations. The European Space Agency is also advancing AI-enabled approaches to space safety, including debris mitigation and active debris removal efforts such as ClearSpace-1.\footnote{\url{https://www.esa.int/Space_Safety/ClearSpace-1}} Sustaining safe and usable orbits will require more capable data fusion, coordination, and policy-aware automation as the number of tracked objects continues to grow.

Communication delays and bandwidth constraints remain fundamental limitations for deep-space missions \cite{frank2014autonomous}. As one-way light-time extends to minutes or longer, spacecraft must act without immediate human intervention. This motivates edge AI for autonomous targeting, diagnosis, and decision-making. For example, the Curiosity rover uses AEGIS to autonomously select science targets when high-bandwidth links are unavailable \cite{francis2017aegis}, illustrating how onboard intelligence can increase science return under operational constraints.

A prominent trend is deploying AI directly on satellites and small spacecraft to process data at the source \cite{wang2020convergence}. ESA’s PhiSat-1 demonstrated onboard filtering of Earth observation imagery by discarding cloud-obscured frames and downlinking only useful data, reducing communication burden and improving responsiveness. In parallel, digital twins and high-fidelity simulation are emerging as a foundation for mission planning and health management: virtual replicas of spacecraft and environments can be updated with telemetry to detect anomalies early, run what-if analyses, and support predictive maintenance and safer operations.

Overall, the current status of the space sector reflects a clear convergence of aerospace engineering and AI. Autonomously landing rockets, self-navigating rovers, and increasingly coordinated satellite constellations are becoming less exceptional and more routine. Looking ahead, continued progress in onboard compute, robust autonomy, and simulation-driven engineering is likely to enable smarter satellite networks, more independent robotic explorers, and ultimately more scalable space infrastructures and operations.

\subsection{Global Regulations and Standards for Space AI}

\begin{table*}[htbp]
\small
\setlength{\tabcolsep}{3pt}
\centering
\caption{Expanded Overview of Global Legal and Technical Frameworks Affecting Space Activities and AI (Part I)}
\label{tab:expanded_space_ai_law}
\begin{tabular}{p{3.3cm}p{2.5cm}p{2.4cm}p{6.4cm}p{2.5cm}}
\toprule
\textbf{Instrument, Framework} & \textbf{Type} & \textbf{Org.} & \textbf{Scope and Relevance to Space or AI} & \textbf{Status} \\
\midrule
Outer Space Treaty (1967) & International Treaty & United Nations (COPUOS) & Foundation of space law. Establishes state responsibility and liability in space, but lacks provisions on AI or autonomy. & In force \\
Liability Convention (1972) & International Treaty & United Nations (COPUOS) & Defines liability for damage caused by space objects. Relevant to AI mishaps but lacks clarity on machine agency. & In force \\
Registration Convention (1976) & International Treaty & United Nations (COPUOS) & Requires states to register space objects. May support traceability of AI systems on orbit. & In force \\
Moon Agreement (1979) & International Treaty & United Nations (COPUOS) & Regulates activities on the Moon; few ratifications. Could become relevant for lunar AI systems. & Partially adopted \\
ITU Radio Regulations & International Regulations & ITU (UN) & Governs radio frequency and orbital slot use; relevant to satellite communications and AI-driven networks. & In force \\
COPUOS LTS Guidelines (2019) & Soft Law Guidelines & United Nations (COPUOS) & Non-binding but widely accepted sustainability practices; covers autonomous and AI-driven space operations. & Adopted \\
ISO 24113: Space Debris Mitigation & Technical Standard & ISO & Debris mitigation protocols, indirectly affect AI navigation, post-mission disposal. & In force \\
ISO 22606: Autonomous Space Systems Guidance & Technical Standard & ISO / ECSS & Draft guidance for autonomy, fault tolerance, and predictability in space systems. & In development \\
CCSDS Mission Ops Standards & Technical Standards & CCSDS (NASA, ESA, JAXA, etc.) & Defines data handling and control protocols; essential for spacecraft AI. & In use \\
NASA Ethical AI Framework & Internal Policy & NASA & Principles for AI transparency, accountability, and mission assurance. & Active \\
ESA “Robots in Space” (2016) & Technical Guideline & ESA & Evaluates risks/benefits of spacecraft autonomy; influences policy. & Informal influence \\
EU AI Act (2024) & Regional Regulation & European Union & Regulates high-risk AI systems including those in critical infrastructure like space. & In force \\
U.S. Commercial Space Launch Act & National Law & USA (FAA/DOT) & Regulates commercial launches; applies to AI-guided launch vehicles. & In force \\
U.S. Land Remote Sensing Act & National Law & USA (NOAA) & Licenses satellite imaging operators; affects AI analytics rights/distribution. & In force \\
\bottomrule
\end{tabular}
\end{table*}

\begin{table*}[htbp]
\ContinuedFloat
\small
\setlength{\tabcolsep}{3pt}
\centering
\caption{Expanded Overview of Global Legal and Technical Frameworks Affecting Space Activities and AI (Part II)}
\begin{tabular}{p{3.3cm}p{2.5cm}p{2.4cm}p{6.4cm}p{2.5cm}}
\toprule
\textbf{Instrument, Framework} & \textbf{Type} & \textbf{Org.} & \textbf{Scope and Relevance to Space or AI} & \textbf{Status} \\
\midrule
U.S. Space Resource Exploration Act (2015) & National Law & USA & Grants rights to space resource utilization; supports AI-driven mining missions. & In force \\
UK Outer Space Act (1986) & National Law & UK & Licensing framework for UK space activities; imposes liability/insurance obligations. & In force \\
UK Space Industry Act (2018) & National Law & UK & Regulates UK launch activities, including AI-equipped vehicles. & In force \\
French Space Operations Act (2008) & National Law & France & Requires licensing, liability, and insurance for French space operators. & In force \\
Germany Sat Data Security Act & National Law & Germany & Limits dissemination of sensitive satellite data; relevant to AI image analytics. & In force \\
Japan Basic Space Law (2008) & National Law & Japan & Establishes strategic and civil space policy; foundational for Japan’s AI missions. & In force \\
Japan Space Activities Act (2016) & National Law & Japan & Licensing regime for launches and satellite ops, including AI-driven systems. & In force \\
Japan Remote Sensing Act (2016) & National Law & Japan & Licensing and data policy for satellite imaging; constrains AI data use. & In force \\
Japan Space Resources Act (2021) & National Law & Japan & Authorizes commercial space resource use; supports autonomous extraction missions. & In force \\
China Interim Space Measures & National Regulations & China (CNSA, COSTIND) & Licensing, registration, and debris mitigation rules; regulate AI-enabled missions. & In force \\
India Space Activities Bill & Draft National Law & India & Proposed commercial licensing and liability framework; includes AI missions. & Draft (2025) \\
UAE Space Law (2019) & National Law & UAE Space Agency & Comprehensive licensing, debris, tourism, and data policies; governs AI space ops. & In force \\
South Africa Space Affairs Act (1993) & National Law & SACSA & Licensing and oversight for space operations; applicable to AI-driven missions. & In force \\
OECD AI Principles (2019) & International Principles & OECD & General AI governance principles (safety, transparency); influence national space AI policies. & Endorsed \\
\bottomrule
\end{tabular}
\end{table*}

As Space AI technologies proliferate, the legal and regulatory framework governing outer space is struggling to keep pace. International space law remains grounded in treaties drafted more than half a century ago, especially the Outer Space Treaty of 1967 \cite{dembling1967evolution}. These instruments established high-level principles for state responsibility and liability, but they did not anticipate autonomous, learning-based systems. AI is not mentioned explicitly, leaving ambiguity around accountability, safety expectations, and ethical constraints for algorithmic decision-making in orbit.

Core provisions such as Article VI (state responsibility for national space activities, including private actors) and Article VII (liability for damage caused by space objects) still apply, but their application becomes unclear when an onboard AI makes an unexpected decision that contributes to an incident. While the legal chain typically returns to a launching state or operator, attributing fault, negligence, or due diligence in the presence of opaque models and stochastic behaviours remains a grey area. In practice, Space AI operates in a partial regulatory vacuum: existing treaties provide a ``skeleton'' of responsibility, but little guidance on AI-specific issues such as transparency, verification, or fail-safe requirements.

Recognising the gap, international bodies and national agencies have begun to explore guidelines and standards. UNCOPUOS\footnote{\url{https://www.unoosa.org/oosa/index.html}} has placed emerging technologies on its agenda and continues discussions through workshops and legal subcommittees. In parallel, expert communities such as the International Institute of Space Law (IISL)\footnote{\url{https://iisl.space/}} have organised focused efforts on legal questions raised by AI in space activities. Space agencies also contribute through internal governance and ethics frameworks: ESA has examined the role of robotics and autonomy in space missions, while NASA has articulated principles for the ethical use of AI, emphasising transparency, accountability, and responsible deployment. Although such policies are not binding internationally, they shape expectations and can seed future norms. Finally, technical standardisation is gaining momentum through organisations such as ISO and CCSDS, where autonomy and data-system standards intersect with AI safety and operational assurance.

Despite these efforts, several regulatory gaps remain. First, liability and accountability for AI-driven incidents are under-specified. International law places liability on states, but it does not clarify how to evaluate fault when an accident is precipitated by model error, distribution shift, or emergent behaviours. This creates uncertainty for insurers, operators, and dispute resolution, and increases pressure for clearer verification and validation expectations pre-launch.

Second, cross-border data governance and privacy are increasingly salient. Space systems collect and analyse high-resolution Earth data across jurisdictions, and onboard AI may filter, transform, or infer sensitive information before downlink. Terrestrial regimes (e.g., data protection, remote sensing controls, or national security restrictions) may not map cleanly onto AI-processed data generated in orbit. Moreover, if AI-generated outputs are biased or incorrect and cause downstream harm, current rules provide limited clarity on accountability for informational damage.

Third, transparency and auditability are difficult to implement in orbit. Many Space AI systems operate with limited introspection and constrained telemetry, making it hard to reconstruct the basis of autonomous decisions after an anomaly. Without standards for logging, explainability, and post-incident audit procedures, investigations may be inconclusive, undermining trust in delegating critical functions to onboard AI.

AI also introduces ethical and security questions that existing treaties address only indirectly. Autonomous actions may implicate planetary protection and harmful contamination obligations, while autonomy in resource utilisation or servicing missions raises concerns about interference, misinterpretation of intent, and escalation risk. These scenarios highlight the need to adapt governance mechanisms to the unique operational realities of autonomous decision-making in space.

A pragmatic near-term path is the development of ``soft law'' instruments: non-binding yet influential guidelines, codes of conduct, and best practices for Space AI. Examples could include requirements for emergency stop modes, conservative authority limits, or reporting norms after autonomous avoidance manoeuvres to support transparency among operators. Over time, states could incorporate such guidance into national licensing and supervision regimes.

Longer-term options include additional protocols that clarify expectations for autonomous systems (e.g., fail-safes and meaningful human oversight for critical functions), and refinements to liability frameworks to better handle AI-involved incidents. In parallel, the community will likely need technical standards for qualifying AI for spaceflight—covering stress testing under realistic distributions, cybersecurity hardening, operational authority constraints, and traceable evidence that the AI will not exceed mission rules. Because space is a shared environment, broad international adoption is essential: weak AI assurance in one operator’s systems can endanger others.

\subsection{Summary and Definition of Space AI as a New Field}

Space AI can be defined as an emerging interdisciplinary field that fuses artificial intelligence with space technology to enable intelligent and autonomous behaviour in off-Earth environments. It sits at the intersection of machine learning, robotics, aerospace engineering, and planetary science, and focuses on endowing spacecraft and space robots with the ability to perceive, reason, and act under limited supervision. In this context, ``intelligent'' typically refers to capabilities such as learning from data, recognising patterns, planning actions, and responding to environmental changes in a goal-directed manner.

Crucially, Space AI is not simply conventional AI deployed in a different location. Space missions impose distinctive constraints—radiation, limited power and compute, extreme environments, and long communication delays—that require AI methods to be adapted for robustness, efficiency, and verifiability. Space AI therefore emphasises architectures and algorithms that can operate reliably on constrained, space-qualified hardware, tolerate uncertainty and faults, and make safe decisions when real-time human intervention is impossible.

The scope of Space AI is broad. A central component is autonomous spacecraft operation: satellites that manage orbit and health, probes that navigate and execute science with minimal commanding, and crewed vehicles where AI supports or automates portions of GNC and mission operations. Examples include trajectory correction under unmodelled perturbations, adaptive bandwidth/power allocation in communications systems, and onboard fault detection and recovery.

A second major area is in-situ robotics and resource operations. This includes intelligent control for robotic excavation and extraction on the Moon or asteroids, as well as autonomous scientific workflows such as onboard target selection and prioritisation. Representative mission concepts span rover autonomy for sampling, landers with automated science pipelines, and future systems that perform in-situ analysis when contact is limited.

A third area is human--AI collaboration in space habitats. In spacecraft, stations, and future lunar/Martian bases, AI-enabled assistants and robots can support astronauts by monitoring systems, assisting with procedures, scheduling tasks, and performing routine or hazardous operations. Free-flying robotic aides (e.g., CIMON, Astrobee) illustrate this direction, acting as interactive partners that can interpret commands, retrieve information, and provide operational support in microgravity.

Space AI must contend with a hazardous and remote operating environment. Radiation and limited compute typically mean space-qualified processors lag far behind terrestrial hardware; thus, methods often require efficiency measures such as compact models, compression, or hybrid designs that combine learning-based perception with more deterministic logic for safety-critical actions. Communication latency is fundamental: Mars operations can face up to $\sim$20 minutes one-way delay, and deep-space missions can experience hours or more, requiring autonomy that can handle contingencies without immediate ground approval.

Space AI also faces data scarcity and epistemic uncertainty. Many tasks cannot rely on massive labelled datasets, and system behaviour may only be observed fully in-flight. This motivates approaches that generalise from limited samples, incorporate physics and domain knowledge, and use rigorous verification/validation, including simulation and stress testing under off-nominal conditions. As a result, hybrid AI—blending machine learning with control theory, rules, redundancy, and safety monitors—remains particularly important in space.

Taken together, these requirements define a coherent research frontier with recurring problem structures: autonomy under delay, constrained onboard compute, fault tolerance, high-stakes safety, and science/engineering decision-making in uncertain environments. Space AI is therefore emerging as a distinct discipline concerned with algorithms, architectures, and assurance practices for intelligent space systems. Its advances are likely to be enabling for ambitious missions and infrastructures—deep-space exploration, sustained lunar/Martian presence, and large-scale orbital systems—and, in many cases, will also feedback to benefit terrestrial safety-critical AI.

An example of human--AI collaboration in space is the CIMON robotic assistant on the International Space Station. This autonomous free-flyer supports astronauts via interactive dialogue and onboard intelligence, illustrating how AI-driven agents can function as cooperative partners in microgravity habitats.

\subsection{Challenges for the Future of Space AI}
The coming decades will confront \textit{Space AI} with a host of challenges that must be addressed to unlock its full potential. These challenges span technical and engineering hurdles, ethical and societal dilemmas, and regulatory and governance issues. As Space AI matures into a distinct field, it will need to overcome the harsh realities of the space environment, gain public trust and ethical footing, and operate within evolving legal frameworks. This section outlines the key challenges ahead, grouped into three broad categories. 

\subsubsection{Technical and Engineering Challenges}

Space is an inherently hostile and resource-constrained setting, posing severe technical challenges for AI systems \cite{seto1998simplex,starek2015spacecraft}. One major hurdle is the \textbf{extreme environment}: spacecraft hardware and AI controllers must endure intense radiation, microgravity, vacuum, and temperature extremes. High-energy cosmic rays can induce bit flips or damage in electronic components, risking AI malfunction. Developing radiation-hardened processors and fault-tolerant algorithms is therefore critical. Unlike terrestrial data centers, computing in space must be ultra-reliable with minimal opportunity for physical maintenance or repair. This necessitates AI architectures that can gracefully handle hardware faults and sensor errors, ensuring continuous operation even under degraded conditions. Another challenge lies in the \textbf{limitations on computing power, memory, and energy} for space-based AI. Spacecraft have tight mass and power budgets, so they cannot carry the massive GPUs or cloud computing resources often used by cutting-edge AI on Earth. Instead, Space AI must run on compact, efficient hardware (often generations behind Earth computing) and with limited electrical power. This greatly constrains model size and complexity. Techniques like model compression, neuromorphic chips, or specialized ASICs for neural networks are being explored to bring powerful AI to the “edge” in space. Moreover, communication bandwidth between Earth and spacecraft is limited, and latency can be significant (up to several minutes for Mars, and hours for deep-space probes). As a result, AI systems in space must make critical decisions autonomously on-board, rather than relying on real-time human input or cloud support. Designing algorithms that can operate robustly with limited data and without human intervention is an ongoing engineering pursuit. A further technical obstacle is \textbf{verification, validation, and safety assurance} of AI in mission-critical roles. Traditional spacecraft software goes through rigorous testing and formal verification to guarantee reliability. However, adaptive AI systems (e.g. machine learning-based) are often opaque (“black boxes”) and can behave unpredictably in novel situations. Ensuring an AI will act safely and as intended across the vast range of off-nominal scenarios in space is extremely difficult. Simulation and field testing for space conditions are limited – for instance, a Mars rover’s autonomous navigation AI can only be fully proven after facing the real Martian terrain. NASA’s Perseverance rover, for example, uses an advanced autonomous driving system (AutoNav) that allows it to traverse hazardous terrain much faster than previous rovers. This achievement underscores both the potential of AI and the years-long development and testing effort required to guarantee such a system works reliably on another planet. In fact, Perseverance’s autonomy had to be validated first on Earth under Mars-like conditions using a rover twin, and even then engineers had to carefully anticipate edge cases. The challenge going forward is to develop new methodologies (perhaps combining formal methods with extensive simulation and machine learning test regimes) to certify AI for use in life-critical and mission-critical applications, such as autonomous landing systems, spacecraft collision avoidance, or crew support robots. Scaling up the complexity of AI systems also introduces \textbf{systems engineering challenges}. Future missions may involve swarms of hundreds of small satellites or cooperative robots on a planetary surface, all governed by AI (sometimes termed \textit{swarm intelligence}). Managing coordination and communication in such multi-agent systems is non-trivial, especially under time delay. Ensuring robust performance of the collective – with no single point of failure – requires new algorithms for distributed decision-making and consensus. Additionally, the integration of AI with legacy spacecraft systems and protocols is a practical engineering issue: many current space missions use heritage software and hardware, so upgrading to AI-driven architectures will demand careful interfacing and incremental testing. In summary, the technical challenges for Space AI revolve around making AI \textbf{rugged, efficient, and trustworthy} under the unique constraints of space. Overcoming these will require innovation in hardware (radiation-proof AI accelerators), software (resilient and explainable AI algorithms), and mission design (architectures that gracefully handle autonomy), ensuring that AI systems can perform reliably over long-duration missions and distant frontiers. 

\subsubsection{Ethical, Social, and Societal Considerations}

Beyond engineering, Space AI faces profound \textbf{ethical and societal questions} about how autonomous systems should act and how their use will impact humans. As we delegate more decision-making to AI in space missions, issues of accountability and moral choice become critical. For example, an AI system controlling a crewed spacecraft or habitat may confront scenarios requiring value-laden decisions (such as how to prioritize life support in a crisis, or how to balance mission goals against astronaut safety). It is essential to embed human values and ethical principles into such autonomous agents. However, codifying ethics into AI – even on Earth – is still an open challenge, and doing so for systems operating millions of kilometers away with no real-time oversight adds extra complexity. Developing \textbf{ethical frameworks and oversight mechanisms} for autonomous space systems will be a key task. This might involve constraining AI behavior with strict rules (akin to Asimov’s laws of robotics, in spirit) or ensuring a human remains “in the loop” for critical decisions whenever possible, despite communication delays. There are also concerns about the \textbf{social impact on Earth} of increasing reliance on AI in space enterprises \cite{garrett2024artificial}. One consideration is the effect on the human workforce. As AI takes over functions like satellite piloting, data analysis, or even spacecraft design (through generative AI), certain jobs in the space industry may be transformed or made redundant. Routine spacecraft operations, for instance, could become highly automated, reducing the need for large teams of human controllers. While this can improve efficiency and lower costs, it necessitates retraining and upskilling programs to prepare the workforce for more AI-centric roles (such as supervising AI or focusing on strategy and creative tasks that AI cannot do). Societally, there is a risk of public resistance or skepticism toward autonomous systems if there are high-profile failures. A mission failure attributed to an AI’s mistake (for example, an autonomous lander making a poor decision) could erode trust in the technology. Therefore, transparency and \textbf{explainability} of AI decisions are important even in space: astronauts and mission controllers should be able to understand why an AI assistant or pilot made a certain choice. Efforts to develop explainable AI and to clearly communicate the role of AI in mission outcomes will help build public trust. Another ethical dimension is the \textbf{peaceful use of AI in space}. The Outer Space Treaty of 1967 declares that space shall be used for peaceful purposes and for the benefit of all humanity. With AI’s powerful capabilities, there is a dual-use dilemma: the same AI that optimizes satellite constellation operations could potentially be used to enable space-based weapons or autonomous military satellites. This raises the fear of an AI-driven arms race in space or the deployment of lethal autonomous spacecraft. The international community will need to remain vigilant that Space AI is developed and used in line with humanitarian principles and does not jeopardize global security. This overlaps with governance (discussed below) but also is an ethical stance: researchers and engineers in Space AI will have to consider the implications of their work and possibly adopt codes of ethics (similar to how AI researchers on Earth have begun adopting guidelines for responsible AI). Finally, we must consider \textbf{equity and societal benefit}: who will benefit from Space AI, and could it widen inequalities? If advanced Space AI is proprietary to a few wealthy nations or corporations, there is a risk that its benefits (such as satellite data or space-derived resources) might not be shared broadly. This could exacerbate the gap between space-faring elites and those who lack access to space technology. On the other hand, if managed well, Space AI could amplify the positive societal impacts of space activities – for example, by greatly improving climate monitoring, disaster response, and global communications for everyone. The challenge is ensuring inclusive access to these benefits. This may involve international policies to share certain data or to provide AI-enhanced services (like GPS, remote sensing data, communication bandwidth) to developing regions at low cost. In summary, the ethical, social, and societal domain challenges us to implement Space AI in a way that is aligned with human values, garners public trust, remains peaceful, and distributes its benefits justly among all people. 

\subsubsection{Regulatory and Governance Challenges}

\begin{table*}[htbp]
\small
\setlength{\tabcolsep}{3pt}
\centering
\caption{Overview of Major Government Space Agencies}
\label{tab:space_agencies}
\begin{tabular}{p{4.5cm}p{2.3cm}p{2.2cm}p{3.5cm}p{4.5cm}}
\toprule
\textbf{Agency Name} & \textbf{Country/ Region} & \textbf{Founding Date} & \textbf{Headquarters} & \textbf{Website} \\
\midrule
NASA (National Aeronautics and Space Administration) & United States & 1958 & Washington, D.C., USA & \url{https://www.nasa.gov} \\
European Space Agency (ESA) & Europe (Multinational) & 1975 & Paris, France & \url{https://www.esa.int} \\
UK Space Agency & United Kingdom & 2010 & Harwell, England & \url{https://www.gov.uk/uk-space-agency} \\
African Space Agency (AfSA) & African Union & 2023 & Cairo, Egypt & \url{https://africanspaceagency.org} \\
China National Space Administration (CNSA) & China & 1993 & Beijing, China & \url{http://www.cnsa.gov.cn} \\
Indian Space Research Organisation (ISRO) & India & 1969 & Bengaluru, India & \url{https://www.isro.gov.in} \\
Roscosmos State Corporation & Russia & 1992 & Moscow, Russia & \url{https://www.roscosmos.ru} \\
Japan Aerospace Exploration Agency (JAXA) & Japan & 2003 & Tokyo, Japan & \url{https://global.jaxa.jp} \\
Canadian Space Agency (CSA) & Canada & 1989 & Quebec, Canada & \url{https://www.asc-csa.gc.ca} \\
CNES (Centre National d'Études Spatiales) & France & 1961 & Paris, France & \url{https://cnes.fr} \\
DLR (German Aerospace Center) & Germany & 1969 & Cologne, Germany & \url{https://www.dlr.de} \\
Italian Space Agency (ASI) & Italy & 1988 & Rome, Italy & \url{https://www.asi.it} \\
Korea Aerospace Research Institute (KARI) & South Korea & 1989 & Daejeon, South Korea & \url{https://www.kari.re.kr} \\
UAE Space Agency (UAESA) & United Arab Emirates & 2014 & Abu Dhabi, UAE & \url{https://www.space.gov.ae} \\
Australian Space Agency & Australia & 2018 & Adelaide, Australia & \url{https://www.space.gov.au} \\
Brazilian Space Agency (AEB) & Brazil & 1994 & Brasília, Brazil & \url{https://www.gov.br/aeb} \\
Israel Space Agency (ISA) & Israel & 1983 & Tel Aviv, Israel & \url{https://www.space.gov.il} \\
\bottomrule
\end{tabular}
\end{table*}

The rapid emergence of AI in space activities has outpaced the development of laws and policies to govern it, creating a gap that poses significant challenges. Current \textbf{international space law} – principally the Outer Space Treaty and related agreements – did not anticipate AI-operated missions \cite{de2023space}. These treaties place responsibility for space activities on nation states (for instance, Article VI of the Outer Space Treaty holds states "internationally responsible" for national activities in space, including those of non-governmental entities, and Article VII along with the Liability Convention (1972) makes states liable for damage caused by their space objects). However, they are silent on how to handle autonomous decision-making systems. If an AI on an autonomous spacecraft makes a decision that results in an accident or violation (such as a collision causing debris or an incident on a celestial body), it is unclear how fault and liability would be assigned. The default is that the state launching the object is liable, but this might be seen as inadequate or unfair if the root cause was a complex AI behavior possibly provided by a private manufacturer from another country. This scenario calls for updated \textbf{liability frameworks} that clarify accountability for AI-driven actions. Determining “who is to blame” – the state, the operator, the AI developer, or the AI itself (as a legal entity) – is a novel question for lawyers and policymakers. Resolving it may require new international agreements or at least guidelines that specifically address AI. \textbf{Regulatory standards} for Space AI are also largely missing. On Earth, there is growing discussion of AI standards (for safety, ethics, interoperability), such as the EU’s proposed AI Act categorizing high-risk AI systems. For space, no equivalent standards body yet dictates how autonomous spacecraft systems should be designed or tested. This poses a challenge: without common standards, different nations or companies might deploy AI of variable quality and safety, potentially endangering the space environment (for example, an inadequately tested autonomous satellite could behave erratically and increase collision risk in orbit). Developing international standards or best practices – possibly under agencies like the International Organization for Standardization (ISO) or through space agencies’ coordination – will be important. These might include requirements for AI reliability, fail-safes (e.g., an “off switch” or manual override for autonomous spacecraft), communication protocols for autonomous coordination, and data sharing norms. A related governance issue is \textbf{space traffic management}. As satellite numbers grow (tens of thousands in mega-constellations), AI will likely be used to autonomously coordinate maneuvers to avoid collisions. However, without a governing body or agreed rules for traffic management, even AI-driven systems could conflict or make suboptimal choices. International governance is needed to set the rules of the road (or orbit) that all autonomous satellites must follow when, say, a collision risk is detected. Another governance challenge is balancing \textbf{innovation with regulation}. Over-regulating too early could stifle the rapid innovation in Space AI by adding heavy compliance burdens, whereas under-regulating increases the risk of accidents or misuse. Policymakers face the challenge of encouraging open experimentation (such as in-orbit AI demonstrations) while still safeguarding public interests. Sandbox environments or pilot programs, where companies can test AI satellites under regulatory waivers but with oversight, might be one solution. Additionally, export control laws and intellectual property issues come into play. Advanced AI algorithms or hardware might be classified as sensitive technology, restricting international collaboration. Governance frameworks will need to navigate these issues to allow the global exchange of ideas and tools in Space AI without compromising national security. Finally, we must consider \textbf{governance beyond Earth orbit}. As Space AI enables activities like lunar mining bases or Mars colonies, jurisdictional questions arise: Which laws govern an AI operating on the Moon? Who regulates a multi-national Mars settlement’s autonomous systems? These are currently unanswered questions. There is a push to develop governance regimes for the Moon (e.g., the Artemis Accords provide some principles for lunar activities), and AI will need to be part of that conversation. Earth-orbit governance (through bodies like the ITU for communications, or COPUOS for debris mitigation guidelines) has some structure, but for planetary bodies and deep space, new agreements may be needed. Proactively establishing \textbf{AI governance principles} – perhaps an international declaration on AI in space – could guide nations and companies in the interim. In summary, the regulatory and governance challenge is to modernize the “rules of space” so that they explicitly address autonomous AI systems. This will likely involve a combination of updating international treaties, creating standards and licensing requirements for AI in spacecraft, and fostering cooperation among spacefaring entities to enforce safe and responsible use of AI in the extraterrestrial domain. 

\subsection{Opportunities and Vision for Space AI}
Despite the challenges outlined above, the future of Space AI is ripe with extraordinary opportunities. As the technology and its governance mature, Space AI is poised to become a transformative force for scientific discovery, economic growth, and international cooperation. Indeed, many of the challenges have flipsides as opportunities: for instance, the need for reliability spurs innovation in robust AI algorithms, and the need for governance opens avenues for global collaboration. This section explores the promising opportunities in several domains – scientific and technological, economic and industrial, and cooperative governance – and culminates with a visionary outlook on Space AI as a critical endeavor for humanity’s long-term future. Embracing these opportunities, Space AI can significantly accelerate our expansion into the solar system and yield benefits that reverberate back to Earth. 

\subsubsection{Scientific and Technological Opportunities}
Space exploration and science stand to be revolutionized by the infusion of advanced AI. One of the foremost opportunities is the dramatic enhancement of \textbf{data analysis and discovery}. Modern space missions generate massive datasets: telescopes survey billions of stars and galaxies, Earth-observing satellites produce continuous high-resolution imagery, and planetary probes send back streams of sensor readings. AI systems (especially machine learning and pattern recognition algorithms) can sift through these vast datasets far more efficiently than humans, uncovering subtle patterns and new phenomena \cite{chen2020applications}. We are already seeing examples of this: machine learning has been used to identify new exoplanets in archival telescope data and to detect gravitational waves and fast radio bursts from noisy signals. In planetary science, AI algorithms can comb through rover images to spot interesting rock formations or signs of past water that scientists might overlook. The opportunity here is that AI becomes a kind of \textit{force multiplier} for scientists – accelerating the rate of discoveries and potentially making connections across data that lead to new knowledge. In the coming years, as missions like the James Webb Space Telescope, lunar orbiters, and Mars sample analysis generate more data than ever, AI will be indispensable to handle the deluge and extract meaningful insights. Another major opportunity is \textbf{autonomous exploration}. AI enables spacecraft and rovers to operate with greater independence and intelligence, which in turn allows exploration of environments that would be impractical with constant human control. For instance, future Mars rovers and planetary drones might use AI to dynamically plan their routes and science targets each day based on what they encounter, maximizing science return without waiting for Earth commands. A probe sent to the outer planets or an ocean-world moon (like Europa) could use onboard AI to make real-time decisions about which samples to analyze or which geysers to fly through, reacting to discoveries on the fly. Autonomy will be crucial for missions farther afield – imagine an orbiter around a distant moon that must manage its own observations during long communication blackouts. AI systems can also perform \textbf{onboard data processing} to send back only the most relevant information, effectively triaging data to cope with limited bandwidth. For example, an Earth observation satellite equipped with AI might analyze images in real-time to detect natural disasters (wildfires, floods) and immediately relay alerts and cropped images of those events, rather than dumping all raw data. This not only saves bandwidth but speeds up the response. In short, AI opens the door for \textbf{smarter missions} that can adapt and self-optimize, reaching targets and accomplishing tasks that would have been impossible under traditional manual control. AI’s integration with robotics offers transformative technological opportunities as well. We foresee \textbf{AI-powered robotic systems} constructing and maintaining infrastructure in space \cite{national2016nasa}. On the International Space Station and future stations, autonomous robotic arms (like the Canadarm descendants) could handle routine tasks and repairs, guided by computer vision and AI planning algorithms. On the Moon or Mars, autonomous robots may build habitats, lay power grids, or excavate soil with minimal human oversight, accelerating the establishment of bases. This is not science fiction – early steps are already being taken with experiments in automated in-situ construction and 3D printing on Earth analog sites. AI will be central to coordinating these complex construction tasks in unpredictable environments (adjusting for terrain, dust, etc.). Another domain is \textbf{in-space servicing and manufacturing}: AI-controlled servicing satellites could routinely refuel, repair, or upgrade other satellites, extending their lifespans and reducing space debris. The success of recent missions that autonomously docked with and serviced satellites demonstrates the feasibility of such operations. As the technology matures, we could see an ecosystem of self-directed service spacecraft keeping the orbital environment sustainable. Furthermore, AI-driven manufacturing units might one day fabricate components in orbit (for example, building large telescope mirrors or antennae that are too big to launch in one piece), unlocking new designs that were previously infeasible. The presence of AI reduces the need for constant human teleoperation, making these processes more efficient and continuous. Pushing the boundaries, AI will enable \textbf{new types of missions and experiments}. Consider the concept of a distributed sensor network across a planet or asteroid belt – hundreds of microprobes working in concert. Coordinating such a swarm for scientific measurements is a complex problem well-suited for AI swarm intelligence algorithms \cite{chao2025swarm}. Likewise, AI can help design optimal mission trajectories and spacecraft configurations by searching vast design spaces (a technique already glimpsed with AI-driven design optimization in engineering). Even in human spaceflight, AI provides opportunities: intelligent assistants for astronauts (such as the CIMON robot on the ISS) can help manage ISS experiments or monitor crew health and mood, potentially improving long-duration mission outcomes. All these technological opportunities emphasize that AI is not just a tool to do the same missions more efficiently; it will qualitatively change what missions we can conceive and execute. It allows us to tackle scientific questions at scale and complexity that were previously unmanageable. The synergy between space science and AI research will likely yield breakthroughs in both domains – with space providing challenging problems that advance AI, and AI providing capabilities that advance space exploration. 

\subsubsection{Economic and Industrial Opportunities}

\begin{table*}[htbp]
\small
\setlength{\tabcolsep}{3pt}
\caption{Global Overview of Private Space Companies (Part I)}
\label{tab:global_private_space_companies}
\centering
\begin{tabular}{p{2.1cm}p{2.1cm}p{1.2cm}p{3.2cm}p{3.2cm}p{4.2cm}}
\toprule
\textbf{Company} & \textbf{Country} & \textbf{Founded} & \textbf{Headquarters} & \textbf{Core Focus} & \textbf{Website} \\
\midrule
SpaceX & USA & 2002 & Starbase, Texas & Launch services, satellite internet, crewed flight & \url{https://www.spacex.com} \\
Blue Origin & USA & 2000 & Kent, Washington & Suborbital tourism, orbital launch, lunar landers & \url{https://www.blueorigin.com} \\
Rocket Lab & USA/NZ & 2006 & Long Beach, California & Small launch vehicles, spacecraft buses & \url{https://www.rocketlabusa.com} \\
Relativity Space & USA & 2015 & Long Beach, California & 3D-printed launch vehicles & \url{https://www.relativityspace.com} \\
Firefly Aerospace & USA & 2017 & Cedar Park, Texas & Launch vehicles, lunar payload delivery & \url{https://www.fireflyspace.com} \\
Axiom Space & USA & 2016 & Houston, Texas & Private space station, human spaceflight & \url{https://www.axiomspace.com} \\
Sierra Space & USA & 2021 & Louisville, Colorado & Spaceplanes, orbital habitats & \url{https://www.sierraspace.com} \\
Virgin Galactic & USA & 2004 & Tustin, California & Suborbital space tourism & \url{https://www.virgingalactic.com} \\
United Launch Alliance (ULA) & USA & 2006 & Centennial, Colorado & Government launch contracts & \url{https://www.ulalaunch.com} \\
Planet Labs & USA & 2010 & San Francisco, California & Earth observation satellites & \url{https://www.planet.com} \\
ICEYE & Finland & 2014 & Espoo, Finland & SAR Earth imaging satellites & \url{https://www.iceye.com} \\
Synspective & Japan & 2018 & Tokyo, Japan & Synthetic aperture radar satellites & \url{https://synspective.com} \\
Astroscale & Japan & 2013 & Tokyo, Japan & Orbital debris removal & \url{https://astroscale.com} \\
ispace & Japan & 2010 & Tokyo, Japan & Lunar landers and robotics & \url{https://ispace-inc.com} \\
\bottomrule
\end{tabular}
\end{table*}

\begin{table*}[htbp]
\ContinuedFloat
\small
\setlength{\tabcolsep}{3pt}
\caption{Global Overview of Private Space Companies (Part II)}
\centering
\begin{tabular}{p{2.1cm}p{2.1cm}p{1.2cm}p{3.2cm}p{3.2cm}p{4.2cm}}
\toprule
\textbf{Company} & \textbf{Country} & \textbf{Founded} & \textbf{Headquarters} & \textbf{Core Focus} & \textbf{Website} \\
\midrule
Skyroot Aerospace & India & 2018 & Hyderabad, India & Small satellite launch vehicles & \url{https://skyroot.in} \\
Pixxel & India & 2019 & Bengaluru, India & Hyperspectral Earth imaging & \url{https://www.pixxel.space} \\
Dhruva Space & India & 2012 & Hyderabad, India & Satellite platforms and integration & \url{https://www.dhruvaspace.com} \\
Kawa Space & India & 2019 & Mumbai, India & Geospatial intelligence from space data & \url{https://www.kawaspace.com} \\
OneWeb & UK & 2012 & London, UK & Satellite broadband constellation & \url{https://oneweb.net} \\
Alén Space & Spain & 2017 & Vigo, Spain & Custom CubeSat and nanosatellite design & \url{https://alen.space} \\
Isar Aerospace & Germany & 2018 & Munich, Germany & Launch services for smallsats & \url{https://www.isaraerospace.com} \\
Orbex & UK & 2015 & Forres, Scotland & Eco-friendly launch systems & \url{https://orbex.space} \\
LandSpace & China & 2015 & Beijing, China & Medium-lift launch vehicles & \url{http://www.landspace.com} \\
Galactic Energy & China & 2018 & Beijing, China & Small/medium launch vehicles & \url{https://www.galactic-energy.cn} \\
Space Pioneer & China & 2019 & Beijing, China & Liquid-fueled reusable rockets & \url{http://www.spacepioneer.cc} \\
SatRev & Poland & 2016 & Wrocław, Poland & Earth observation CubeSats & \url{https://satrevolution.com} \\
EnduroSat & Bulgaria & 2015 & Sofia, Bulgaria & Nanosatellite platforms & \url{https://www.endurosat.com} \\
Horizon Space Technologies & South Africa & 2018 & Johannesburg, South Africa & Low-cost CubeSats for emerging markets & \url{https://horizonst.co.za} \\
Orbital Space & Kuwait & 2018 & Kuwait City, Kuwait & Educational and experimental CubeSats & \url{https://www.orbital-space.com} \\
TIERRA & Argentina & 2021 & Buenos Aires, Argentina & Smart agriculture via CubeSat data & \url{https://www.tierrasat.com.ar} \\
\bottomrule
\end{tabular}
\end{table*}

The fusion of AI with space activities also carries immense economic promise. The global space industry is expanding rapidly (with estimates of its value approaching half a trillion USD in the 2020s) and is projected to reach into the trillions in the coming decades. AI can act as a catalyst for this growth by \textbf{lowering costs, increasing efficiency, and enabling new services}. One immediate economic impact is through the reduction of operational costs for space missions. AI-based automation can streamline satellite operations: satellites that adjust their own orbits, manage their power and thermal systems optimally, and diagnose problems without awaiting ground commands require far fewer human operators. This means a single ground controller might supervise dozens of satellites via an AI-assisted dashboard, as opposed to teams of engineers per satellite in the past. Launch operations and mission planning can similarly be optimized by AI, saving time (and thus money) in complex scheduling tasks. For example, AI algorithms are being used to choose optimal launch windows, detect anomalies in rocket sensor data, and even assist in autonomous flight termination systems for rockets. As a result, launch providers and satellite operators can achieve higher cadence and reliability at lower cost, making space ventures more economically viable. AI is also an \textbf{enabler of new commercial space services}. One burgeoning area is Earth observation analytics: companies are deploying constellations of imaging satellites and using AI to process the imagery into actionable intelligence for customers (in agriculture, finance, urban development, climate monitoring, etc.) \cite{tuia2024artificial}. Here, AI is the differentiator that turns raw data into value – e.g., identifying crop health and predicting yields, or monitoring supply chains and natural resource use in near-real-time. This kind of high-frequency, AI-driven insight was not possible a decade ago and represents a significant market. Another emerging service is satellite communications optimized by AI. As constellations like SpaceX’s Starlink or OneWeb grow, AI is used to dynamically allocate network resources, steer digital beams, and route internet traffic in space, maximizing throughput and user coverage. This improves the performance of satellite internet, which could connect billions of new users and thus create massive economic impact. Additionally, navigation services (like GPS/GNSS) can be augmented with AI for better accuracy and reliability, supporting the expanding ecosystem of autonomous vehicles on Earth. AI might also allow entirely new kinds of space-based businesses: for example, real-time hyperspectral imaging combined with AI could enable detailed mineral exploration or environmental compliance monitoring from orbit, services that industries on Earth would pay for. Perhaps the most visionary economic opportunities lie in \textbf{resource utilization and space infrastructure}, which AI will be key to unlocking. The concept of asteroid mining or lunar resource extraction has long been discussed as a trillion-dollar opportunity – harvesting metals, water ice, or Helium-3 fuel from space. The logistical challenges and costs are enormous, but AI-controlled robotic miners and processing facilities could make it feasible. By operating autonomously on the Moon’s surface or an asteroid, AI systems could extract resources continuously without needing a human presence, dramatically reducing labor costs and safety risks. This could jump-start an off-planet mining industry, supplying materials for in-space construction (fuel, building materials) or even bringing precious metals back to Earth’s markets. In-space resources, in turn, would facilitate further economic activity (creating a positive feedback loop): for instance, readily available propellant from the Moon could refuel satellites or rockets in Earth orbit, spawning fuel depot services and lowering the cost for spacecraft to reach higher orbits or deep space. All of these complex operations rely on advanced AI for autonomy and coordination, since real-time human control in such environments is impractical. AI is also transforming the \textbf{manufacturing and design process} in the space industry. Companies are using AI-driven generative design to create lighter and stronger rocket and spacecraft components, which are then often 3D-printed. This cuts down development time and can produce innovative designs that human engineers might not have conceived. In satellite manufacturing, AI and robotics together enable assembly-line style production (as seen in some small satellite “factories”), increasing throughput and reducing unit cost. Economically, this means space hardware is no longer bespoke and exorbitantly expensive, but rather more standardized and affordable, allowing more actors to participate in space ventures. Furthermore, improved reliability from AI diagnostics can reduce insurance costs for launches and satellites (a significant factor in commercial missions). The reduction in risk and cost opens the door for startups and emerging countries’ space programs, democratizing economic participation in space. Finally, the integration of Space AI could lead to \textbf{economic spillovers on Earth}. Technologies and solutions originally developed for autonomous spacecraft or extraterrestrial habitats might find lucrative applications in terrestrial industries. For example, algorithms for managing life support systems in a closed lunar base could be applied to smart building systems on Earth for energy savings. Robotics developed for planetary exploration could be repurposed for mining, agriculture, or disaster response on Earth (extending the value of that R\&D). Even the discipline of operating complex AI-driven systems remotely (like a Mars rover) teaches lessons in human-machine collaboration and remote operations that can translate to sectors like offshore energy or undersea exploration. These spillovers amplify the return on investment in Space AI beyond the space sector itself. In summary, the economic and industrial opportunity presented by Space AI is multifaceted: it makes current space activities cheaper and more effective, it enables entirely new industries such as space resource utilization and mega-constellation services, and it yields innovations that benefit Earth’s economy. As such, Space AI will be a key driver in the growth of the space economy, helping to turn ambitious visions (like a thriving cislunar economy or routine interplanetary travel) into plausible and sustainable enterprises. 

\subsubsection{International Collaboration and Governance}

Space has always been an arena that prompts international cooperation – from the Apollo-Soyuz handshake to the International Space Station – and Space AI offers new avenues to strengthen global collaboration. One major opportunity is using AI as a \textbf{tool for shared global challenges}, encouraging countries to work together. A prime example is space debris mitigation and \textbf{space traffic management}: the proliferation of satellites and debris affects all spacefaring nations, and AI will be essential to monitor objects in orbit, predict conjunctions (near-collisions), and coordinate avoidance maneuvers. No single country can solve this issue alone; thus, there is strong incentive for nations to pool tracking data and develop common AI systems to manage orbital traffic safely. Collaborative projects could create shared AI-driven catalogs of orbital objects or joint alert systems, fostering transparency and trust. Similarly, planetary defense (detecting and potentially deflecting hazardous asteroids) is an inherently international concern – AI is used to search astronomical data for new near-Earth objects and to model impact probabilities, and any mission to deflect an asteroid would likely involve multiple space agencies. Working together on AI for such missions can build international goodwill and ensure that all of humanity is protected. Another collaborative opportunity lies in \textbf{scientific data sharing and research}. AI thrives on data, and the more diverse and rich the dataset, the better the insights. Different countries’ space missions produce complementary datasets – for instance, one country’s Earth observation satellite may excel at radar imaging while another’s provides hyperspectral data. By sharing data and applying AI jointly, scientists worldwide can gain a more complete picture, for example, of climate change patterns or disaster events. International consortia could develop AI platforms that allow researchers from any nation to contribute algorithms and analyze combined datasets from many satellites and space telescopes. This open approach would accelerate scientific progress and ensure that even nations without large space programs can benefit from global space data (they could contribute expertise in AI instead of hardware). We already see precursors to this in global efforts like the Group on Earth Observations (GEO), which encourages sharing of Earth-monitoring data. Incorporating AI into such frameworks – perhaps an international “AI lab” for space data – could be a game-changer in how science is done, making it more inclusive and efficient. In terms of governance, the necessity to address Space AI challenges can itself become an opportunity for \textbf{strengthening international law and institutions}. Nations could come together to establish guidelines or even formal agreements on the responsible use of AI in space. For example, building on the Artemis Accords (which outline principles for peaceful Moon exploration), countries might agree on provisions specific to AI – such as transparency about autonomous systems on spacecraft, sharing of lessons learned from AI anomalies, or commitments not to weaponize AI in orbit. Crafting these norms collaboratively can reduce suspicion and prevent misunderstandings in space operations. Moreover, involving a broad set of stakeholders (established and emerging space nations, private sector, academia) in dialogues about Space AI governance ensures the resulting frameworks are more widely accepted. This inclusive process could rejuvenate international forums like the United Nations Committee on the Peaceful Uses of Outer Space (COPUOS), expanding their mandate to cover cutting-edge issues like AI. In doing so, humanity has a chance to proactively govern a technology before it becomes a source of conflict. If successful, it could serve as a template for other global AI governance efforts, showing that even great-power rivals can find common ground in managing powerful technologies in shared domains like outer space. Collaboration can also accelerate the \textbf{development of Space AI itself} by pooling expertise and resources. Not every nation has specialized experts in both aerospace and AI; partnerships allow cross-pollination of knowledge. For instance, a country with advanced AI research might partner with a country with a lunar rover program to develop the rover’s autonomy together. This way, each side contributes their strength and both benefit from the outcome. International challenge programs or competitions (similar to the DARPA challenges, but international) could be organized to solve specific Space AI problems – for example, an international competition to design the best AI for a Mars aerial drone, with the winning algorithms tested on a real mission. Such friendly competitions promote innovation and share results openly. Additionally, working together on Space AI can distribute the costs of research. Large endeavors, such as building a simulated Moon base on Earth to test AI systems for habitat management, might be too expensive for one agency but feasible as a joint project among many. Lastly, Space AI in the civilian realm can indirectly foster \textbf{peace and diplomacy}. Joint missions and AI projects build relationships and interdependence, making conflict less likely. If satellite collision avoidance systems are interconnected globally, for example, countries will communicate regularly about orbital movements, increasing transparency and reducing the chance of accidental confrontations in space. In a hopeful scenario, space could remain a largely cooperative domain even as AI technology advances, precisely because the complexity of managing AI-driven activities will push nations to communicate more, not less. In conclusion, the challenges of regulating Space AI may turn into an opportunity to enhance international governance mechanisms, and the technology itself offers a powerful incentive and means for countries to collaborate on shared goals. Through these collaborations, Space AI could become a unifying endeavor, much like the ISS has been – but on an even larger scale, integrating not just space agencies but also AI institutes, companies, and international organizations all working together for the betterment of humanity’s presence in space. 

\subsubsection{Vision for Space AI as a Critical Human Endeavor}

Looking ahead, Space AI is poised to become not just a field of research and development, but a \textbf{pillar of humanity’s future}, deeply entwined with our long-term survival and prosperity. In this final vision, we consider Space AI as a critical human endeavor – one that could shape the trajectory of our civilization much as fire, agriculture, or electricity did in the past. The ambition driving Space AI is more than solving technical problems; it is about expanding the human horizon. With AI as our partner, the species stands on the cusp of a new era of exploration and growth, from settling other planets to safeguarding our home world. A core aspect of this vision is enabling a \textbf{multi-planetary and interplanetary existence}. Human exploration of the Moon, Mars, and beyond will heavily rely on intelligent systems. In the coming decades, we can imagine AI-managed lunar bases where autonomous systems run life support, grow food in hydroponic farms, and mine water ice for fuel – creating self-sustaining outposts that humans can periodically visit and eventually inhabit. On Mars, before the first astronauts arrive, fleets of AI-controlled robots might have already built landing pads, habitats, and power stations. Such a scenario vastly increases the safety and feasibility of human missions, as the heavy groundwork is done by machines that can operate 24/7 in harsh conditions. As the famed technologist Elon Musk and others have championed, becoming a multi-planetary species is insurance for humanity’s long-term survival. Space AI is the linchpin to make that possible: without advanced autonomy, maintaining a habitat on Mars (with its thin atmosphere, radiation, and distance from Earth) would be nearly impossible. Furthermore, as we contemplate sending humans further out – perhaps one day to the moons of Jupiter or Saturn – the communications delay and extreme environment mean those missions will only be viable if AI systems can handle most contingencies independently. In essence, wherever humans go, AI will be there as an indispensable support, and wherever humans cannot yet go, AI will go first as our vanguard. Even uncrewed exploration will reach astounding heights with AI. We might send \textbf{interstellar probes} equipped with AI to neighboring star systems. Given the immense distances (years or decades of travel), these probes must be highly autonomous scientists, able to make complex decisions on their own and even reprogram themselves as needed. They would be, in a sense, extensions of humanity – carrying our intellect in silicon form far beyond our physical reach. Such probes could identify exoplanets, analyze their atmospheres for signs of life, and perhaps send back unprecedented knowledge that reshapes our understanding of life in the universe. This is a bold vision, but not out of the question: initiatives like Breakthrough Starshot are already conceptualizing sending ultrafast probes to Alpha Centauri. Marrying those concepts with AI capabilities suggests a future where humanity, through AI proxies, becomes an interstellar explorer. In this way, Space AI contributes to fulfilling one of humanity’s oldest dreams – to know what lies beyond the horizon – on a galactic scale. Back on Earth, the grand endeavor of Space AI will yield profound \textbf{benefits for our civilization}. The technologies developed to survive on other planets (closed-loop life support, solar energy storage, radiation shielding, AI-driven healthcare in isolated environments) can directly help us build a more sustainable and resilient society on our own planet. For instance, managing resources efficiently in a Mars base parallels the challenge of living sustainably on Earth; AI optimizations in that context could translate to better renewable energy management or recycling systems at home. Moreover, the vantage point of space, enhanced by AI, will allow us to tackle global problems with greater efficacy. Continuous Earth observation with AI analysis can help us mitigate climate change by tracking emissions, monitor ecosystems to prevent biodiversity loss, and improve disaster preparedness by spotting early signs of hurricanes or earthquakes. In effect, Space AI acts as a guardian for Earth, enabling a level of planetary management and insight that was previously unattainable. Thus, even those who never go to space will experience tangible improvements in quality of life thanks to advances in Space AI. Considering the societal and inspirational dimension, Space AI as a collective enterprise can \textbf{unite humanity and inspire future generations}. Much like the Apollo era inspired a generation of scientists and engineers, the quest to develop AI that can explore the oceans of Enceladus or navigate the caves of Mars can captivate minds around the world. It blends the allure of space – the final frontier – with the cutting-edge excitement of AI. Educational programs and international STEM initiatives can rally around this theme, recognizing that it sits at the intersection of multiple disciplines (computer science, aerospace engineering, planetary science, ethics, etc.). The inclusivity of this field means a student in any country, even those without a space agency, can contribute via software and AI to humanity’s space endeavors. Over time, as the fruits of Space AI become apparent (new worlds mapped, new resources found, Earth’s environment better managed), public support for science and exploration is likely to grow. Space AI could thus play a role in fostering a more scientifically literate and future-focused global culture. Finally, there is a philosophical aspect to viewing Space AI as a critical human endeavor. It compels us to confront the relationship between human intelligence and artificial intelligence in a very profound way. By pushing AI to its extremes – to operate in alien places, to act as our surrogate in environments we ourselves cannot survive – we are effectively forging a new tool of human agency. We will need to imbue these AIs with our aspirations and our caution, trusting them yet setting boundaries. In doing so, we learn more about ourselves: our values, our ingenuity, our willingness to explore the unknown. If one day an AI agent on a distant world discovers evidence of alien life or even just sends back a breathtaking panorama of an unexplored landscape, that moment will belong to all of humanity, achieved by our extended “mind” in the form of AI. It will show that our ability to transcend our limitations is not only through physical evolution but through technological innovation guided by wisdom. In conclusion, the future of Space AI is as challenging as it is exhilarating. It will require seriousness of purpose, global cooperation, and steadfast commitment to ethical principles to navigate the difficulties. But the payoff is enormous: a deeper scientific understanding of the universe, a thriving space economy linking Earth and the heavens, a more united international community, and a pathway to becoming a multi-planetary species. Space AI stands at the forefront of human progress – a critical endeavor that could well define the 21st century and beyond. By mastering Space AI, we are not only solving technical problems; we are taking purposeful strides toward a future where humanity’s presence expands into the solar system (and eventually the stars) in a sustainable and beneficial manner, with artificial and human intelligence working hand in hand. This vision of Space AI as a bridge between Earth and the cosmos, between present and future, underscores its transformative potential. It is an endeavor worthy of our highest ambitions, one that carries the promise of securing humanity’s long-term survival and unlocking wonders that today we can only imagine.

{
\bibliographystyle{ieee_fullname}
\bibliography{egbib}
}

\end{document}